\documentclass[twocolumn,floatfix,prb,aps,showpacs]{revtex4}
\usepackage{graphicx,amsmath,amssymb,color}
\usepackage{nicefrac}
\usepackage[titletoc,title]{appendix}

\newcommand{\be}{\begin{equation}}
\newcommand{\ee}{\end{equation}}

\newcommand{\ba}{\begin{eqnarray}}
\newcommand{\ea}{\end{eqnarray}}

\begin{document}

\title{Phase Diagram of Fractional Quantum Hall Effect of Composite Fermions in Multi-Component Systems}
\author{Ajit C. Balram,$^1$ Csaba T\H oke,$^2$ A. W\'ojs,$^3$ and J. K. Jain,$^1$}

\affiliation{
   $^{1}$Department of Physics, 104 Davey Lab, Pennsylvania State University, University Park, PA 16802, USA}
\affiliation{
   $^{2}$BME-MTA Exotic Quantum Phases ``Lend\"ulet" Research Group, Budapest University of Technology and Economics,
Institute of Physics, Budafoki \'ut 8., H-1111 Budapest, Hungary}
\affiliation{
   $^{3}$Department of Theoretical Physics, Wroclaw University of Technology, Wybrzeze Wyspianskiego 27, 50-370 Wroclaw, Poland}

\begin{abstract} 
While the integer quantum Hall effect of composite fermions manifests as the prominent fractional quantum Hall effect (FQHE) of electrons, the FQHE of composite fermions produces further, more delicate states, arising from a weak residual interaction between composite fermions. We study the spin phase diagram of these states, motivated by the recent experimental observation by Liu {\em et al.}  \cite{Liu14a,Liu14b} of several spin-polarization transitions at 4/5, 5/7, 6/5, 9/7, 7/9, 8/11 and 10/13 in GaAs systems. We show that the FQHE of composite fermions is much more prevalent in multicomponent systems, and consider the feasibility of such states for systems with ${\cal N}$ components for an SU(${\cal N}$) symmetric interaction. Our results apply to GaAs quantum wells, wherein electrons have two components, to AlAs quantum wells and graphene, wherein electrons have four components (two spins and two valleys), and to an H-terminated Si(111) surface, which can have six components.  The aim of this article is to provide a fairly comprehensive list of possible incompressible fractional quantum Hall states of composite fermions, their SU(${\cal N}$) spin content, their energies, and their phase diagram as a function of the generalized ``Zeeman" energy.  We obtain results at three levels of approximation: from ground state wave functions of the composite fermion theory, from composite fermion diagonalization, and, whenever possible, from exact diagonalization. Effects of finite quantum well thickness and Landau level mixing are neglected in this study. We compare our theoretical results with the experiments of Liu {\em et al.}  \cite{Liu14a,Liu14b} as well as of Yeh {\em et al.} \cite{Yeh99} for a two component system.
\end{abstract}
\maketitle

\section{Introduction}

The fractional quantum Hall effect\cite{Tsui82} (FQHE) is one of the most remarkable phenomena in condensed matter physics arising from interelectron interactions. It refers to the observation, in two dimensional electron systems exposed to a strong magnetic field, of precisely quantized plateaus in the Hall resistance at $R_{\rm H}=h/fe^2$, and associated minima in the longitudinal resistance at filling factors $\nu=f$. A large number of fractions have so far been observed: $\sim$70 in the lowest Landau level (LL) and $\sim$15 in the second LL. The number of FQHE states is even larger, because FQHE states with different spin polarizations have been seen at many of these fractions \cite{Eisenstein89,Du95,Kukushkin99,Kukushkin00,Melinte00,Tiemann12,Yeh99,Liu14a}. In recent years, FQHE has been observed in systems with valley degeneracies, such as graphene \cite{Xu09,Bolotin09,Dean11,Feldman12,Feldman13,Amet14}, AlAs quantum wells \cite{Bishop07,Padmanabhan09}, and  an H-terminated Si(111) surface\cite{Kott14}, further adding to the richness of the phenomenon. These systems allow, in principle, the possibility of FQHE states that involve more than two components.  [We note that four component physics can also be accessed in wide quantum wells of GaAs with high electron densities, where LLs belonging to different subbands can cross one another. This allows formation of four-component (two subbands and two spins) FQHE states, as reported in Ref.~\onlinecite{Shabani10} However, these systems do not satisfy the SU(${\cal N}$) symmetry because the interaction is not subband index independent. Therefore, our results below, which assume SU(${\cal N}$) symmetry, are not directly applicable to these experiments.]

The prominent features of the FQHE are understood in terms of weakly interacting composite fermions \cite{Jain89,Lopez91,Halperin93,Jain07,Heinonen98,Jain98,Lopez07,Simon98,DasSarma07,Halperin07,Stormer07,Jain00,Halperin03,Smet98a,Murthy03}. Composite fermions (CFs) are topological bound states of electrons and an even number of quantized vortices. They experience a reduced effective magnetic field, and form Landau-like levels called $\Lambda$ levels ($\Lambda$Ls). Composite fermions carrying $2p$ vortices are denoted $^{2p}$CFs. The integer quantum Hall effect (IQHE) of weakly interacting composite fermions manifests as FQHE at fractions of the form $\nu=n/(2pn\pm 1)$ and their hole conjugates, which are the prominently observed fractions. The CF theory also provides an account of the spin physics of the FQHE. Specifically, it predicts the possible spin polarizations at each fraction, and also the critical Zeeman energies where transitions between differently spin polarized states are expected to occur\cite{Wu93,Park98,Park98b,Park99,Park01,Davenport12}. The measured spin polarizations and the phase transitions as a function of the Zeeman energy \cite{Eisenstein89,Du95,Yeh99,Kukushkin99,Kukushkin00,Melinte00,Tiemann12,Feldman12,Feldman13} or the valley splitting \cite{Bishop07,Padmanabhan09} are in satisfactory agreement with theory. The values of critical Zeeman energies depend on very small energy differences between the competing states, and thus serve as a sensitive test of the quantitative accuracy of the CF theory. The IQHE states of composite fermions for an SU(4) system have also also been studied \cite{Toke07,Toke07b,Goerbig07,Sodemann14,Abanin13}. 

This article deals with the physics beyond the IQHE of composite fermions, namely the FQHE of composite fermions,  which arises as a result of the weak residual interaction between composite fermions. The CF FQHE states are much more delicate, and more readily obscured by disorder and temperature, than the IQHE states of composite fermions. This is analogous to the situation for electrons, where only the IQHE states would be seen for non-interacting electrons, but interelectron interactions cause further structure that appears in the form of the FQHE. Many CF FQHE states in the vicinity of $\nu=1/3$, e.g. at 4/11, 5/13, were observed by Pan {\em et al.} \cite{Pan03}. These motivated theoretical studies of FQHE of fully spin polarized composite fermions \cite{Wojs04,Mukherjee14} as well as partially spin polarized FQHE of composite fermions  \cite{Park00b,Chang03b,Wojs07}. The spin polarization of these states has not been measured experimentally so far, however.

The primary motivation for our theoretical study presented in this paper comes from the recent experiment of Liu {\em et al.},\cite{Liu14a} who have observed spin-polarization transitions for several CF FQHE states in the filling factor region $2/3<\nu<4/3$, specifically for the FQHE states at 4/5, 5/7, 6/5, 9/7, 7/9, 8/11 and 10/13, as a function of the Zeeman energy.  (It is worth clarifying a point here taking the example of $\nu=4/5$. The fully spin polarized FQHE state at this fraction can be understood either as the $\nu^*=4/3$ FQHE state of $^2$CFs or as the $\nu^*=1$ IQHE state of $^4$CFs made of holes in the lowest LL (LLL) -- these interpretations are equivalent, in the sense that the states occur at the same quantum numbers and the actual wave functions obtained from the two interpretations are practically identical. However, the non-fully spin polarized states at $\nu=4/5$ can only be understood in term of $\nu^*=4/3$ FQHE of $^2$CFs. This is discussed in more detail in Ref.~\onlinecite{Liu14a} and below.)  With this understanding, certain spin polarization transitions observed previously by Yeh {\em et al.}\cite{Yeh99}, whose origin was not understood at the time, can also be explained in terms of FQHE of composite fermions.  Given ongoing improvements in experimental conditions as well as availability of new two-dimensional electron systems that exhibit FQHE, we have undertaken an exhaustive study of FQHE of composite fermions in multi-component systems. 

Specifically, this article reports on: 

\begin{enumerate}

\item A fairly exhaustive enumeration of FQHE states of composite fermions for multicomponent systems; 

\item Thermodynamic energies of many prominent states; and 

\item Critical values of the ``Zeeman" energies where transitions between different states are predicted to take place (i.e. the phase diagram of the CF FQHE states).

\end{enumerate}

\noindent (The term Zeeman energy is used in a general sense here as an energy that introduces a preference for one of the components.) We have also included, for completeness, some previously known results.

Interestingly, the FQHE of composite fermions is more prevalent for multicomponent systems, for reasons that can be understood as follows. For a single component system, the FQHE of composite fermions occurs, typically, in the second or higher $\Lambda$ levels, where very few states can be stabilized. (The FQHE of $^2$CFs in the lowest $\Lambda$ level can generally be understood as IQHE of $^4$CFs.) With multiple components, it becomes possible to consider states in which $^2$CFs form an IQHE state in one or more components, but a FQHE state in the {\em lowest} $\Lambda$L in one of the components; such a state does not lend itself to an interpretation as an IQHE of composite fermions. Many FQHE states of composite fermions thus become available in multicomponent systems.

The spin physics of FQHE state can be studied most conveniently through variations in the Zeeman energy, which causes transitions between these states. Such spin transitions between the CF-FQHE states provide an extremely rigorous test of our theoretical understanding of the FQHE, and in particular, of the residual interaction between composite fermions. At a qualitative level, such experiments can confirm if the number of available states is consistent with that expected from the CF theory. Further, the actual values of the critical Zeeman energies where spin transitions occur depend sensitively on the very small energy differences between the two competing states with different spin polarizations, and thus constitute a quantitative test of the theory. In many cases only a small fraction of composite fermions flip their spins at the transition, which requires multiplying the energy difference by a large integer (e.g. 4 for the transition from partially polarized to a fully polarized state at $\nu=4/11$) to obtain the critical Zeeman energy, which further enhances the impact of any error in the theoretical energy difference. 

We make many simplifying assumptions in our study. We assume an SU(${\cal N}$) symmetric interaction. A Zeeman-type term can be added straightforwardly. Our considerations allow for a spontaneous breaking of the SU(${\cal N}$) symmetry, but do not consider any an interaction that explicitly breaks the SU(${\cal N}$) symmetry.  We have not included finite width, LL mixing or disorder, mainly because of the large parameter space. It has been shown elsewhere that these make significant corrections to the critical Zeeman energies \cite{Liu14}, because the critical Zeeman energies depend sensitively on the rather small energy differences between various incompressible states. These corrections should be considered for specific experimental parameters whenever a detailed comparison is sought, but our results at least enumerate the various possible states and an estimate for where transitions between them are expected. 

We obtain results from the CF theory at two levels of approximation. In the zeroth order approximation, we construct a single wave function for the ground state, which we refer to as the ``CF wave function," and evaluate its average Coulomb energy. In a more accurate approximation, we perform diagonalization within a small basis of CF basis functions to obtain the ground state energy; this is referred to as ``CF diagonalization."  These methods are described in somewhat greater detail below. We have also given results from exact diagonalization studies wherever possible. A comparison between the CF and the exact results also gives an idea of the reliability of the predictions of the CF theory. For many fractions, the wave function from the CF theory is very difficult to evaluate for technical reasons, because of the need for {\em reverse} flux attachment (for which we do not have an equally accurate method). In such cases, we draw our quantitative conclusions primarily from exact diagonalization studies (these studies do confirm the spin and the angular momentum quantum numbers of the incompressible FQHE states predicted by the CF theory). 

For single component FQHE, particle-hole symmetry relates $\nu$ to $1-\nu$. For an ${\cal N}$ component system, particle-hole symmetry relates $\nu$ to ${\cal N}-\nu$. We therefore only give results for fractions up to ${\cal N}/2$. It was predicted that no spin transitions occur \cite{Park99} (and none have been seen) at fractions of the form $n/(4n+1)=$1/5, 2/9, {\em etc.}, but we see below that spin transitions are possible (and some have been seen) for fractions of the form 4/5, 5/9 {\em etc.} There is no contradiction, because the states at $n/(4n\pm 1)$ and $1-n/(4n\pm 1)$ are not related by particle-hole symmetry unless they are both fully spin polarized. 

We compare our results to available experiments. Spin transitions for CF FQHE states at $\nu=4/11$, $5/13$ \cite{Pan03} have not yet been observed directly, but indirect information on that issue is available from Raman experiments \cite{Balram14} which show a change in the character of the excitations that can be associated with a change in the spin polarization of the ground state. We compare our theoretical results with the spin polarization transitions observed at 4/5, 5/7, 6/5, 9/7, 7/9, 8/11 and 10/13 \cite{Liu14a,Liu14b,Yeh99} in Section \ref{six}, and find that the measured critical Zeeman energies are in reasonable agreement with those predicted theoretically in most cases. A remaining puzzle is the experimental evidence \cite{Liu14a} for two transitions at 5/7 and 9/7, even though only a single transition is expected in the simplest theoretical model in which one allows for FQHE only in a single component of composite fermions. (States in which FQHE occurs in two or more components of composite fermions are expected to be weaker and are not considered in this paper; the only ``double" CF FQHE states considered are 5/13 and 5/7, discussed in the Appendix.)

We have not considered excitations in this work. As seen in previous studies, an enormously rich structure is obtained when excitations of composite fermions across different components is considered \cite{Mandal01,Majumder09,Rhone11,Wurstbauer11,Toke12,Balram14,Majumder14}. We will only be interested in the nature of the ground state and the phase diagram as a function of parameters. 

The plan of our paper is as follows. In Section \ref{two}, we list the large number of incompressible states predicted by the CF theory. We carefully define a unique notation of possible FQHE states of composite fermions, and give the corresponding wave function. All of the states constructed here satisfy the Fock conditions \cite{Hamermesh62}. Section \ref{three} lists all of the states that have been studied previously, along with the original references. Sections \ref{four} and \ref{ED} give an outline of the methods of exact and CF diagonalization, respectively, which have been used extensively in the our calculations, followed by Sec.~\ref{dens-corr} that lists some technical details. Section \ref{five} discusses many specific states, listing all possible ``spin" polarizations at numerous fractions. Section \ref{six} mentions the experimental status of many of these states, and we conclude in Section \ref{seven}. 

For convenience of the readers who are not interested in the details of the calculations but only in the final results, we note that the theoretical phase diagrams for the FQHE of two-component composite fermions are shown below in Fig.~\ref{fig:2}. Figure~\ref{fig:3} contains a summary of the experimentally observed transitions for various CF FQHE states, along with the theoretical predictions at zero width. These figures summarize some of the most important results obtained in our article for two component systems. We note that some of these results are slightly different from those reported in the earlier literature using the same calculations; the difference arises because we extrapolate the energies in this article after performing the so-called density correction to the finite system energies (see Sec.~\ref{dens-corr}), which we believe provides more accurate numbers.

\section{Fractional quantum Hall effect of composite fermions}
\label{two}

We illustrate the construction of various possible FQHE states of composite fermions in this section. We find it convenient to use two notations to denote a FQHE state. The notation
\be
(\nu_1, \nu_2, \cdots \nu_{{\cal N}})
\ee 
with 
\be
\nu=\sum_{\alpha=1}^{\cal N}\nu_{\alpha}
\ee
displays the occupation of each spin component, where the word ``spin" is used in a general sense; only the non-zero occupations will be shown, and the convention $\nu_1\geq \nu_2 \geq \cdots \nu_{\cal N}$ will be assumed.  While this is a convenient notation for reading off the ``spin" polarization (denoted by $\gamma$), it is important to remember that this notation does not specify the actual state. In particular, one must take care to remember that this is in general {\em not} a product state of the type $\Psi_{\nu_1}\otimes \Psi_{\nu_2}\otimes \cdots \otimes \Psi_{\nu_{\cal N}}$; such a state is, in general, not a valid state for a system with SU(${\cal N}$) symmetry because it does not satisfy the so-called Fock conditions. The actual state is much more complex, and, in some cases, more than one possible state can produce the same spin content at a given filling factor. For that reason, we introduce another notation, defined below, which will precisely specify the CF structure of the state. We will see below how to combine IQHE and FQHE states of composite fermions in such a manner that the resulting wave function has proper SU(${\cal N})$ symmetry and consequently conforms to the Fock conditions.

\subsection{IQHE of composite fermions: $\Lambda$ levels inside Landau levels}

The simplest states are IQHE states of composite fermions of the form 
\be
[n_1, n_2,n_3,\cdots]_{\pm 2p} \leftrightarrow \left({n_1\over n} \nu,  {n_2\over n} \nu, \cdots \right)
\label{IQHE}
\ee
at the Jain fractions
\be
\nu={n\over 2pn\pm 1},\; n\equiv \sum_{\alpha}n_{\alpha}
\label{Jain_sequence}
\ee
which are obtained by attaching $2p$ vortices to each electron in the IQHE states $[n_1, n_2,n_3,\cdots]$. The $+$ ($-$) sign denotes vortex attachment in the same (opposite) direction as that of the external magnetic field and this process is termed ``parallel flux attachment'' (``reverse flux attachment'').  We assume $n_1\geq n_2\geq \cdots$, which can always be arranged by a relabeling of the component index. The Jain wave function for this $\mathcal{N}$-component state of $N$ electrons is given by 
\be
\Psi_{[n_1, n_2,n_3,\cdots]_{2p}}=J^{2p-2}{\cal P}_{\rm LLL} \prod_{\alpha=1}^\mathcal{N} \Phi^{\alpha}_{n_{\alpha}} J^{2}
\label{paral-flux}
\ee
and 
\be
\Psi_{[n_1, n_2,n_3,\cdots]_{2p}}=J^{2p-2}{\cal P}_{\rm LLL} \prod_{\alpha=1}^\mathcal{N} \left[\Phi^{\alpha}_{n_{\alpha}}\right]^* J^{2}, 
\label{rev-flux}
\ee
where 
\be
J= \prod_{1\leq j<k \leq N}(z_j-z_k),
\ee
$\Phi^{\alpha}_{n_{\alpha}}$ is the Slater determinant of $n_\alpha$ filled LLs for electrons in the $\alpha^{\text{th}}$ sector, and $z_j$ is the coordinate of the $j^{\text{th}}$ electron. These are straightforward generalizations of the Jain wave functions for the single component FQHE states \cite{Jain89}. We use the Jain-Kamilla method \cite{Jain97,Jain97b} for performing the LLL projection.

The validity of these wave functions has been ascertained by comparison to exact diagonalization studies which can produce reliable numbers for certain simple FQHE states. For parallel flux-attached Jain states, the above wave functions produce critical Zeeman energies that are  accurate at the level of 10-15\%. For the reversed-flux-attached states, on the other hand, the above wave functions correctly predict the energy ordering of the states with different spin polarization, but produce critical Zeeman energies that can be off by approximately a factor of two from the exact value. In the treatment below of the states that involve reverse flux attachment at any stage of their construction, we  use either exact diagonalization (which can be performed for only very small systems) or the Jain-Kamilla projection; one should keep in mind that our conclusions for these states are quantitatively less reliable than for the states that do not involve reverse-flux attachment. 

[It is noted that the above-mentioned deviation for reverse-flux-attached states is not a deficiency of the CF theory, but of the projection method. One can define a ``hard-core projection" \cite{Wu93} as
\be
\Psi_{[n_1, n_2,n_3,\cdots]_{2p}}=J^{2p-1}{\cal P}_{\rm LLL} \prod_{\alpha=1}^\mathcal{N} \left[\Phi^{\alpha}_{n_{\alpha}}\right]^* J
\label{rev-HC}
\ee
Here the external factor $J^{2p-1}$ guarantees that the wave function vanishes when any two electrons coincide independent of their spin even for $2p=2$, unlike the corresponding wave function in Eq.~\ref{rev-flux}. The hard-core projection can be explicitly evaluated for small systems and has been found to produce extremely accurate wave functions \cite{Wu93}. Unfortunately, no methods currently exist for dealing with it for large systems.]

The above wave functions reduce to certain previous wave functions for the special cases when they involve only $n_j=1$ and do not require any lowest LL projection. The single component wave function $[1]_{2p} \leftrightarrow \left( {1\over 2p+1}\right)$ reproduces the Laughlin wave function \cite{Laughlin83}. The multicomponent wave functions $[1, 1, \cdots]_{2p}$ were earlier proposed by Halperin \cite{Halperin83} for multicomponent systems. 

As an example, at $4/9$, we can construct the states 
\be
[1,1,1,1]_2 \leftrightarrow \left(   {1\over 9}, {1\over 9}, {1\over 9}, {1\over 9}  \right) \nonumber 
\ee

\be
[2,1,1]_2 \leftrightarrow \left(   {2\over 9}, {1\over 9}, {1\over 9}  \right) \nonumber 
\ee
\be
 [3,1]_2 \leftrightarrow \left(   {3\over 9}, {1\over 9}  \right) \nonumber
 \ee
 \be
 [2,2]_2 \leftrightarrow \left(   {2\over 9}, {2\over 9}  \right) \nonumber
 \ee
 \be
 [4]_2 \leftrightarrow \left(   {4\over 9}  \right) \nonumber
\ee
which involve, respectively, 4, 3, 2, 2, and 1 components. In GaAs only two spin components are available, and hence only the last three states are relevant. In graphene and AlAs quantum wells four components (two spins and two valleys) are available and thus all five states may be relevant (depending on parameters). 

A straightforward generalization of the above states is given by 
\be
[\{ n_{\alpha}\}, [\{ m_{\beta}\}]_{\pm 2q}]
\label{interm}
\ee
where some filled LLs have been added in certain components. In other words, only the partially filled LLs in some of the components split into $\Lambda$ levels. These are essentially the same as the IQHE of composite fermions.

Another straightforward generalization is particle-hole conjugation. We denote the hole conjugate of a state $[\cdots]$ by $\overline{[\cdots]}$. For example, a two-component state with filling factor $(\nu_{\uparrow},\nu_{\downarrow})$ transforms into its particle-hole conjugate state $\overline{(\nu_{\uparrow},\nu_{\downarrow})} \equiv (1-\nu_{\uparrow},1-\nu_{\downarrow})$. This allows us to write down wave functions at filling factors $(1-\nu_{\uparrow},1-\nu_{\downarrow})$ from the corresponding state at $(\nu_{\uparrow},\nu_{\downarrow})$. The generalization to states involving an arbitrary number of components is straightforward.

\subsection{FQHE of composite fermions: $\Lambda$ levels within $\Lambda$ levels}

\begin{figure}
\begin{center}
\includegraphics[width=0.5\textwidth]{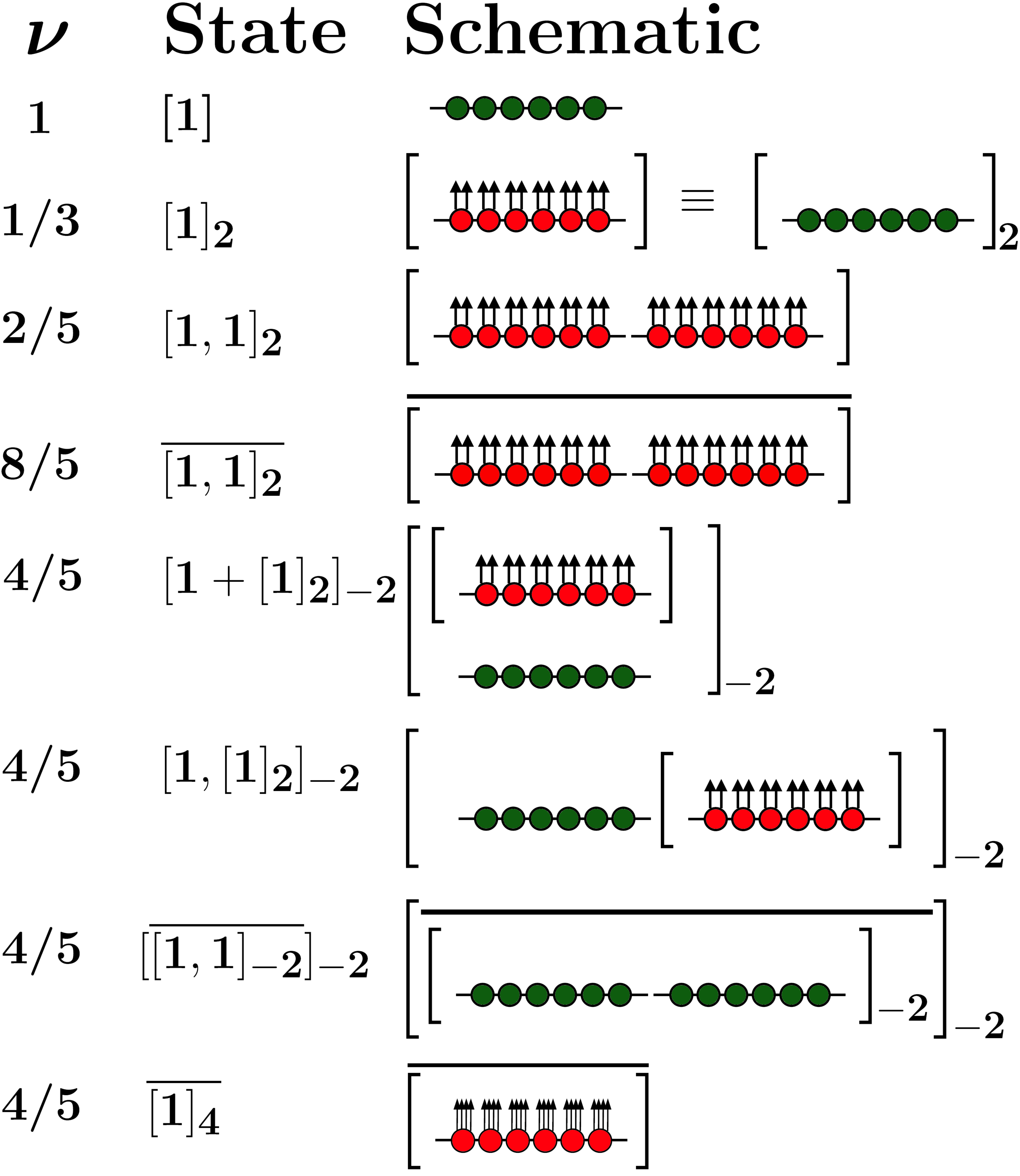}
\caption{(color online) Some examples to illustrate the notation used in Eq. \ref{IFQHE}. The green solid dots represent electrons while the red solid dots with $2p$ arrows are composite fermion particles carrying $2p$ vortices. The overline notation is used to indicate a particle-hole transformation while $[~]_{2p}$ and $[~]_{-2p}$ denote composite fermionization with parallel and reverse flux attachment, respectively. Note that even though the two fully polarized $\nu=4/5$ states shown above, namely $[1,[1]_2]_{-2}$ and $\overline{[1]_4}$, look different, they represent the same state, as explained in the text.}
\label{fig:1}
\end{center}
\end{figure}

We next consider the FQHE states of composite fermions, which have the form
\be
[\{ n_{\alpha}\}, [\{ m_{\beta}\}]_{\pm 2q}]_{\pm 2p}
\label{IFQHE}
\ee
Here, $\Lambda$ levels split into further $\Lambda$ levels in some components. These CF-FQHE states are expected to be much less robust, as measured by the excitation gaps, than the IQHE states of composite fermions considered in the previous subsection. The filling factor of the resulting state is given by
\be
\nu={n+ {m\over 2qm\pm  1}   \over 2p (n+ {m\over 2qm\pm  1}) \pm 1 }
\ee
where $n=\sum_\alpha n_{\alpha}$ and $m=\sum_\beta m_\beta$. The wave functions for $N$ electrons is given by
\be
\Psi_{[\{ n_{\alpha}\}, [\{ m_{\beta}\}]_{\pm 2q}]_{2p}}={\cal P}_{\rm LLL} \prod_{\alpha} \Phi^{\alpha}_{n_{\alpha}} \psi_{[\{m_{\beta} \}]_{\pm 2q}} J^{2p}
\label{fqhe_CFs_paral-flux}
\ee
\be
\Psi_{[\{ n_{\alpha}\}, [\{ m_{\beta}\}]_{\pm 2q}]_{- 2p}}={\cal P}_{\rm LLL} \left[\prod_{\alpha} \Phi^{\alpha}_{n_{\alpha}}\psi_{[\{m_{\beta} \}]_{\pm 2q}} \right] ^* J^{2p}
\label{fqhe_CFs_rever-flux}
\ee
(The Jastrow factor in the wave function $\psi_{[\{m_{\beta} \}]_{\pm 2q}}$ only involves electrons in the components labeled by $\beta$.) In addition to $n_1\geq n_2 \cdots$ we also assume $m_1\geq m_2 \cdots$ without any loss of generality.  We do not consider states of the form $[[\{ m_{\beta}\}]_{\pm 2q}]_{\pm 2p}$ because these are same as the states $[\{ m_{\beta}\}]_{\pm 2q\pm 2p}$. For an SU(4) system, no more than four integers may be used. To illustrate the states corresponding to the notation used here we show some examples in Figure \ref{fig:1}.

\subsection{Moore-Read Pfaffian and W\'ojs-Yi-Quinn unconventional states}

So far we have constructed states that ultimately arise from IQHE-like states of composite fermions. Moore and Read \cite{Moore91} have proposed that composite fermions can also form a paired Pfaffian (Pf) state at $\nu=1/2$, which will be referred to as $1/2^{\rm Pf}$ below. Its hole partner, namely the anti-Pfaffian (APf) \cite{Levin07,Lee07} wave function will be referred to as $$\overline{1/2^{\rm Pf}} \equiv 1/2^{\rm APf}.$$ W\'ojs, Yi and Quinn (WYQ) proposed \cite{Wojs04} a state at 1/3 which is topologically distinct from $[1]_2$; this will be referred to as $1/3^{\rm WYQ}$ and its hole partner as $$\overline{1/3^{\rm WYQ}}\equiv2/3^{\rm WYQ}.$$ These allow construction of many additional FQHE states of composite fermions,  which have the form:
\be
[\{ n_{\alpha}\}, 1/2^{\rm Pf}]_{\pm 2p},\; [\{ n_{\alpha}\}, 1/2^{\rm APf}]_{\pm 2p}, 
\ee
\be
[n_1+1/2^{\rm Pf}, n_2, \cdots]_{\pm 2p},\; [n_1+1/2^{\rm APf}, n_2, \cdots]_{\pm 2p},
\ee
\be
[\{ n_{\alpha}\}, 1/3^{\rm WYQ}]_{\pm 2p}, \; [\{ n_{\alpha}\}, 2/3^{\rm WYQ}]_{\pm 2p},
\ee
and 
\be
[n_1+1/3^{\rm WYQ}, n_2, \cdots]_{\pm 2p},\; [n_1+2/3^{\rm WYQ}, n_2, \cdots]_{\pm 2p},
\ee
with the integers chosen so that the final wave functions satisfies Fock conditions (next subsection).

\subsection{Fock conditions}

The energy spectrum for a given system consists of degenerate SU(${\cal N}$) multiplets. It is convenient to work with a maximal weight state of a given multiplet, from which all other states of the multiplet can be constructed by repeated applications of appropriate ladder operators. For a two component system, the maximal weight state has $S_z=S$, the ladder operator is the spin lowering operator, and the degeneracy is $2S+1$. (Because we are only interested in the energy, we do not need to explicitly construct other states of a multiplet; it suffices to know that they are all degenerate due to the SU(${\cal N}$) symmetry.) The maximal weight states satisfy the generalized Fock condition, \cite{Hamermesh62} i.e., they vanish when we further attempt to antisymmetrize an electron in the $\alpha^{\text{th}}$ component with respect to those in the $\beta<\alpha$ component.  All wave functions constructed above satisfy the generalized Fock condition. This can be seen as follows. For the states given in Eq.~\ref{IQHE}, the Fock condition is obviously satisfied because every single particle orbital occupied in $\alpha$ is also occupied in $\beta$ (with $\beta<\alpha$). This property is preserved under composite fermionization, because the Jastrow factor is symmetric under the exchange of all coordinates. For the more complicated states given in Eq.~\ref{interm}, every single particle orbital occupied in the $m$-sector is necessarily occupied in the $n$ sectors. Again, composite fermionization of the wave function to produce states in Eq.~\ref{IFQHE} preserves this property. We do not consider states of the type $$[[m_1, m_2]_{\pm 2p}, [m_3,m_4]_{\pm 2q}]$$ as these do not satisfy the Fock condition, and do not possess proper SU(4) quantum numbers. Such states may become relevant when the SU(4) symmetry is broken, but we will not consider that case here. For further details, the reader is referred to the discussions in Refs.~\onlinecite{Jain07,Toke07}. 

\subsection{Further generalization: ``Double" FQHE of composite fermions}
One may also construct states which involve at least two fractional fillings of composite fermions. In general, a state of the form 
\be
[\{ n_{\alpha}\}_{2q}, [\{ m_{\beta}\}]_{\pm 2q'}]_{2p}
\ee
does not satisfy Fock conditions, because an occupied state in a given component index is not necessarily occupied in all previous component indices. However, one may construct states of the type
\be
\left[1+{n\over 2qn\pm 1}, 1, [m_1,m_2]_{2q'}\right]_{2p}
\ee
where in the first component the lowest LL is full and a FQHE state is created in the second LL. Such states satisfy Fock conditions, and some examples will be considered below.

\section{Connection to previous states}
\label{three}

The IQHE states of composite fermions for single and multi-component systems have been considered previously. For a single component, the state $[1]_{2p}$ at $\nu=1/(2p+1)$ is the Laughlin state \cite{Laughlin83} and the state $[n]_{\pm 2p}$ at $\nu=n/(2pn\pm 1)$ are the Jain states \cite{Jain89}. The two component CF-IQHE states of the type $[n_1,n_2]_{2p}$ and $[n_1,n_2]_{-2p}$ have been considered by Wu {\em et al.}\cite{Wu93}, Park and Jain \cite{Park98,Park99,Park01}, W\'ojs {\em et al.}\cite{Wojs07} and Davenport and Simon \cite{Davenport12}. Multicomponent states of the type $[n_1, n_2, n_3,n_4]_2$ were studied by T\H oke and Jain \cite{Toke06,Toke07,Toke07b} in the context of graphene. These are all IQHE states of composite fermions, in that the CF filling in each sector is an integer.  We will not consider these states in this article.

Many FQHE states of composite fermions have also been considered previously. The states of the form $[1, 1, \cdots, [1, 1, \cdots]_{2p}]_{2q}$ are the Halperin states \cite{Halperin83} that satisfy the Fock condition; these were introduced as multi-component generalizations of the Laughlin wave functions. Two component states of the form $[1, [n]_2]_2$ were considered by Park and Jain \cite{Park00b} and Chang {\em et al.}\cite{Chang03b}. The state $1/3^{\rm WYQ}$ was proposed by W\'ojs, Yi and Quinn \cite{Wojs04} and the state $1/2^{\rm Pf}$ by Moore and Read \cite{Moore91}; $1/2^{\rm APf}$ is its particle-hole conjugate. The states $[1+1/3^{\rm WYQ}]_2$ and $[1+2/3^{\rm WYQ}]_2$ were considered by Mukherjee {\em et al.}\cite{Mukherjee14} as candidates of fully spin polarized (i.e. one component) FQHE at 4/11 and 5/13. The states $[1+1/2^{\rm Pf}]_2$ and $[1+1/2^{\rm APf}]_2$ were considered by Mukherjee {\em et al.} \cite{Mukherjee12} for fully spin polarized FQHE at 3/8. Two component states $[1,1/2^{\rm Pf}]$ and $[1,1/2^{\rm APf}]$ at $\nu=3/8$ have been studied by Mukherjee, Jain and Mandal \cite{Mukherjee14c}. Ref.~\onlinecite{Sodemann14} considered the $\nu=5/3$ state $(1,1/3,1/3)$ which is $[ 1, [1,1]_{-2}]$, where $[1,1]_{-2}$ is the 2/3 ``spin-singlet" state in the two relevant components. (Note that the wave function $\Psi_{2/3}={\cal P}_{\rm LLL}[\Phi_{1}\Phi_{1}]^*J^2$, studied in Ref.~\onlinecite{Davenport12} is less accurate than the ``hard-core projected" $\Psi_{2/3}=J{\cal P}_{\rm LLL}[\Phi_{1}\Phi_{1}]^*J$ studied previously by Wu {\em et al.}\cite{Wu93}. This is not an intrinsic deficiency of the CF theory, but is related to the technical issue of using the Jain-Kamilla projection method\cite{Jain97,Jain97b} for dealing with unpolarized states requiring reverse flux attachment). For completeness, we will reproduce results of the aforementioned states in this work.

\section{CF Diagonalization}
\label{four}

In may cases, we will produce results obtained from CF diagonalization, outlined below. This approach has several advantages. First, the CF basis has a much smaller dimension than the full basis, and thus allows a study of much larger systems than would be possible from exact diagonalization. Second, it gives very accurate results for the lowest LL FQHE. This is particularly useful for Jain states involving reverse flux attachment, for which no good method currently exists to evaluate the ``hard-core projected" wave function given in Eq.~\ref{rev-flux}. Finally, this method also provides an independent test of whether the actual ground state indeed occurs at the $(L,S)$ quantum numbers predicted by the CF theory. 

The Coulomb interaction is taken as the Hamiltonian and the energies of the states are evaluated using the Monte-Carlo method as follows: First, $L^2$ eigenstates are created in the corresponding IQHE system. Composite fermionisation, i.e multiplying by $J^{2p}$ and projection into the lowest LL, of this state gives the required $L^2$ eigenstate, since $J^{2}$ has zero angular momentum and therefore $L^2$ operator commutes with it. The set of basis states $\{\Psi_i\}$'s are constructed by taking all possible $L^2$ eigenstates. The Hamiltonian matrix is given by:
\begin{eqnarray}
\mathcal{H}(\Psi_1,\Psi_2)= \int \Psi_1^{*} \Bigg(\sum_{i<j}\frac{1}{|z_{i}-z_{j}|}\Bigg) \Psi_2~dz
\end{eqnarray}
where $dz$ stands for the collective set of coordinates i.e, $dz \equiv dz_{1}dz_{2}...dz_{N}$.
Note that the CF wave functions are in general not orthogonal to each other. To implement the Gram-Schmidt orthogonalization procedure one calculates the overlap matrix defined as:
\begin{eqnarray}
\mathcal{O}(\Psi_1,\Psi_2)= \int \Psi_1^{*}\Psi_2~d^2\vec{\Omega}
\end{eqnarray}

To evaluate the above quantities, we need to do multi-dimensional integrals. We use the Metropolis Monte Carlo algorithm to evaluate such integrals. The algorithm works by approximating the integral as a sum and then sampling different configurations of the set of coordinates $\{\Omega_i\}$ drawn from a probability distribution which is weighted by the absolute square of the wave function. Both $\mathcal{O}$ and $\mathcal{H}$ are evaluated in a single Monte Carlo run. The energies of the states are given by the eigenvalues of $\mathcal{O}^{-1}\mathcal{H}$ (see Appendix \ref{appendix:C}). To make the computation more efficient we do the calculation within the subspace of $L^2$ eigenstates. The lowest LL projection, as stated above, is carried out by the Jain-Kamilla method \cite{Jain97b}, details of which can be found in \cite{Jain07}. This procedure of evaluation of energies of CF wave functions is termed composite fermion diagonalization (CFD) \cite{Mandal02}. 

As noted above, Jain's CF wave functions can be constructed for states involving parallel flux attachment ($2p$ positive); for such cases it is possible to go to very large systems to obtain accurate thermodynamic limits. For states involving reverse flux attachment ($2p$ negative), no good method exists for implementing Jain's CF wave functions. Here, we obtain the ground state energies by performing CF diagonalization, which often allows us to go to larger systems than exact diagonalization.

\section{Exact Diagonalization}
\label{ED}
For one-component states we carry out standard exact diagonalization (in the full space spanned by all $N$-electron configurations), i.e. apply Lanczos algorithm to diagonalize the Hamiltonian matrix $\mathcal{H}$ in the subspace of minimum total orbital angular momentum projection $L_z=0$ in search of the absolute ground state, for which we then compute the expectation value of $L^2$ to verify that it indeed has $L=0$. 
 
For two-component states with arbitrary polarization corresponding to a total spin $S$ we perform diagonalization in the subspace of $S_z=S$ and $L_z=0$ and then compute the expectation values of $S^2$ and $L^2$ to verify that the ground state indeed has $L=0$ and the assumed spin $S$ (by taking $S_z=S$ we disregard the possibility that the absolute ground state may have spin lower than $S$, but by checking average $S^2$ we confirm that no state with spin larger than the assumed value $S$ has a lower energy).
 
The critical part of the Lanczos procedure is efficient on-the-fly computation of the $\mathcal{H}$ matrix elements. Configuration basis is generated once and stored in the form of binary numbers $B$ with consecutive bits representing consecutive orbitals (0-empty, 1-occupied), which is equivalent to the tabulation and storage of the index-to-configuration mapping $B(i)$. For each row $i$, the initial configuration $I=B(i)$ is immediately assigned, then all possible final configurations $F=\mathcal{H}I$ and the corresponding scattering amplitudes $\left<F|\mathcal{H}|I\right>$ are generated by explicit action on $I$ of the second-quantized two-body Hamiltonian $\mathcal{H}$. The column indexes $f=B^{-1}(F)$ must then be obtained from configuration search, as storing the inverse (configuration-to-index) mapping $B^{-1}$ is generally unfeasible. The ordering of configurations, i.e., the monotonicity of $B(i)$, allows for efficient bisection search, further accelerated by the {\em partial} storage of $B^{-1}$ (all configurations classified by a certain number of leading bits, allowing bisection only within the relevant class, i.e. range of configurations).
 
Our optimized OpenMP-parallel code (the two most critical parts being the generation of $F=\mathcal{H}I$ and the column search $f=B^{-1}(F)$) permits us diagonalization of Hamiltonians with dimensions up to a few billion, at the speed of at least a few Lanczos iterations per day (on a single cluster node) for the largest systems.

We note that all exact diagonalization studies are performed for the $N$ and $2Q$ values guided by the CF theory. The spin and angular momentum quantum numbers are consistent with that predicted by the CF theory, and the energies and wave functions are also in excellent agreement wherever a comparison can be made. 

\section{Some technical details}
\label{dens-corr}

In all calculations that follow, we use the spherical geometry \cite{Haldane83}, where in $N$ electrons reside on the surface of a sphere, with a Dirac monopole at the center which generates a radial magnetic field that produces a total magnetic flux of $2Q\phi_0$ through the surface of the sphere, where $\phi_{0}=hc/e$ is referred to as a flux quantum. Using the mapping postulated by CF theory, the effective flux seen by composite fermions is given by $2Q^{*}=2Q-2p(N-1)$. States are characterized by the orbital angular momentum quantum number $L$ (appropriate in this geometry) and the spin-angular momentum quantum number $S$. Ground state are seen to be incompressible i.e., have $L=0$. The total energies include contributions from the electron-background and the background-background interactions. These are determined by assuming that the neutralizing background positive charge of strength $Ne$ is distributed uniformly on the surface of the sphere. The total energy $E$ for $N$ particles is given by: $E_{N}=E_{el-el}-\frac{N^{2}}{2\sqrt{Q}}\frac{e^2}{\epsilon\ell}$, where the first term on the right hand side is the electron-electron interaction energy, which is evaluated by exact and/or CF diagonalization, and the second term incorporates interaction with a positively charged neutralizing background. Comparison with experiments requires the thermodynamic values of the energies of the ground states. To account for the slight density dependences on the number of particles in the spherical geometry, we make a ``density correction" to the finite system energies before extrapolation to the thermodynamic limits. The density corrected energy is defined as\cite{Morf86} $E_{N}^{'}=\sqrt{\frac{\rho_{\infty}}{\rho_{N}}}E_{N}=\sqrt{\frac{2Q\nu}{N}}E_{N}$.  We find that extrapolation with and without density correction produces slightly different critical Zeeman energies.  All of the thermodynamic limits quoted below are obtained with density correction. 

We note that some small systems admit two different interpretations. For example, the fully polarized system of $9$ particles at a flux of $12$ can be thought of either as 5/7 or as 7/9 FQHE state. However, because the density correction depends on the filling factor $\nu$, the density corrected energies are different for this system in Tables \ref{tab:Table1} and \ref{tab:Table2}. Of course, such aliasing does not occur for larger systems, which are needed for the determination of the thermodynamically extrapolated energies.

\section{Prominent CF-FQHE states for multi-component systems}
\label{five}

The IQHE states of two component composite fermions have been investigated in great detail before \cite{Wu93,Park98,Park99,Davenport12}. The earlier experiments in GaAs\cite{Eisenstein89,Du95,Yeh99,Kukushkin99,Kukushkin00,Melinte00} fully confirm this physics. In particular, the measured spin polarizations\cite{Kukushkin99} agree with the theoretical predictions, and the observed critical Zeeman energies are roughly consistent with theory, although a very precise agreement is not expected as the theory does not include corrections due to LL mixing, finite thickness and the ubiquitous disorder. The two-component systems in AlAs quantum wells, where the two components are valleys, are also in good agreement with the CF theory  \cite{Bishop07,Padmanabhan09}. The fractions seen in graphene \cite{Xu09,Bolotin09,Dean11,Feldman12,Feldman13,Amet14} and in an H-terminated Si(111) surface \cite{Kott14} are also consistent with one or two component IQHE of composite fermions. 

Our focus below is on {\em fractional} QHE of composite fermions, which we define as those states in which composite fermions in at least one component show FQHE. (The states where all components are integers are defined as IQHE of composite fermions, not considered here.) Many such states have already been observed, and many more are predicted to occur. We mention the experimental status of each fraction below, while also listing the number of possible states and their generalized spin contents.

\subsection{$\nu=4/11$ (parent state $\nu^{*}=4/3$)}

\subsubsection{One component fully polarized $4/11$}

The fully polarized one component 4/11 state 
$$[1+1/3^{\rm WYQ}]_2 \leftrightarrow(4/11)$$
corresponds to $\nu^*=4/3$, which is obtained by filling the lowest $\Lambda$L completely and forming a 1/3 state in the second $\Lambda$L; explicit calculation shows that the residual interaction between composite fermions in the second $\Lambda$L is of the form as to favor the WYQ 1/3 state \cite{Wojs04,Mukherjee14}. 

\subsubsection{Two component partially polarized $4/11$}
The partially polarized two-component 4/11 state 
$$[1,[1]_2]_{2} \leftrightarrow (3/11,1/11)$$ 
is obtained by composite-fermionizing the partially polarized 4/3 state $$[1,[1]_2]\leftrightarrow(1,1/3).$$ It corresponds to filling the lowest spin-up $\Lambda$L completely and forming a 1/3 state in the lowest spin-down $\Lambda$L \cite{Chang03b}. 

\subsubsection{Two component singlet $4/11$}
The state 
$$[\overline{[1,1]_{-2}}]_2 \leftrightarrow (2/11,2/11)$$
was first proposed in Ref. \cite{Park00b}. It is obtained from the 2/3 spin singlet state 
$${[1,1]_{-2}} \leftrightarrow (1/3,1/3)$$
by taking its particle hole conjugate to produce a singlet state at 4/3 of the form $(2/3, 2/3)$ \cite{Wu93}, and then composite-fermionizing it. We stress that the $(2/3, 2/3)$ is {\em not} a direct product of two one-component 2/3 states in two spin sectors; that state with a wave function $\Psi_{2/3} \Psi_{2/3} = [\overline{[1]_2},\overline{[1]_2}]$ does not satisfy the Fock condition and thus does not have proper symmetry properties.  
\\

In Table \ref{tab:Table_4_11} we show the thermodynamic energies of these states and in Table \ref{tab:critical_Ez} we show the critical Zeeman energies for the transitions among these states.  (Slight difference from the value in Ref.~\onlinecite{Mukherjee14} arises because that article did not make density correction while obtaining the thermodynamic limits of the various energies.)

\subsubsection{Three and more component $4/11$}

It is not possible to construct a wave function of the type considered here with three or more components that satisfies Fock conditions. For example, a naive wave function 
$$[[1]_2,[1]_2,[1]_2,[1]_2]_2 \leftrightarrow (1/3, 1/3, 1/3, 1/3)$$ is not a valid wave function for an SU(4)-symmetric interaction. Thus, a 4/11 FQHE state is likely to be a one or two component state; the stabilization of a 4/11 FQHE with three or more components will require physics that is beyond what is considered in this work.

The one component $[1+1/3^{\rm WYQ}]_2$ was considered by W\'ojs, Yi and Quinn \cite{Wojs04} and by Mukherjee {\em et al.}\cite{Mukherjee14}, and the two component $[1,[1]_2]_{2}$ was considered in Refs.~\cite{Chang03b,Wojs07}. These studies did not identify the spin singlet 4/11 state $[\overline{[1,1]_{-2}}]_2$. 

Since the $4/11$ FQHE has been observed \cite{Pan03} (the spin polarization of the state has not yet been measured), and this is the first time that the spin singlet 4/11 has been investigated in detail, we give more analysis of its spectrum in Appendix \ref{4_11_SS}. We see there that the CF theory is in good agreement with the spectrum obtained from exact diagonalization. The spectrum also indicates the presence of an unconventional spin wave with a spin roton minimum, as found previously for the fully spin polarized 2/5 and 3/7 states \cite{Mandal01}. Furthermore, a charged collective mode is also identified.

\begin{table*}
\begin{center}
\begin{tabular}{|c|c|c|c|c|c|c|c|}
\hline
\multicolumn{1}{|c|}{$\nu$} & \multicolumn{2}{|c|}{$[1+1/3^{\rm WYQ}]_2 \leftrightarrow(4/11)$} & \multicolumn{3}{|c|}{$[1,[1]_2]_{2} \leftrightarrow (3/11,1/11)$} & \multicolumn{2}{|c|}{$[\overline{[1,1]_{-2}}]_2 \leftrightarrow (2/11,2/11)$} \\ \hline
	    & exact	  		& CFD 		   & exact	& CFD 	     & CF w. f.	      & exact	    & CFD	 \\ \hline
4/11	    & -0.4166(0)		& -0.4160(1)	   & -0.4218(0)	& -0.4203(0) & -0.42054(0)    & -0.4224(0)  & -0.4220(0) \\ \hline
\end{tabular}
\end{center}
\caption {The Coulomb energies of the states obtained from $\nu^{*}=4/3$ with $p=1$ and parallel flux attachment. We quote both the exact and CF energies (both obtained from CFD and from just the CF wave function) for fully polarized, partially polarized and spin-singlet states. The numbers are obtained by extrapolating finite size results to the thermodynamic limit and the fitting errors are shown. All energies are quoted in units of $e^2/\epsilon \ell$. }  
\label{tab:Table_4_11} 
\end{table*}

\subsection{$\nu=4/5$ (parent state $\nu^{*}=4/3$)}

\subsubsection{One component fully polarized $4/5$}
The fully polarized one component 4/5 state 
$$[1+[1]_{2}]_{-2} \leftrightarrow(4/5)$$
corresponds to $\nu^*=4/3$, which is obtained by filling the lowest $\Lambda$L completely and forming a 1/3 state in the second $\Lambda$L. One might think that another candidate for a fully polarized FQHE at $4/5$ is 
$$\overline{[1]_{4}} \leftrightarrow(4/5)$$
which is the hole conjugate of the 1/5 state. However, in spite of the superficial difference, the two wave functions are equivalent, i.e. represent the same state, as can be seen by noting that they occur at the same flux and have the same excitation spectrum. The situation is analogous to $2/3$ for which two wave functions can be written, namely $\overline{[1]_2}$ and $[2]_{-2}$, but these two were shown to be essentially identical by explicit calculation \cite{Wu93}. However, one of those two points of view is more useful in consideration of the spin phase transitions. For example, for 2/3, its understanding as two filled $\Lambda$ levels of composite fermions immediately reveals the possibility of a spin singlet state, and gives an intuitive understanding of the spin phase transition as a level crossing transition as the Zeeman energy is varied. For similar reasons, the mapping of 4/5 into a state at $\nu^*=4/3$ is more useful in bringing out the physics of spin transitions. (We also note that for the single component 4/5 state, the interpretation $\overline{[1]_{4}}$ views it as an IQHE of composite fermions, whereas $[1+[1]_{2}]_{-2}$ as a FQHE state of composite fermions. This is only a difference of nomenclature, however. Both states involve $^4$CFs.)

\subsubsection{Two component partially polarized $4/5$}
The partially polarized two-component 4/5 state 
$$[1,[1]_{2}]_{-2} \leftrightarrow (3/5,1/5)$$ 
is obtained from the partially polarized 4/3 state
$$[1,[1]_2] \leftrightarrow(1,1/3)$$ 
which corresponds to filling the lowest $\Lambda$L of spin-up completely and forming a 1/3 state in the spin-down lowest $\Lambda$L. 

\subsubsection{Two component singlet $4/5$}
The state 
$$[\overline{[1,1]_{-2}}]_{-2} \leftrightarrow (2/5,2/5)$$
is obtained from the 2/3 spin singlet state \cite{Wu93}
$${[1,1]_{-2}} \leftrightarrow (1/3,1/3)$$
by taking its particle hole conjugate to produce a singlet state at 4/3 and then composite-fermionizing it. 

\subsubsection{Three and more component $4/5$}
It is not possible to construct a wave function of the type considered here with three or more components that satisfies Fock conditions for the same reasons as mentioned above for $4/11$.

In Table \ref{tab:Table_4_5} we show the thermodynamic energies of these states and in Table \ref{tab:critical_Ez} we show the critical Zeeman energies for the transitions among these states. 

\begin{table*}
\begin{center}
\begin{tabular}{|c|c|c|c|c|c|c|}
\hline
\multicolumn{1}{|c|}{$\nu$} & \multicolumn{2}{|c|}{$[1+[1]_{2}]_{-2} \equiv \overline{[1]_{4}} \leftrightarrow(4/5)$} & \multicolumn{2}{|c|}{$[1,[1]_{2}]_{-2} \leftrightarrow (3/5,1/5)$} & \multicolumn{2}{|c|}{$[\overline{[1,1]_{-2}}]_{-2} \leftrightarrow (2/5,2/5)$} \\ \hline
	    & exact	  		& CF 		    		& exact	  	& CF 		    & exact	  & CF	      \\ \hline
4/5	    & -0.5504(7)    		& - 		    		& -0.5601(0)	& - 		    & -0.5637(5)  & -		      \\ \hline
\end{tabular}
\end{center}
\caption {The Coulomb energies of the states obtained from $\nu^{*}=4/3$ with $p=1$ and reverse flux attachment.}  
\label{tab:Table_4_5} 
\end{table*}

\subsection{$\nu=5/13$ (parent state $\nu^{*}=5/3$)}
\subsubsection{One component fully polarized $5/13$}
The fully polarized one component 5/13 state 
$$[1+2/3^{\rm WYQ}]_{2} \leftrightarrow(5/13)$$
corresponds to $\nu^*=5/3$, which is obtained by filling the lowest $\Lambda$L completely and forming a 2/3 WYQ state in the second $\Lambda$L.

\subsubsection{Two component partially polarized $5/13$}
The partially polarized two-component 5/13 state 
$$[1,[2]_{-2}]_{2} \leftrightarrow (3/13,2/13)$$ 
is obtained from the partially polarized 5/3 state
$$[1,[1]_{-2}] \leftrightarrow(1,2/3)$$ 
which in turn is obtained by filling the lowest $\Lambda$L of spin-up completely and forming a 2/3 state in the spin-down lowest $\Lambda$L.

\subsubsection{Three and more component $5/13$}
A three component state of the following kind can be constructed:
$$[1,[1,1]_{-2}]_{2} \leftrightarrow (3/13,1/13,1/13)$$ 
where in the lowest $\Lambda$L of one of the components is fully filled and a spin-singlet 2/3 state is formed in any of the two other components.

In Table \ref{tab:Table_5_13} we show the Coulomb energies for these states extrapolated to the thermodynamic limit and in Table \ref{tab:critical_Ez} we give the critical Zeeman energies for the transitions among these states. 

\begin{table}
\begin{center}
\begin{tabular}{|c|c|c|c|c|}
\hline
\multicolumn{1}{|c|}{$\nu$} & \multicolumn{2}{|c|}{$[1+2/3^{\rm WYQ}]_{2} \leftrightarrow(5/13)$} & \multicolumn{2}{|c|}{$[1,[2]_{-2}]_{2} \leftrightarrow (3/13,2/13)$}  \\ \hline
		    & exact	  		& CFD 		    		& exact	  		& CFD 		     \\ \hline
5/13 		    & -0.4243(7)	 	& -0.4243(1)		 	& -0.4317(0)	  	& -0.4303(0)	     \\ \hline
\end{tabular}
\end{center}
\caption {The Coulomb energies of the states obtained from $\nu^{*}=5/3$ with $p=1$ and parallel flux attachment.}  
\label{tab:Table_5_13} 
\end{table}

\subsection{$\nu=5/7$ (parent state $\nu^{*}=5/3$)}
\subsubsection{One component fully polarized $5/7$}
The fully polarized one component 5/7 state 
$$[1+[2]_{-2}]_{-2} \leftrightarrow(5/7)$$
corresponds to $\nu^*=5/3$, which is obtained by filling the lowest $\Lambda$L completely and forming a 1/3 state in the second $\Lambda$L. This state is equivalent to 
$$\overline{[2]_{-4}}$$
i.e., the state obtained from the 2/7 state by particle-hole transformation.

\subsubsection{Two component partially polarized $5/7$}
The partially polarized two-component 5/7 state 
$$[1,[2]_{-2}]_{-2} \leftrightarrow (3/7,2/7)$$ 
is obtained from the partially polarized 5/3 state
$$[1,[2]_{-2}] \leftrightarrow(1,2/3)$$ 
which in turn is obtained by filling the lowest $\Lambda$L of spin-up completely and forming a 2/3 state in the spin-down lowest $\Lambda$L.

\subsubsection{Three and more component $5/7$}
A three component state of the following kind can be constructed:
$$[1,[1,1]_{-2}]_{-2} \leftrightarrow (3/7,1/7,1/7)$$ 
where in the lowest $\Lambda$L of one of the components is fully filled and a spin-singlet 2/3 state is formed in any of the two other components.

In Table \ref{tab:Table_5_7} we show the Coulomb energies for these states extrapolated to the thermodynamic limit and in Table \ref{tab:critical_Ez} we give the critical Zeeman energies for the transitions among these states. 

\begin{table}
\begin{center}
\begin{tabular}{|c|c|c|c|c|}
\hline
\multicolumn{1}{|c|}{$\nu$} & \multicolumn{2}{|c|}{$[1+[2]_{-2}]_{-2} \equiv \overline{[2]_{-4}} \leftrightarrow(5/7)$} & \multicolumn{2}{|c|}{$[1,[2]_{-2}]_{-2} \leftrightarrow (3/7,2/7)$}  \\ \hline
		    & exact	  		& CF 		    		& exact	  		& CF 		     \\ \hline
5/7	            & -0.5294(0)     		& - 		    		& -0.5389(0)  		& - 		     \\ \hline
\end{tabular}
\end{center}
\caption {The Coulomb energies of the states obtained from $\nu^{*}=5/3$ with $p=1$ and reverse flux attachment.}  
\label{tab:Table_5_7} 
\end{table}

\subsection{$\nu=7/19$ (parent state $\nu^{*}=7/5$)}
\subsubsection{One component fully polarized $7/19$}
The fully polarized one component 7/19 state 
$$[1+[2]_{2}]_{2} \leftrightarrow(7/19)$$
corresponds to $\nu^*=7/5$, which is obtained by filling the lowest $\Lambda$L completely and forming a 2/5 state in the second $\Lambda$L. 

\subsubsection{Two component partially polarized $7/19$}
\paragraph{Obtained from fully polarized $3/5$ of holes}
This partially polarized two-component 7/19 state is obtained from fully polarized $\nu_{h}=3/5$ state:
$$[1,[2]_{2}]_{2} \equiv [\overline{[3]_{-2}}]_{2} \leftrightarrow (5/19,2/19)$$ 
\paragraph{Obtained from partially polarized $3/5$ of holes}
This partially polarized two-component 7/19 state is obtained from partially polarized $\nu_{h}=3/5$ state:
$$[\overline{[1,2]_{-2}}]_{2} \leftrightarrow (4/19,3/19)$$ 

Table \ref{tab:Table_7_19} shows the Coulomb energies for these states extrapolated to the thermodynamic limit and in Table \ref{tab:critical_Ez} we give the critical Zeeman energies for the transitions among these states. 

\begin{table*}
\begin{center}
\begin{tabular}{|c|c|c|c|c|c|c|c|}
\hline
\multicolumn{1}{|c|}{$\nu$} & \multicolumn{2}{|c|}{$[1+[2]_{2}]_{2} \leftrightarrow(7/19)$} & \multicolumn{3}{|c|}{$[1,[2]_{2}]_{2} \equiv [\overline{[3]_{-2}}]_{2} \leftrightarrow (5/19,2/19)$} & \multicolumn{2}{|c|}{$[\overline{[1,2]_{-2}}]_{2} \leftrightarrow (4/19,3/19)$}  \\ \hline
	    & exact  	& CFD 		& 	exact	& CFD 		& CF w. f.	 & exact	& CFD 		     	\\ \hline
7/19 	    & 	-	& -0.4169(0)	& 	  -	& -0.4227(2)    & -0.42258(4) 	&	-	& -0.4242(3)		\\ \hline
\end{tabular}
\end{center}
\caption {The Coulomb energies of the states obtained from $\nu^{*}=7/5$ with $p=1$ and parallel flux attachment.} 
\label{tab:Table_7_19} 
\end{table*}

\subsubsection{Three component partially polarized $7/19$}
A three-component state\cite{Halperin83,Goerbig07} can be written as
$$[1,[1,1]_2]_2 \leftrightarrow (5/19,1/19,1/19).$$
Its parent state is the spin-singlet CF ground state wave function at $\nu^*=2/5$,
$$[1,1]_2 \leftrightarrow (1/5,1/5).$$

\begin{table}
\begin{center}
\begin{tabular}{|c|c|c|c|}
\hline
\multicolumn{1}{|c|}{$\nu$} &
\multicolumn{3}{|c|}{$[1,[1,1]_{2}]_{2} \leftrightarrow (5/19,1/19,1/19)$}
\\ \hline
		    & exact	  & CFD  & CF w. f.		     \\ \hline
7/19 		    &  	-	  &  -   & -0.422640(7)		      \\ \hline
\end{tabular}
\end{center}
\caption {The Coulomb energies of the states obtained from $\nu^{*}=7/5$ with $p=1$ and parallel flux attachment.} 
\label{tab:Table_7_19b} 
\end{table}

\subsection{$\nu=7/9$ (parent state $\nu^{*}=7/5$)}
\subsubsection{One component fully polarized $7/9$}
The fully polarized one component 7/9 state 
$$[1+[2]_{2}]_{-2} \leftrightarrow(7/9)$$
corresponds to $\nu^*=7/5$, which is obtained by filling the lowest $\Lambda$L completely and forming a 2/5 state in the second $\Lambda$L. An equivalent state $$\overline{[2]_{4}}$$  is obtained by first constructing a $\nu_{h}=2/9$ and taking its particle-hole conjugate state.

\subsubsection{Two component partially polarized $7/9$}
\paragraph{Obtained from fully polarized $3/5$ of holes:}
This partially polarized two-component 7/9 state is obtained from fully polarized $\nu_{h}=3/5$ state:
$$[1,[2]_{2}]_{-2} \equiv [\overline{[3]_{-2}}]_{-2} \leftrightarrow (5/9,2/9)$$ 
\paragraph{Obtained from partially polarized $3/5$ of holes:}
This partially polarized two-component 7/9 state is obtained from partially polarized $\nu_{h}=3/5$ state:
$$[\overline{[1,2]_{-2}}]_{-2} \leftrightarrow (4/9,3/9)$$ 

Table \ref{tab:Table_7_9} shows the Coulomb energies for these states extrapolated to the thermodynamic limit and in Table \ref{tab:critical_Ez} we give the critical Zeeman energies for the transitions among these states. 

\begin{table*}
\begin{center}
\begin{tabular}{|c|c|c|c|c|c|c|}
\hline
\multicolumn{1}{|c|}{$\nu$} & \multicolumn{2}{|c|}{$[1+[2]_{2}]_{-2} \equiv \overline{[2]_{4}} \leftrightarrow(7/9)$} & \multicolumn{2}{|c|}{$[1,[2]_{2}]_{-2} \equiv [\overline{[3]_{-2}}]_{-2} \leftrightarrow (5/9,2/9)$} & \multicolumn{2}{|c|}{$[\overline{[1,2]_{-2}}]_{-2} \leftrightarrow (4/9,3/9)$}  \\ \hline
		    & exact  		& CF 		& 	exact	& 	CF 		 & exact	  		& CF 		     	\\ \hline
7/9	            & -0.5450(1)	& - 		&    -0.5541(0)	&  	-		 &   		-		& -			 \\ \hline
\end{tabular}
\end{center}
\caption {The Coulomb energies of the states obtained from $\nu^{*}=7/5$ with $p=1$ and reverse flux attachment.} 
\label{tab:Table_7_9} 
\end{table*}

\subsubsection{Three component partially polarized $7/9$}
A three-component state can be obtained as
$$[1,[1,1]_2]_{-2} \leftrightarrow (5/9,1/9,1/9).$$
The parent state is the spin-singlet CF ground state wave function at $\nu^*=2/5$,
$$[1,1]_2 \leftrightarrow (1/5,1/5).$$

\subsection{$\nu=8/21$ (parent state $\nu^{*}=8/5$)}

\subsubsection{One component fully polarized $8/21$}
The fully polarized one component 8/21 state 
$$[1+[3]_{-2}]_{2} \leftrightarrow(8/21)$$
corresponds to $\nu^*=8/5$, which is obtained by filling the lowest $\Lambda$L completely and forming a 3/5 state in the second $\Lambda$L. 

\subsubsection{Two component partially polarized $8/21$}
The partially polarized two-component 8/21 state is obtained from fully polarized $\nu_{h}=2/5$ state:
$$[1,[3]_{-2}]_{2} \equiv [\overline{[2]_{2}}]_{2} \leftrightarrow (5/21,3/21)$$ 
The energy in the thermodynamic limit obtained from the unperturbed CF wave function is -0.4278(2).

\subsubsection{Two component spin singlet $8/21$}
The singlet two-component 8/21 state is obtained from spin singlet $\nu_{h}=2/5$ state:
$$[\overline{[1,1]_{2}}]_{2} \leftrightarrow (4/21,4/21)$$ 

\ref{tab:critical_Ez} we give the critical Zeeman energies for the transitions among these states. 

\subsubsection{Three component partially polarized $8/21$}
The state
$$[1,[2,1]_{-2}]_{2} \leftrightarrow (5/21,2/21,1/21)$$ 
can be derived by the composite fermionization of the from the partially polarized state at 3/5,
$$[2,1]_{-2} \leftrightarrow (2/5,1/5)$$ 
using parallel flux attachment.

\subsubsection{Four component partially polarized $8/21$}
The state
$$[1,[1,1,1]_{-2}]_{2} \leftrightarrow (5/21,1/21,1/21,1/21)$$ 
is obtained from a parent state at 3/5,
$$[1,1,1]_{-2}] \leftrightarrow (1/5,1/5,1/5),$$ 
which would be SU(3) singlet in a three-component system.

\subsection{$\nu=8/11$ (parent state $\nu^{*}=8/5$)}

\subsubsection{One component fully polarized $8/11$}
The fully polarized one component 8/11 state 
$$[1+[3]_{-2}]_{-2} \leftrightarrow(8/11)$$
corresponds to $\nu^*=8/5$, which is obtained by filling the lowest $\Lambda$L completely and forming a 3/5 state in the second $\Lambda$L. An exactly equivalent construction via the $\nu_{h}=3/5$ state exists and we denote it by:
$$\overline{[3]_{-4}} \leftrightarrow(8/11)$$

\subsubsection{Two component partially polarized $8/11$}
The partially polarized two-component 8/11 state is obtained from fully polarized $\nu_{h}=2/5$ state:
$$[1,[3]_{-2}]_{-2} \equiv [\overline{[2]_{2}}]_{-2} \leftrightarrow (5/11,3/11)$$ 

\subsubsection{Two component spin singlet $8/21$}
The singlet two-component 8/11 state is obtained from spin singlet $\nu_{h}=2/5$ state:
$$[\overline{[1,1]_{2}}]_{-2} \leftrightarrow (4/11,4/11)$$ 

Table \ref{tab:Table_8_11} shows the Coulomb energies for these states extrapolated to the thermodynamic limit and in Table \ref{tab:critical_Ez} we give the critical Zeeman energies for the transitions among these states. 

\begin{table*}
\begin{center}
\begin{tabular}{|c|c|c|c|c|c|c|}
\hline
\multicolumn{1}{|c|}{$\nu$} & \multicolumn{2}{|c|}{$[1+[3]_{-2}]_{-2} \equiv \overline{[3]_{-4}} \leftrightarrow(8/11)$} & \multicolumn{2}{|c|}{$[1,[3]_{-2}]_{-2} \equiv [\overline{[2]_{2}}]_{-2} \leftrightarrow (5/11,3/11)$} & \multicolumn{2}{|c|}{$[\overline{[1,1]_{2}}]_{-2} \leftrightarrow (4/11,4/11)$}  \\ \hline
		    & exact  		& CF 		& 	exact	& 	CF 		 & exact	  		& CF 		     	\\ \hline
8/11	            & -0.5328(0)	& - 		&    		&  	-		 &  -0.5429(0)			& -			 \\ \hline
\end{tabular}
\end{center}
\caption {The Coulomb energies of the states obtained from $\nu^{*}=8/5$ with $p=1$ and reverse flux attachment.} 
\label{tab:Table_8_11} 
\end{table*}

\subsubsection{Three component partially polarized $8/11$}
The state
$$[1,[2,1]_{-2}]_{-2} \leftrightarrow (5/11,2/11,1/11)$$ 
can be derived by the composite fermionization of the from the partially polarized state at 3/5,
$$[2,1]_{-2} \leftrightarrow (2/5,1/5)$$ 
using reverse flux attachment.

\subsubsection{Four component partially polarized $8/11$}
The state
$$[1,[1,1,1]_{-2}]_{-2} \leftrightarrow (5/11,1/11,1/11,1/11)$$ 
is obtained from a parent state at 3/5,
$$[1,1,1]_{-2}] \leftrightarrow (1/5,1/5,1/5),$$ 
which would be SU(3) singlet in a three-component system.

\subsection{$\nu=10/27$ (parent state $\nu^{*}=10/7$)}

\subsubsection{One component fully polarized $10/27$}
The fully polarized one component 10/27 state 
$$[1+[3]_{2}]_{2} \leftrightarrow(10/27)$$
corresponds to $\nu^*=10/7$, which is obtained by filling the lowest $\Lambda$L completely and forming a 3/7 state in the second $\Lambda$L. 

\subsubsection{Two component partially polarized $10/27$}
The partially polarized two-component 10/27 state is obtained from fully polarized $\nu_{h}=4/7$ state:
$$[1,[3]_{2}]_{2} \equiv [\overline{[4]_{-2}}]_{2} \leftrightarrow (7/27,3/27)$$ 

\subsubsection{Two component spin singlet $10/27$}
The singlet two-component 10/27 state is obtained from spin singlet $\nu_{h}=4/7$ state:
$$[\overline{[2,2]_{-2}}]_{2} \leftrightarrow (5/27,5/27)$$ 

Table \ref{tab:Table_10_27} shows the Coulomb energies for these states extrapolated to the thermodynamic limit and in Table \ref{tab:critical_Ez} we give the critical Zeeman energies for the transitions among these states. 

\begin{table*}
\begin{center}
\begin{tabular}{|c|c|c|c|c|c|c|c|}
\hline
\multicolumn{1}{|c|}{$\nu$} & \multicolumn{2}{|c|}{$[1+[3]_{2}]_{2} \leftrightarrow(10/27)$} & \multicolumn{3}{|c|}{$[1,[3]_{2}]_{2} \equiv [\overline{[4]_{-2}}]_{2} \leftrightarrow (7/27,3/27)$} & \multicolumn{2}{|c|}{$[\overline{[2,2]_{-2}}]_{2} \leftrightarrow (5/27,5/27)$}  \\ \hline
		    & exact  		& CFD 		& 	exact	& 	CFD 		& CF w. f. 	& exact	& CF 	\\ \hline
10/27 		    & 		 	&  -0.4170(2)	& 	  	& 	-0.4231(1)     	& -0.42346(6)   &	&	\\ \hline
\end{tabular}
\end{center}
\caption {The Coulomb energies of the states obtained from $\nu^{*}=10/7$ with $p=1$ and parallel flux attachment.} 
\label{tab:Table_10_27} 
\end{table*}

\subsubsection{Three component partially polarized $10/27$}
This 10/27 state is obtained from the partially polarized $\nu=3/7$ state:
$$[1,[2,1]_{2}]_{2} \leftrightarrow (7/27,2/27,1/27).$$ 

\subsubsection{Four component partially polarized $10/27$}
This 10/27 state is obtained from the three-component $\nu=3/7$ state:
$$[1,[1,1,1]_{2}]_{2} \leftrightarrow (7/27,1/27,1/27,1/27).$$

\begin{table}
\begin{center}
\begin{tabular}{|c|c|c|c|c|}
\hline
\multicolumn{1}{|c|}{$\nu$} &
\multicolumn{2}{|c|}{$[1,[2,1]_{2}]_{2} \leftrightarrow (\frac{7}{27},\frac{2}{27},\frac{1}{27})$} &
\multicolumn{2}{|c|}{$[1,[1,1,1]_{2}]_{2} \leftrightarrow (\frac{7}{27},\frac{1}{27},\frac{1}{27},\frac{1}{27})$} \\ \hline
		    & exact 	   & CF w. f.		& exact & CF w. f.      \\ \hline
10/27 		    & -		   & -0.42350(4)  	&       &-0.42353(4)    \\ \hline
\end{tabular}
\end{center}
\caption {The Coulomb energies of the states obtained from $\nu^{*}=10/7$ with $p=1$ and parallel flux attachment.} 
\label{tab:Table_10_27b} 
\end{table}

\subsection{$\nu=10/13$ (parent state $\nu^{*}=10/7$)}

\subsubsection{One component fully polarized $10/13$}
The fully polarized one component 10/13 state 
$$[1+[3]_{2}]_{-2} \leftrightarrow(10/13)$$
corresponds to $\nu^*=10/7$, which is obtained by filling the lowest $\Lambda$L completely and forming a 3/7 state in the second $\Lambda$L. An exactly equivalent state is constructed from $\nu_{h}=3/13$ and is denoted by:
$$\overline{[3]_{4}}$$

\subsubsection{Two component partially polarized $10/13$}
The partially polarized two-component 10/13 state is obtained from fully polarized $\nu_{h}=4/7$ state:
$$[1,[3]_{2}]_{-2} \equiv [\overline{[4]_{-2}}]_{-2} \leftrightarrow (7/13,3/13)$$ 

\subsubsection{Two component spin singlet $10/13$}
The singlet two-component 10/13 state is obtained from spin singlet $\nu_{h}=4/7$ state:
$$[\overline{[2,2]_{-2}}]_{-2} \leftrightarrow (5/13,5/13)$$

Table \ref{tab:Table_10_13} shows the Coulomb energies for these states extrapolated to the thermodynamic limit and in Table \ref{tab:critical_Ez} we give the critical Zeeman energies for the transitions among these states. 

\begin{table*}
\begin{center}
\begin{tabular}{|c|c|c|c|c|c|c|}
\hline
\multicolumn{1}{|c|}{$\nu$} & \multicolumn{2}{|c|}{$[1+[3]_{2}]_{-2} \equiv \overline{[3]_{4}} \leftrightarrow(10/13)$} & \multicolumn{2}{|c|}{$[1,[3]_{2}]_{-2} \equiv [\overline{[4]_{-2}}]_{-2} \leftrightarrow (7/13,3/13)$} & \multicolumn{2}{|c|}{$[\overline{[2,2]_{-2}}]_{-2} \leftrightarrow (5/13,5/13)$}  \\ \hline
		    & exact  		& CF 		& 	exact	& 	CF 		 & exact	  		& CF 		     	\\ \hline
10/13	            & -0.5414(5)	& - 		&    -0.5519(0)	&  	-		 &  				& -			 \\ \hline
\end{tabular}
\end{center}
\caption {The Coulomb energies of the states obtained from $\nu^{*}=10/7$ with $p=1$ and reverse flux attachment.} 
\label{tab:Table_10_13} 
\end{table*}

\subsubsection{Three component partially polarized $10/13$}
This 10/13 state is obtained from the partially polarized $\nu=3/7$ state by reverse flux attachment:
$$[1,[2,1]_{2}]_{-2} \leftrightarrow (7/13,2/13,1/13).$$ 

\subsubsection{Four component partially polarized $10/27$}
This 10/13 state is obtained from the three-component $\nu=3/7$ state by reverse flux attachment:
$$[1,[1,1,1]_{2}]_{-2} \leftrightarrow (7/13,1/13,1/13,1/13).$$ 

\subsection{$\nu=11/29$ (parent state $\nu^{*}=11/7$)}

\subsubsection{One component fully polarized $11/29$}
The fully polarized one component 11/29 state 
$$[1+[4]_{-2}]_{2} \leftrightarrow(11/29)$$
corresponds to $\nu^*=11/7$, which is obtained by filling the lowest $\Lambda$L completely and forming a 4/7 state in the second $\Lambda$L.

\subsubsection{Two component partially polarized $11/29$}
\paragraph{Obtained from fully polarized $3/7$ of holes:}
This partially polarized two-component 11/29 state is obtained from fully polarized $\nu_{h}=3/7$ state:
$$[1,[4]_{-2}]_{2} \equiv [\overline{[3]_{2}}]_{2} \leftrightarrow (7/29,4/29)$$ 
\paragraph{Obtained from partially polarized $3/7$ of holes:}
This partially polarized two-component 11/29 state is obtained from partially polarized $\nu_{h}=3/7$ state:
$$[\overline{[1,2]_{2}}]_{2} \leftrightarrow (6/29,5/29)$$ 

Table \ref{tab:Table_11_29} shows the Coulomb energies for these states extrapolated to the thermodynamic limit and in Table \ref{tab:critical_Ez} we give the critical Zeeman energies for the transitions among these states. 

\begin{table*}
\begin{center}
\begin{tabular}{|c|c|c|c|c|c|c|c|}
\hline
\multicolumn{1}{|c|}{$\nu$} & \multicolumn{2}{|c|}{$[1+[4]_{-2}]_{2} \leftrightarrow(11/29)$} & \multicolumn{3}{|c|}{$[1,[4]_{-2}]_{2} \equiv [\overline{[3]_{2}}]_{2} \leftrightarrow (7/29,4/29)$} & \multicolumn{2}{|c|}{$[\overline{[1,2]_{2}}]_{2} \leftrightarrow (6/29,5/29)$}  \\ \hline
		    & exact  	& CFD 		& 	exact	& CFD 		& CF w. f.	& exact & CF 	\\ \hline
11/29 		    & 		& -0.4213(0) 	& 	  	& -0.4279(0) 	& -0.4281(3)  	&	&	\\ \hline
\end{tabular}
\end{center}
\caption {The Coulomb energies of the states obtained from $\nu^{*}=11/7$ with $p=1$ and parallel flux attachment.} 
\label{tab:Table_11_29} 
\end{table*}

\subsubsection{Three component state partially polarized $11/29$}
\paragraph{Obtained from the partially polarized $\nu=4/7$ state:}
This state is
$$[1,[3,1]_{-2}]_{2} \leftrightarrow (7/29,3/29,1/29).$$ 
\paragraph{Obtained obtained from the unpolarized $\nu=4/7$ state:}
This state is
$$[1,[2,2]_{-2}]_{2} \leftrightarrow (7/29,2/29,2/29).$$ 

\subsubsection{Four component partially polarized $11/29$}
This state is obtained from one of the partially polarized $\nu=4/7$ states:
$$[1,[2,1,1]_{-2}]_{2} \leftrightarrow (7/29,2/29,1/29,1/29).$$ 

\subsubsection{Five component partially polarized $11/29$}
This state is obtained from the $\nu=4/7$ state that would be SU(4) singlet in a four-component system:
$$[1,[1,1,1,1]_{-2}]_{2} \leftrightarrow (7/29,1/29,1/29,1/29,1/29).$$ 

\subsection{$\nu=11/15$ (parent state $\nu^{*}=11/7$)}

\subsubsection{One component fully polarized $11/15$}
The fully polarized one component 11/15 state 
$$[1+[4]_{-2}]_{-2} \leftrightarrow(11/15)$$
corresponds to $\nu^*=11/7$, which is obtained by filling the lowest $\Lambda$L completely and forming a 4/7 state in the second $\Lambda$L. An exactly identical state is obtained by taking the particle-hole conjugate of the state at $\nu_{h}=4/15$. We denote this state by:
$$\overline{[4]_{-4}}$$

\subsubsection{Two component partially polarized $11/15$}
\paragraph{Obtained from fully polarized $3/7$ of holes:}
This partially polarized two-component 11/15 state is obtained from fully polarized $\nu_{h}=3/7$ state:
$$[1,[4]_{-2}]_{-2} \equiv [\overline{[3]_{2}}]_{-2} \leftrightarrow (7/15,4/15)$$ 
The energy in the thermodynamic limit obtained from exact diagonalization is -0.5336(0).
\paragraph{Obtained from partially polarized $3/7$ of holes:}
This partially polarized two-component 11/15 state is obtained from partially polarized $\nu_{h}=3/7$ state:
$$[\overline{[1,2]_{2}}]_{-2} \leftrightarrow (6/15,5/15)$$ 

In Table \ref{tab:critical_Ez} we give the critical Zeeman energies for the transitions among these states. 

\subsubsection{Three component partially polarized $11/15$}
\paragraph{Obtained from the partially polarized $\nu=4/7$ state:}
This state is
$$[1,[3,1]_{-2}]_{-2} \leftrightarrow (7/15,3/15,1/15).$$ 
\paragraph{Obtained from the unpolarized $\nu=4/7$ state:}
This state is
$$[1,[2,2]_{-2}]_{-2} \leftrightarrow (7/15,2/15,2/15).$$ 

\subsubsection{Four component partially polarized $11/15$}
This state is obtained from one of the partially polarized $\nu=4/7$ states:
$$[1,[2,1,1]_{-2}]_{-2} \leftrightarrow (7/15,2/15,1/15,1/15).$$ 

\subsubsection{Five component partially polarized $11/15$}
This state is obtained from the $\nu=4/7$ state that would be SU(4) singlet in a four-component system:
$$[1,[1,1,1,1]_{-2}]_{-2} \leftrightarrow (7/15,1/15,1/15,1/15,1/15).$$ 

\subsection{$\nu=3/8$ (parent state $\nu^{*}=3/2$)}
\subsubsection{One component fully polarized $3/8$}
The fully polarized one component 3/8 state 
$$[1+1/2^{\rm APf}]_{2} \leftrightarrow(3/8)$$
corresponds to $\nu^*=3/2$, which is obtained by filling the lowest $\Lambda$L completely and forming an anti-Pfaffian (APf) state in the second $\Lambda$L. Reference \cite{Mukherjee12} has shown that the APf state is favored over the Moore-Read Pfaffian state in the second $\Lambda$L.

\subsubsection{Two component partially polarized $3/8$}
The partially polarized two-component 3/8 state 
$$[1,1/2^{\rm APf}]_{2} \leftrightarrow (2/8,1/8)$$ 
is obtained from the partially polarized 3/2 state
$$[1,1/2^{\rm APf}] \leftrightarrow(1,1/2)$$ 
which in turn is obtained by filling the lowest $\Lambda$L of spin-up completely and forming an APf state in the spin-down lowest $\Lambda$L.  Reference \cite{Mukherjee14c} shows that this state provides an almost exact realization of the APf state.

\subsubsection{Three or more component $3/8$}
It is not possible to construct a wave function of the type considered here with three or more components that satisfies Fock conditions.

Table \ref{tab:Table_3_8} shows the Coulomb energies for these states extrapolated to the thermodynamic limit and in Table \ref{tab:critical_Ez} we give the critical Zeeman energies for the transitions among these states. The numbers for fully polarized and partially polarized states are reproduced from references \cite{Mukherjee12} and \cite{Mukherjee14c} respectively.

\begin{table}
\begin{center}
\begin{tabular}{|c|c|c|c|c|}
\hline
\multicolumn{1}{|c|}{$\nu$} & \multicolumn{2}{|c|}{$[1+1/2^{\rm APf}]_{2} \leftrightarrow(3/8)$} & \multicolumn{2}{|c|}{$[1,1/2^{\rm APf}]_{2} \leftrightarrow (2/8,1/8)$}  \\ \hline
		    & exact	  		& CFD 		    		& exact	  		& CFD 		     \\ \hline
3/8 		    & -0.4215(0) 		& -0.4195(2) 			& -	  	  	& -0.4256(1)	     \\ \hline
\end{tabular}
\end{center}
\caption {The Coulomb energies of the states obtained from $\nu^{*}=3/2$ with $p=1$ and parallel flux attachment.}  
\label{tab:Table_3_8} 
\end{table}

\begin{table*}
\begin{center}
\begin{tabular}{|c|c|c|c|c|}
\hline
\multicolumn{1}{|c|}{$\nu$} & \multicolumn{1}{|c|}{Transition} 								& \multicolumn{2}{|c|}{$E_{Z}^{c}$}	\\ \hline
	&														&	exact		& 	CFD	\\ \hline
4/11	& $[1,[1]_{2}]_{2} \leftrightarrow [\overline{[1,1]_{-2}}]_{2}$ 						& 	0.0026		&	0.0070	\\ \hline
4/11	& $[1+1/3^{\rm WYQ}]_{2} \leftrightarrow [1,[1]_{2}]_{2}$ 							& 	0.0208		&	0.0171	\\ \hline
4/5	& $[1,[1]_{2}]_{-2} \leftrightarrow [\overline{[1,1]_{-2}}]_{-2}$ 						& 	0.0117		&	-	\\ \hline
4/5	& $[1+[1]_{2}]_{-2}\equiv \overline{[1]_{4}} \leftrightarrow [1,[1]_{2}]_{-2}$ 					& 	0.0388		&	-	\\ \hline
5/13	& $[1+2/3^{\rm WYQ}]_{2} \leftrightarrow [1,[2]_{-2}]_{2}$ 							& 	0.0183		&	0.0149	\\ \hline
5/7	& $[1+[2]_{-2}]_{-2} \equiv \overline{[2]_{-4}} \leftrightarrow [1,[2]_{-2}]_{-2}$ 				& 	0.0238		&	-	\\ \hline
7/19	& $[1,[2]_{2}]_{2} \equiv [\overline{[3]_{-2}}]_{2} \leftrightarrow [\overline{[1,2]_{-2}}]_{2}$		& 	-		&	0.0095	\\ \hline
7/19	& $[1+[2]_{2}]_{2} \leftrightarrow [1,[2]_{2}]_{2} \equiv [\overline{[3]_{-2}}]_{2}$				& 	-		&	0.0194	\\ \hline
7/9	& $[1,[2]_{2}]_{-2} \equiv [\overline{[3]_{-2}}]_{-2} \leftrightarrow [\overline{[1,2]_{-2}}]_{-2}$		& 	-		&	-	\\ \hline
7/9	& $[1+[2]_{2}]_{-2} \equiv \overline{[2]_{4}} \leftrightarrow [1,[2]_{2}]_{-2} \equiv [\overline{[3]_{-2}}]_{-2}$& 	0.0320		&	-	\\ \hline
8/21	& $[1,[3]_{-2}]_{2} \equiv [\overline{[2]_{2}}]_{2} \leftrightarrow [\overline{[1,1]_{2}}]_{2}$			& 	-		&	-	\\ \hline
8/21	& $[1+[3]_{-2}]_{2} \leftrightarrow [1,[3]_{-2}]_{2} \equiv [\overline{[2]_{2}}]_{2}$				& 	-		&	-	\\ \hline
8/11	& $[1,[3]_{-2}]_{-2} \equiv [\overline{[2]_{2}}]_{-2} \leftrightarrow [\overline{[1,1]_{2}}]_{-2}$		& 	-		&	-	\\ \hline
8/11	& $[1+[3]_{-2}]_{-2} \equiv \overline{[3]_{-4}}\leftrightarrow [1,[3]_{-2}]_{-2} \equiv [\overline{[2]_{2}}]_{-2}$& 	-		&	-	\\ \hline
10/27	& $[1,[3]_{2}]_{2} \equiv [\overline{[4]_{-2}}]_{2} \leftrightarrow [\overline{[2,2]_{-2}}]_{2}$		& 	-		&	-	\\ \hline
10/27	& $[1+[3]_{2}]_{2} \leftrightarrow [1,[3]_{2}]_{2} \equiv [\overline{[4]_{-2}}]_{2}$ 				& 	-		&	0.0203	\\ \hline
10/13	& $[1,[3]_{2}]_{-2} \equiv [\overline{[4]_{-2}}]_{-2} \leftrightarrow [\overline{[2,2]_{-2}}]_{-2}$		& 	-		&	-	\\ \hline
10/13	& $[1+[3]_{2}]_{-2} \equiv \overline{[3]_{4}} \leftrightarrow [1,[3]_{2}]_{-2} \equiv [\overline{[4]_{-2}}]_{-2}$& 	0.0349		&	-	\\ \hline
11/29	& $[1,[4]_{-2}]_{2} \equiv [\overline{[3]_{2}}]_{2} \leftrightarrow [\overline{[1,2]_{2}}]_{2}$			& 	-		&	-	\\ \hline
11/29	& $[1+[4]_{-2}]_{2} \leftrightarrow [1,[4]_{-2}]_{2} \equiv [\overline{[3]_{2}}]_{2}$ 				& 	-		&	0.0181	\\ \hline
11/15	& $[1,[4]_{-2}]_{-2} \equiv [\overline{[3]_{2}}]_{-2} \leftrightarrow [\overline{[1,2]_{2}}]_{-2}$		& 	-		&	-	\\ \hline
11/15	& $[1+[4]_{-2}]_{-2} \equiv \overline{[4]_{-4}} \leftrightarrow [1,[4]_{-2}]_{-2} \equiv [\overline{[3]_{2}}]_{-2}$& 	-		&	-	\\ \hline
3/8	& $[1+1/2^{\rm APf}]_{2} \leftrightarrow [1,1/2^{\rm APf}]_{2}$							& 	-		&	0.0183	\\ \hline
\end{tabular}
\end{center}
\caption {The critical Zeeman energy $E_{Z}^{c}$ in units of $e^2/\epsilon \ell$ for various transitions. For $E_{Z}>E_{Z}^{c}$ ($E_{Z}<E_{Z}^{c}$), the state on the left (right) is favored over the state on the right (left). The first column gives values obtained from extrapolating exact diagonalization results, the second one gives results obtained from CFD and the last column gives results obtained from calculating energies of CF wave functions. The critical Zeeman energies are extremely sensitive to the ground state energies as well as the extrapolation to the thermodynamic limit, so these numbers should only be taken as ball park estimates.}  
\label{tab:critical_Ez} 
\end{table*}

In Table \ref{tab:multi} we show energies extrapolated to the thermodynamic limit of states with atleast three components. 

\begin{figure*}[htbp]
\begin{center}
\includegraphics[width=0.62\columnwidth,keepaspectratio]{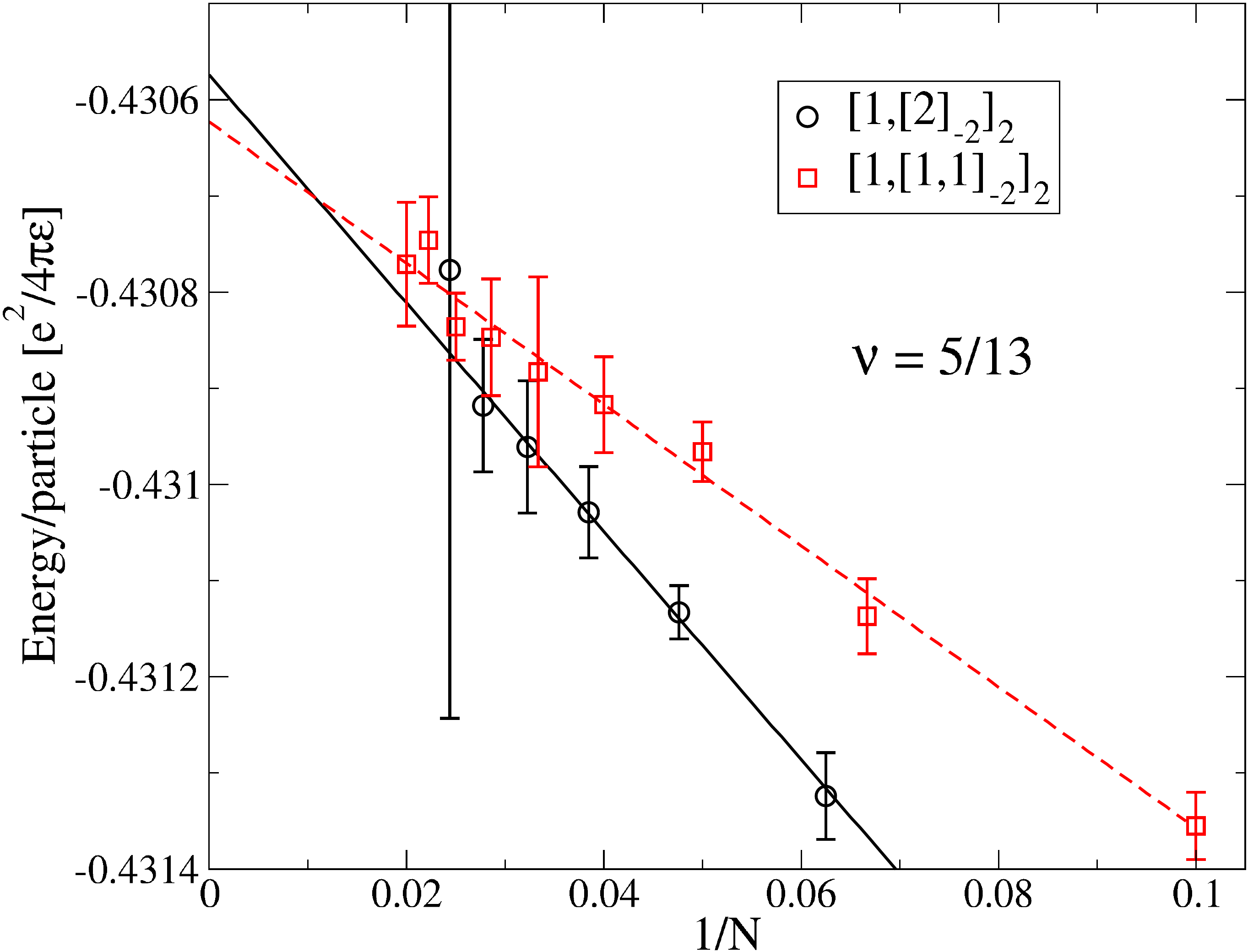}
\includegraphics[width=0.62\columnwidth,keepaspectratio]{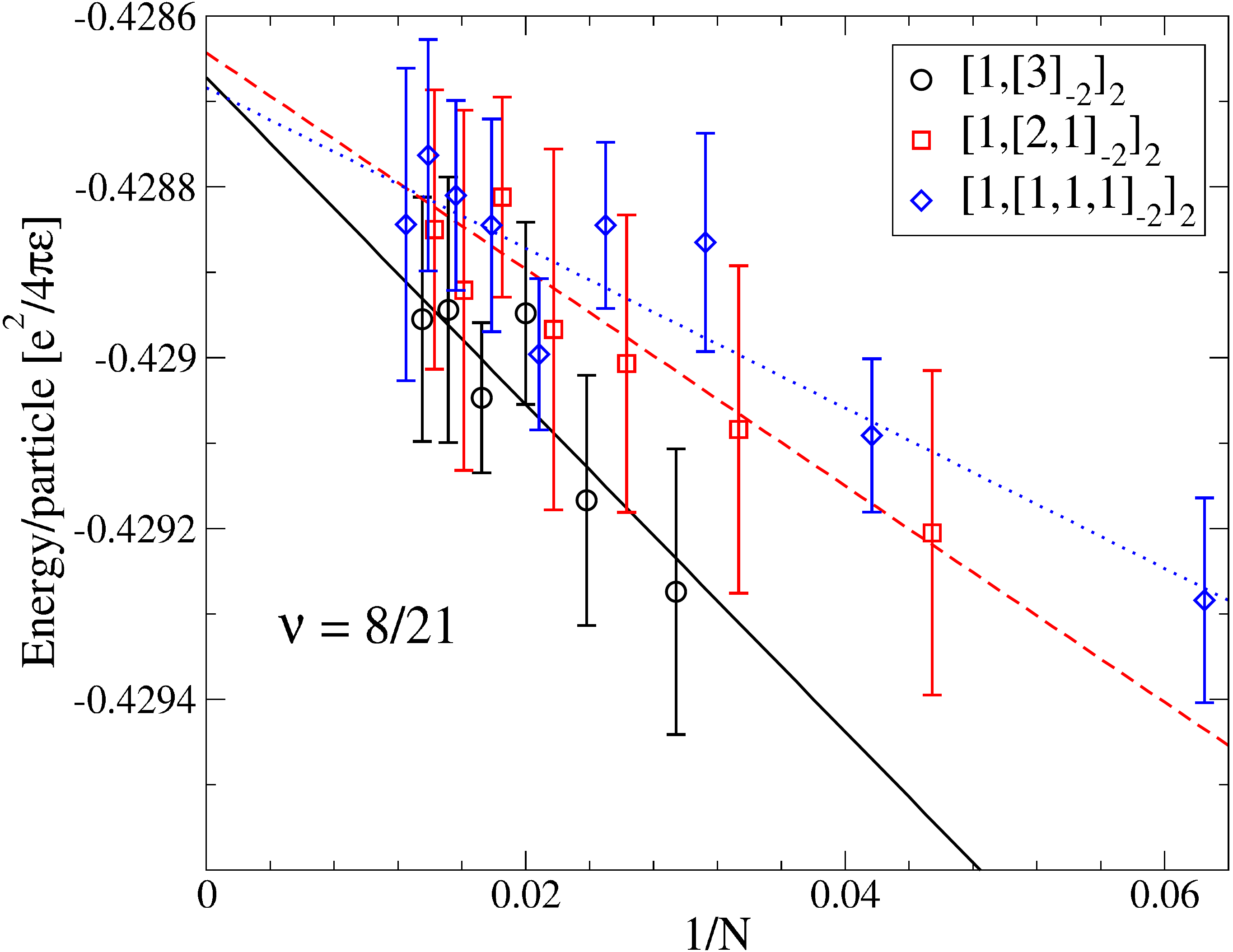}
\includegraphics[width=0.62\columnwidth,keepaspectratio]{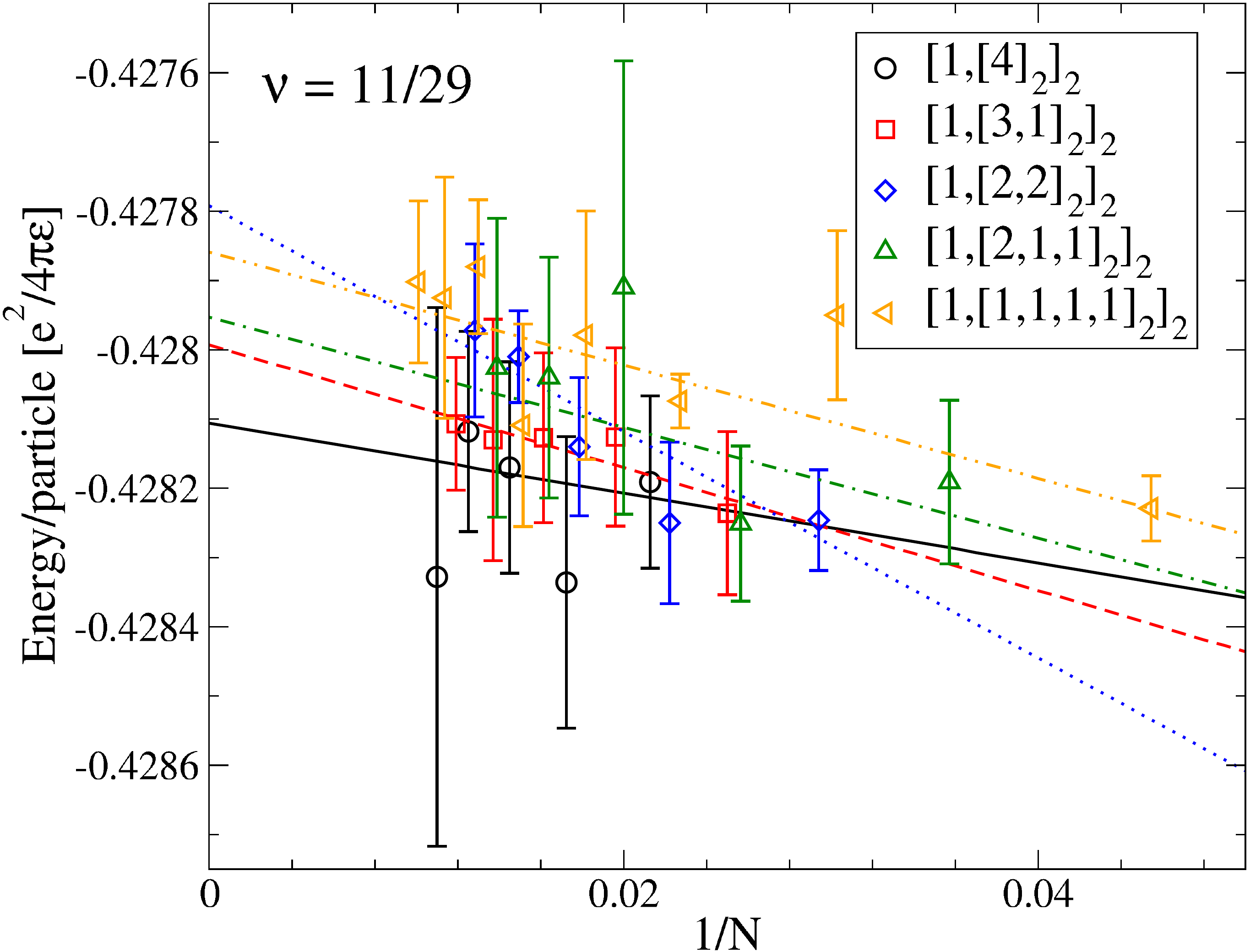}
\end{center}
\caption{\label{extra1}
Extrapolation of the ground state energy to the thermodynamic limit, assuming zero thickness.
The density correction has been applied.}
\end{figure*}

\begin{figure*}[htbp]
\begin{center}
\includegraphics[width=0.62\columnwidth,keepaspectratio]{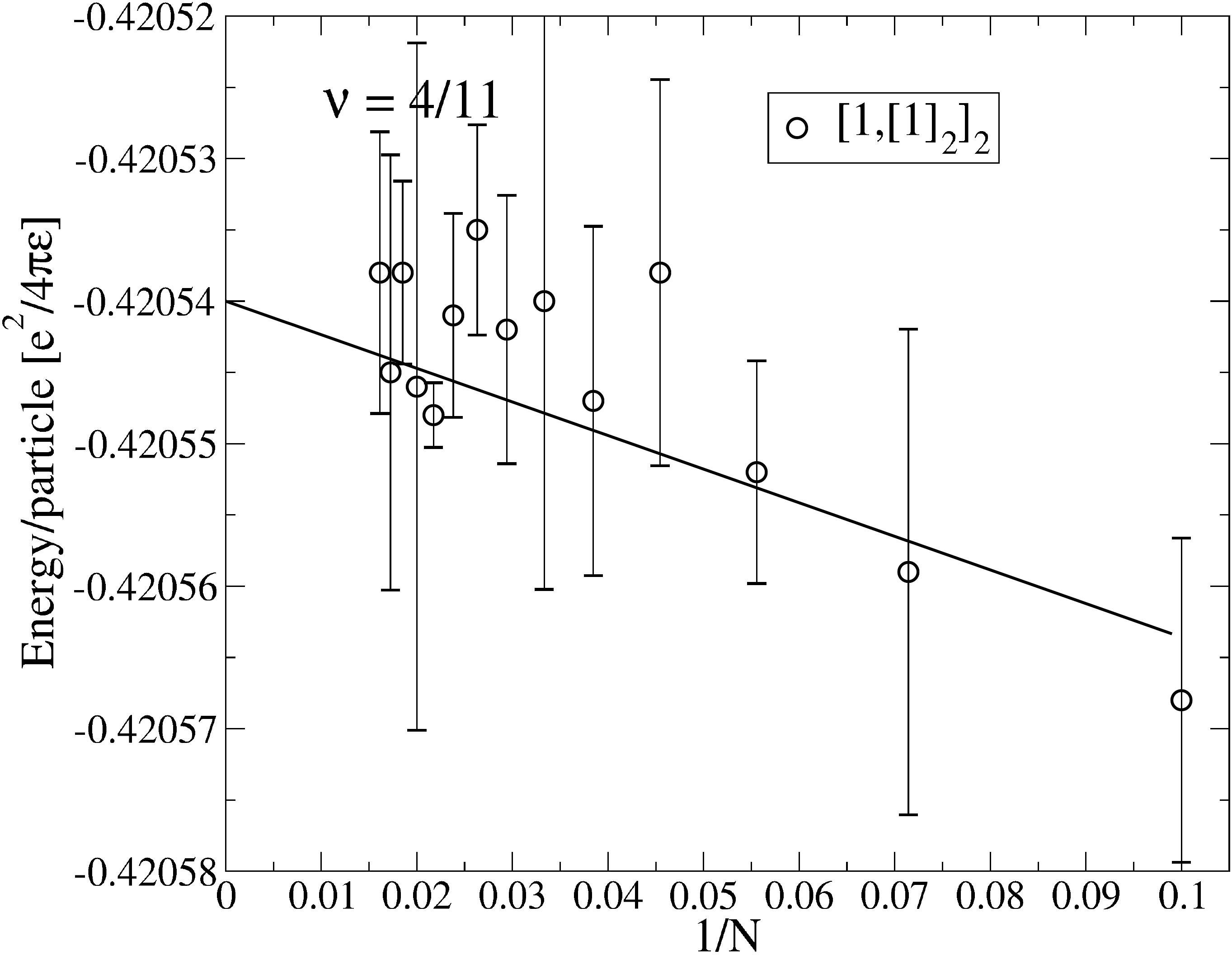}
\includegraphics[width=0.62\columnwidth,keepaspectratio]{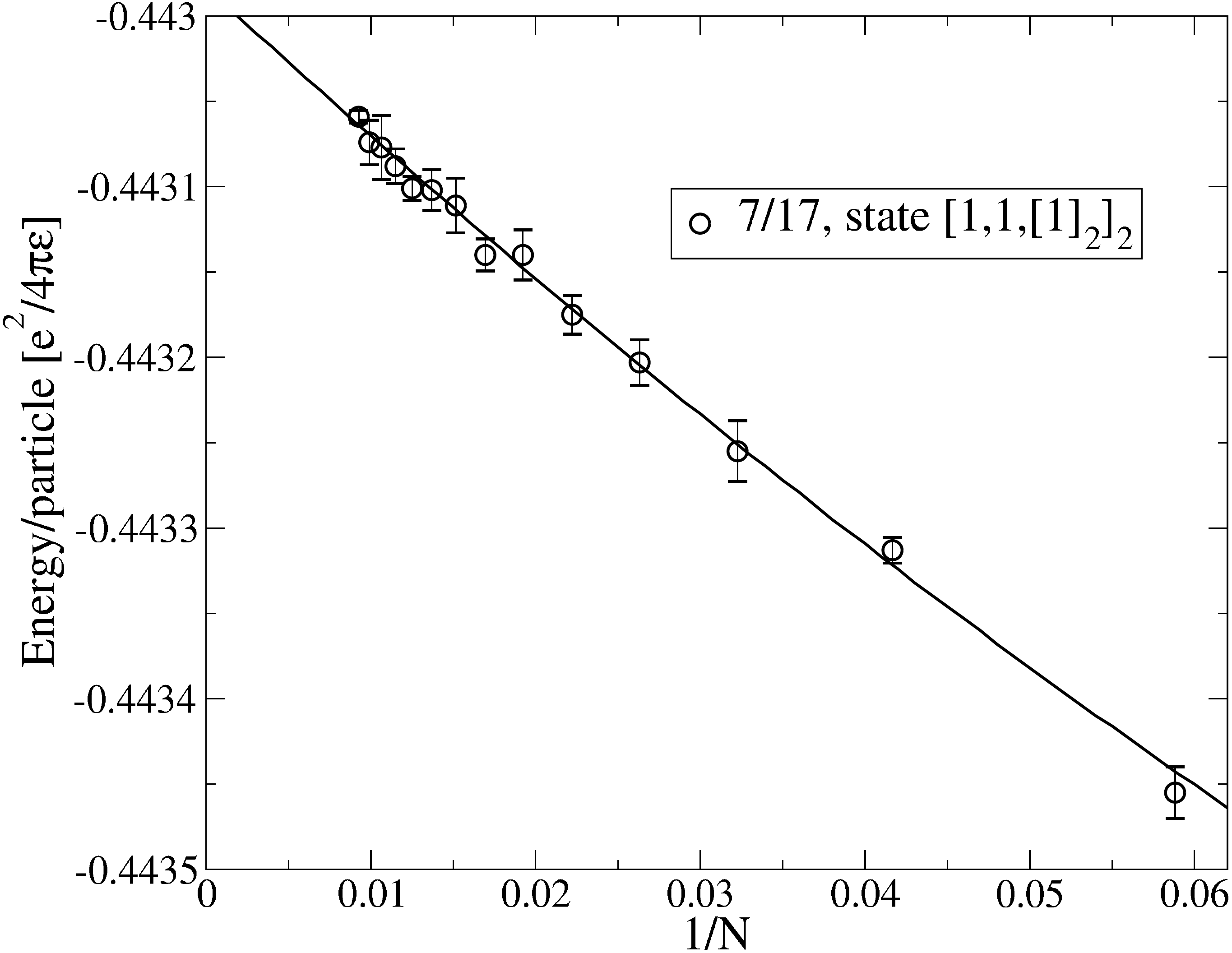}
\includegraphics[width=0.62\columnwidth,keepaspectratio]{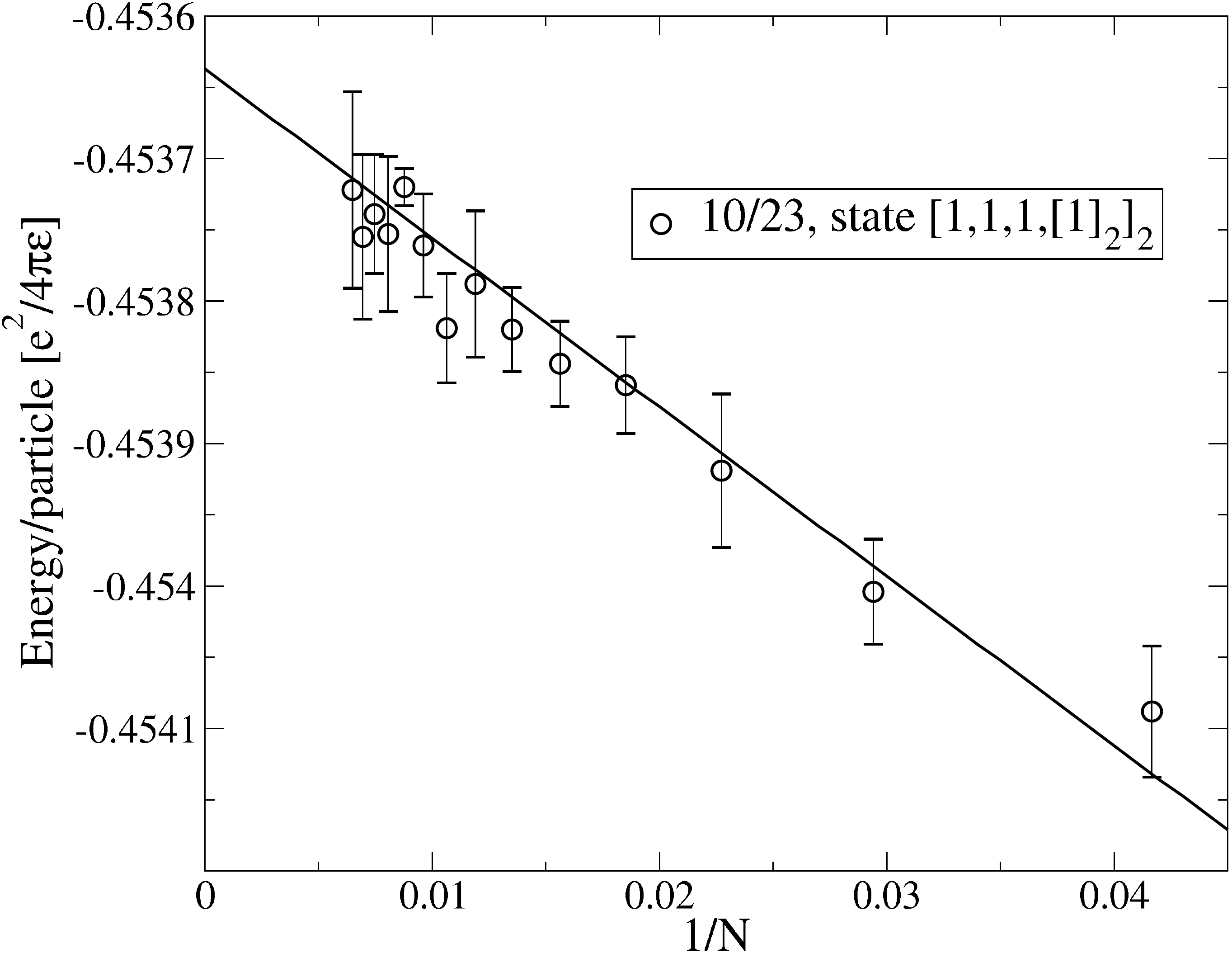}

\includegraphics[width=0.62\columnwidth,keepaspectratio]{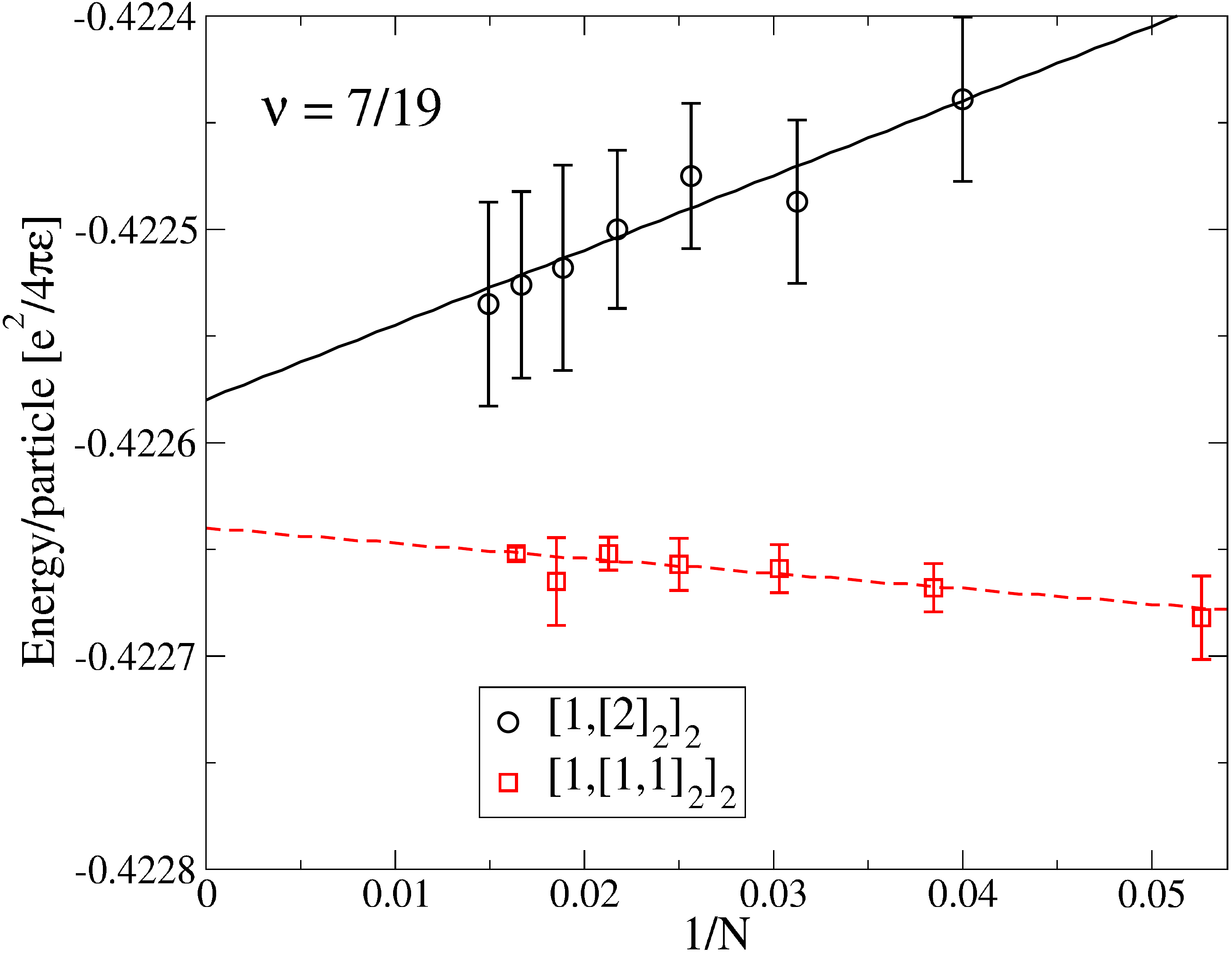}
\includegraphics[width=0.62\columnwidth,keepaspectratio]{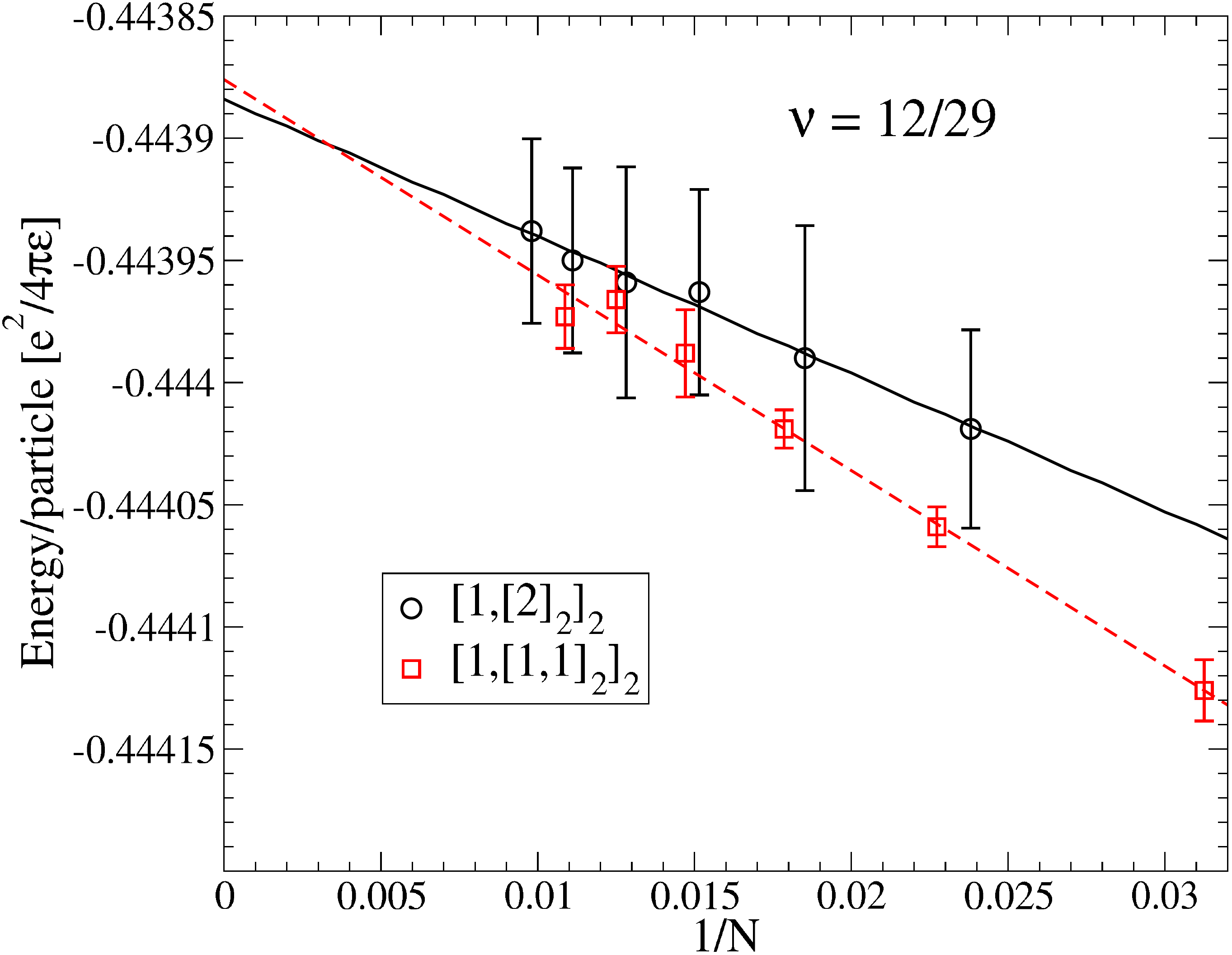}
\includegraphics[width=0.62\columnwidth,keepaspectratio]{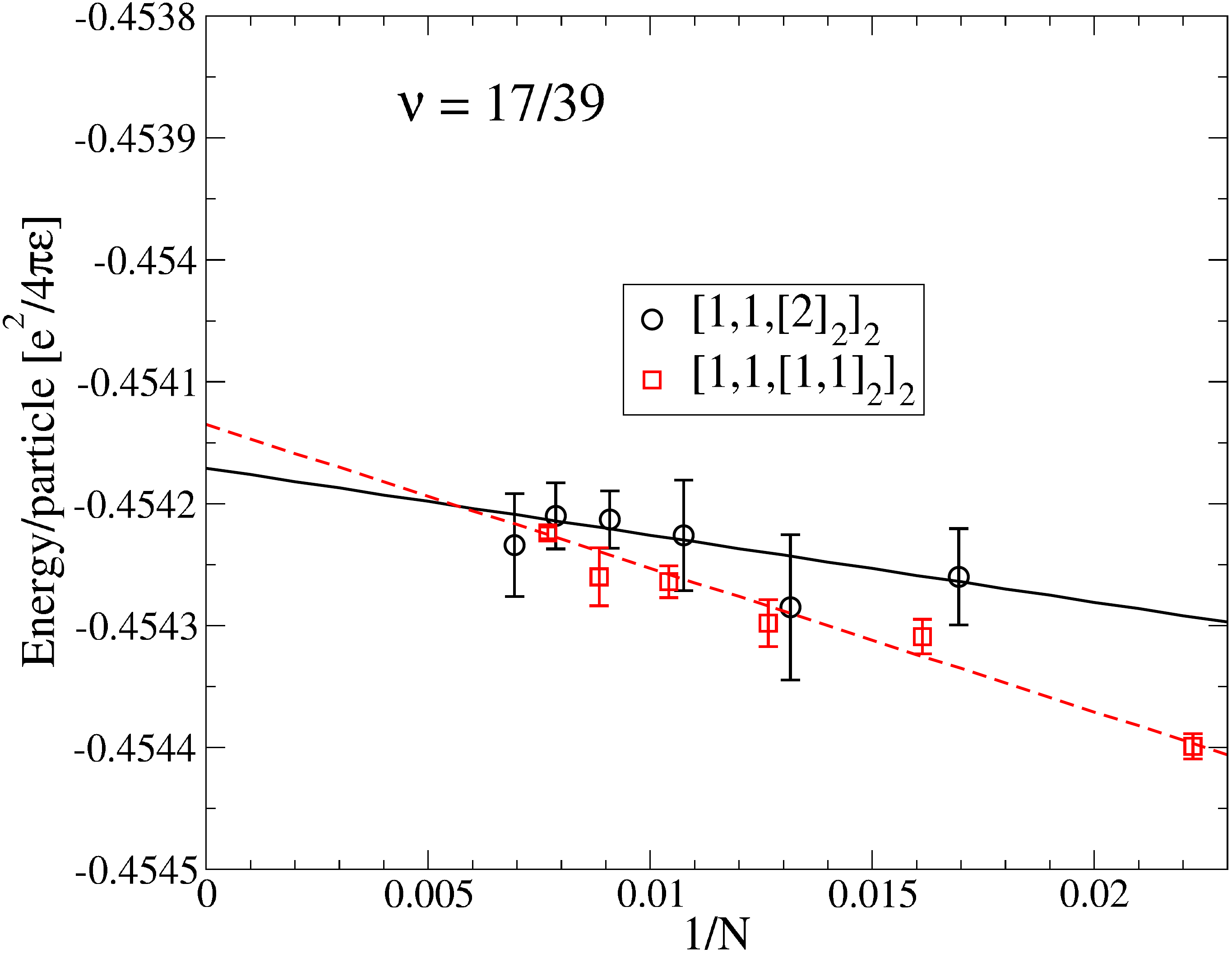}

\includegraphics[width=0.62\columnwidth,keepaspectratio]{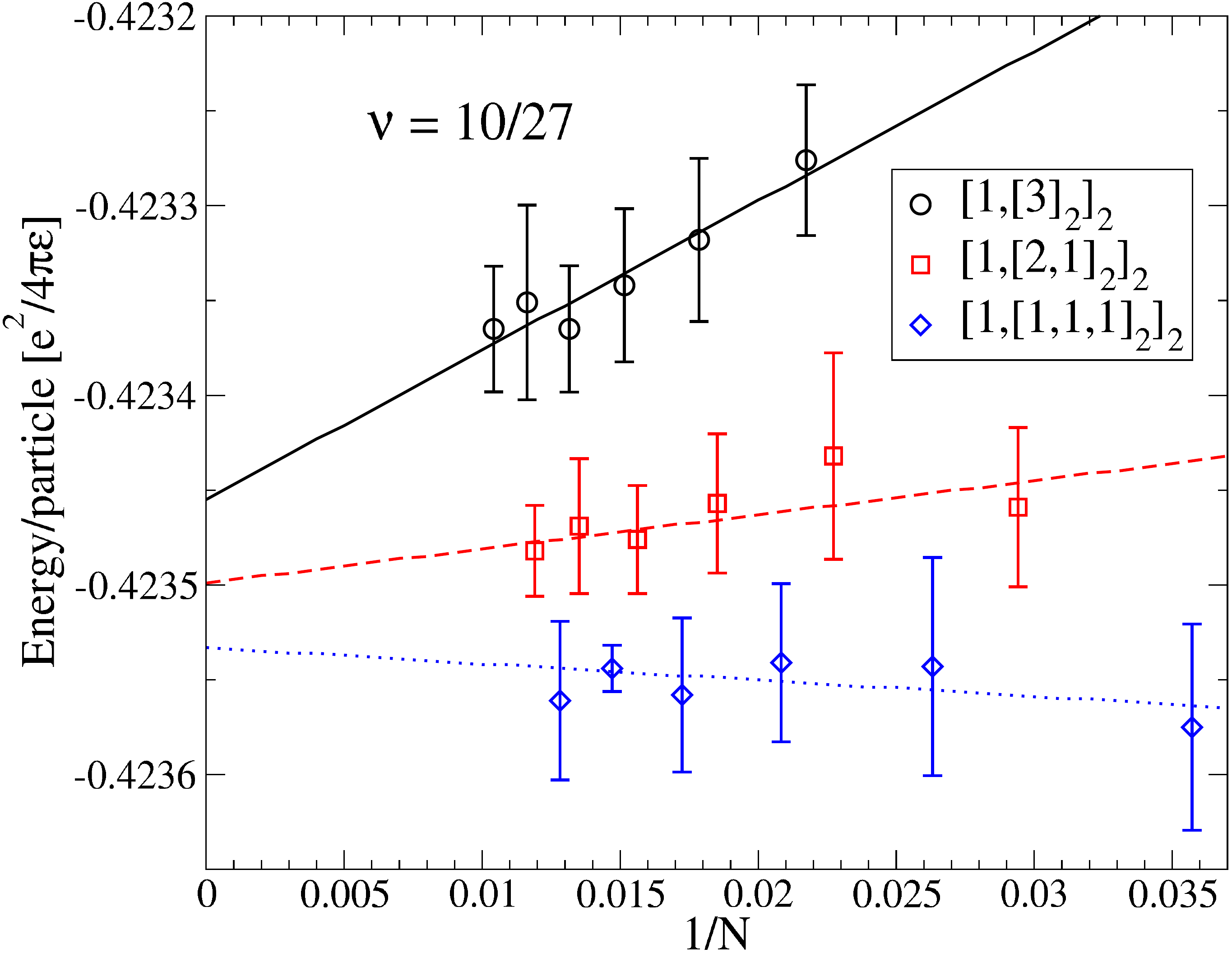}
\includegraphics[width=0.62\columnwidth,keepaspectratio]{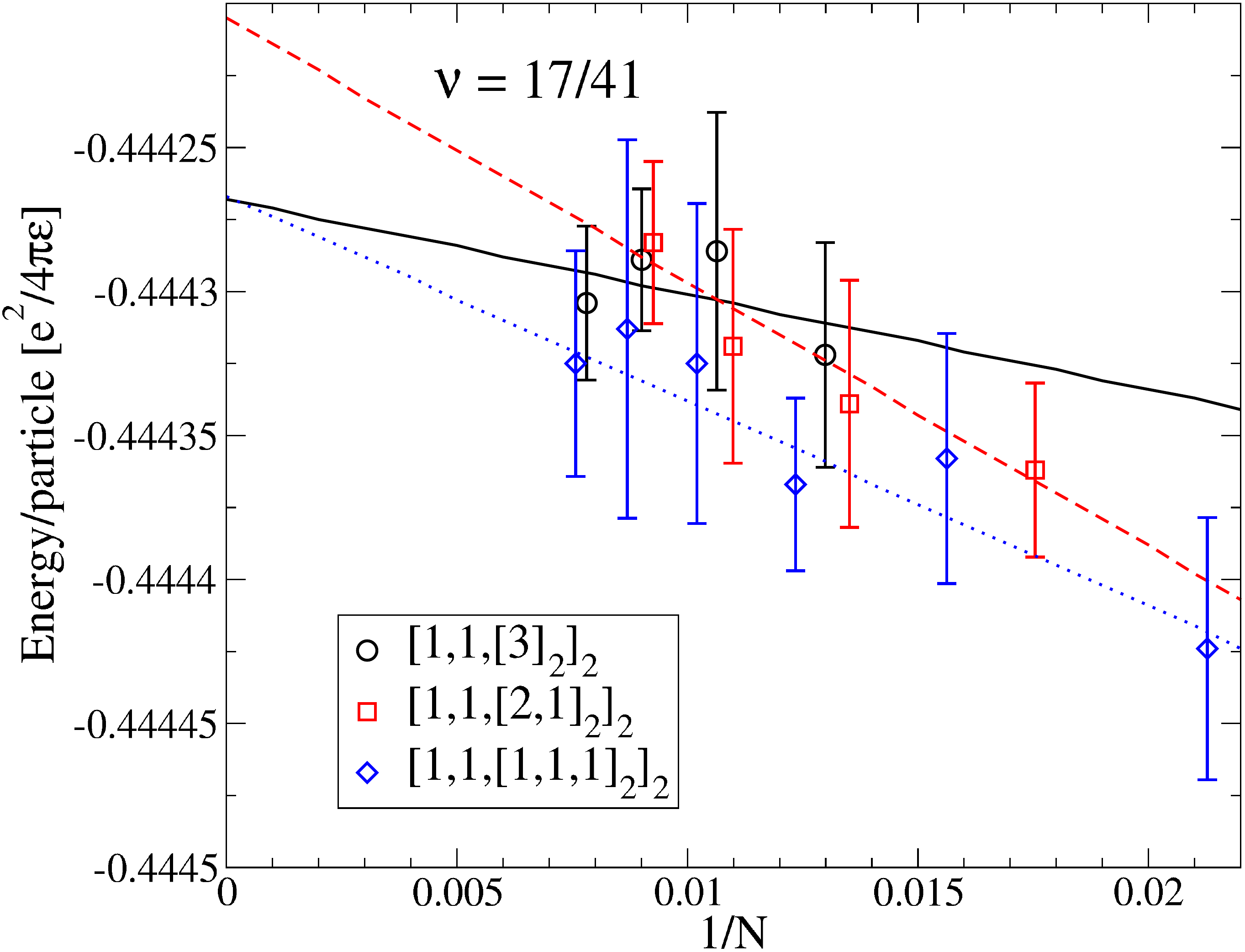}

\includegraphics[width=0.62\columnwidth,keepaspectratio]{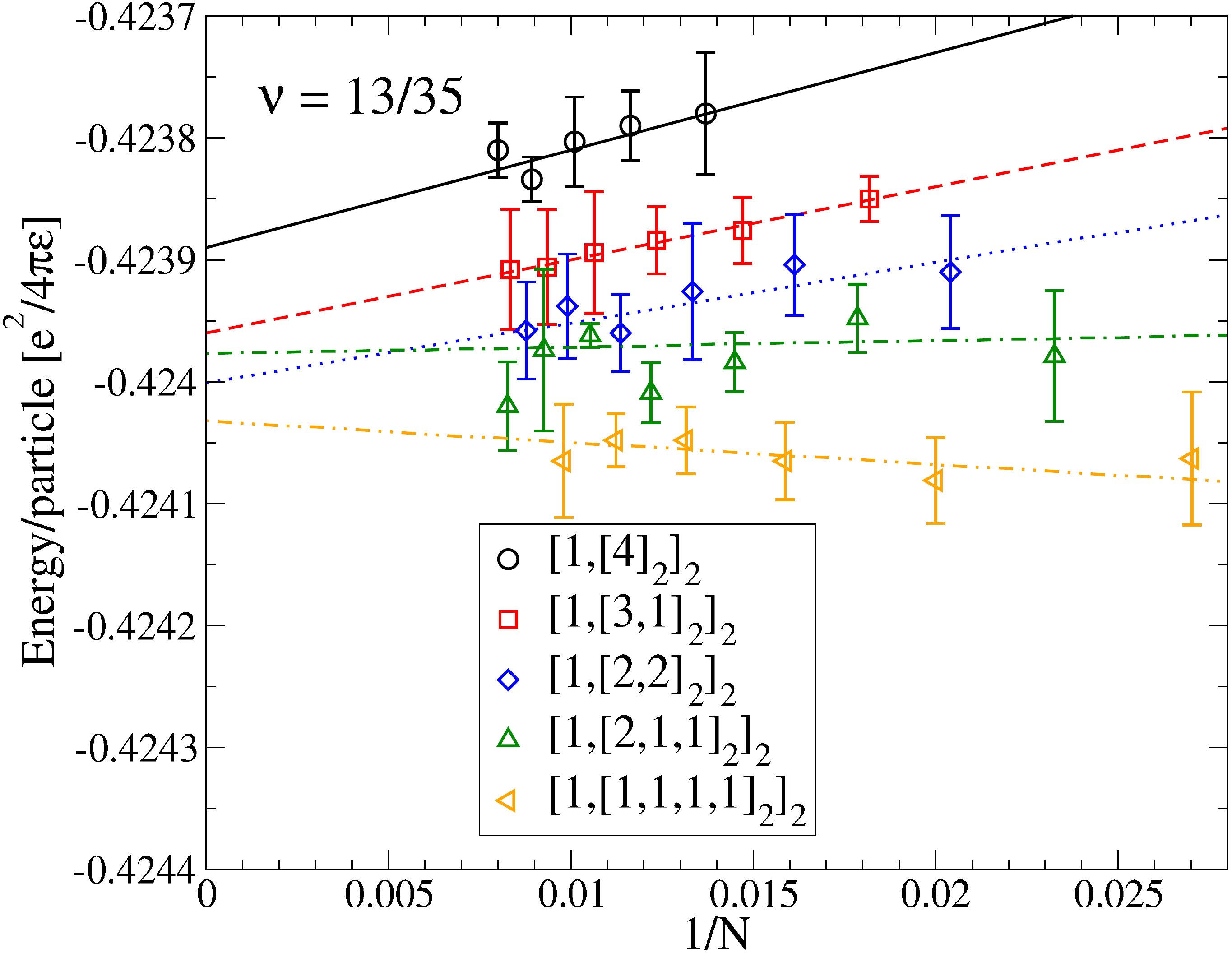}
\includegraphics[width=0.62\columnwidth,keepaspectratio]{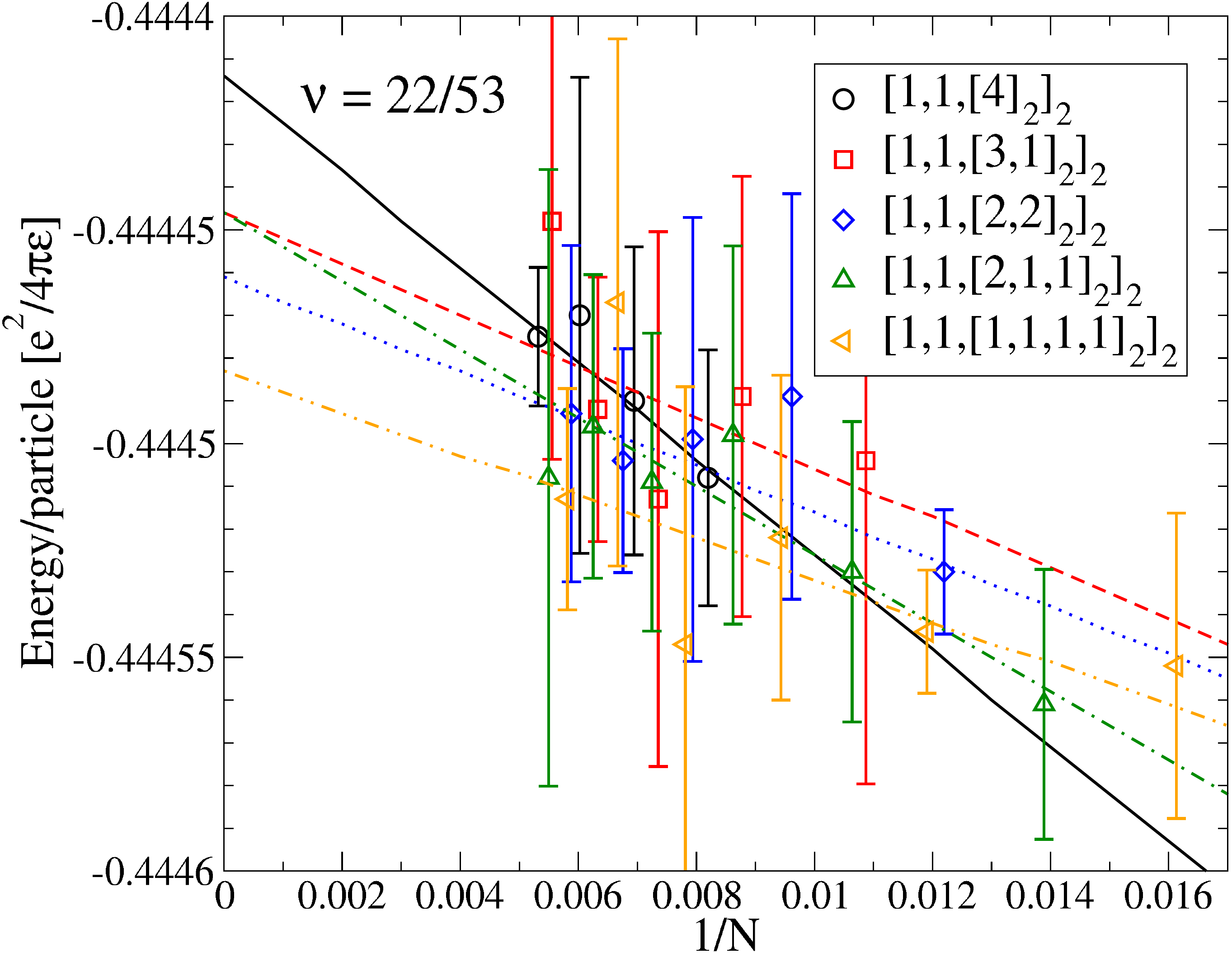}
\includegraphics[width=0.62\columnwidth,keepaspectratio]{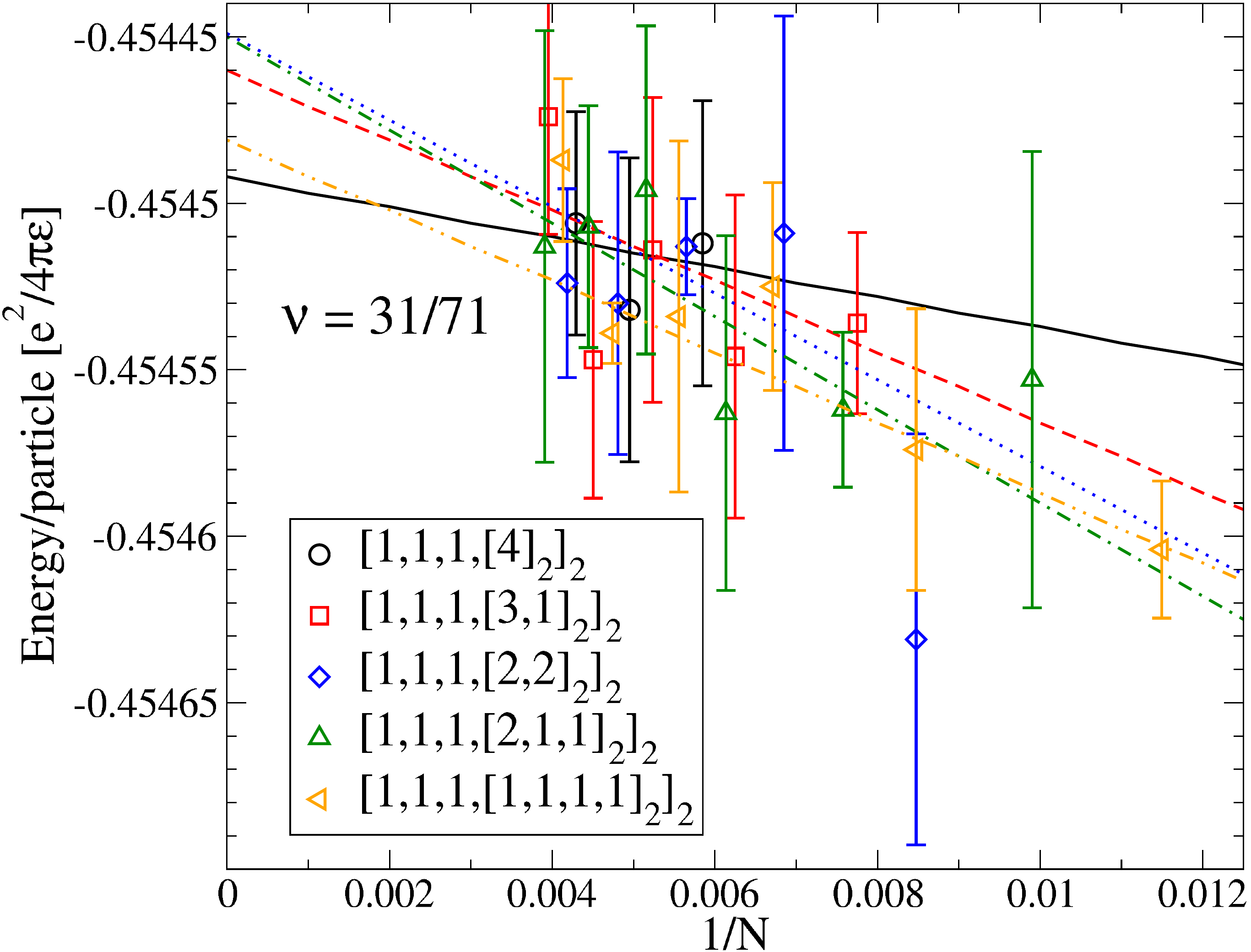}
\end{center}
\caption{\label{extra}
Extrapolation of the ground state energy to the thermodynamic limit, assuming zero thickness. The density correction has been applied. Among the fractions that allow several multicomponent states, the data is convincing only at $\nu=7/19$ and $\nu=10/27$; even here it is difficult to conclude anything beyond error bars (see Table \ref{tab:multi}).}
\end{figure*}

\begin{table*}
\begin{center}
\begin{tabular}{c|c|c|c|c|l|l}
\hline\hline
$\nu$ & Construction & $N_\text{min}$ & $N_\text{max}$ & Data points & energy & $\chi^2_\text{red}$ \\
\hline
$\frac{4}{11}$ & $[1,[1]_2]_2$ & 18 & 62 & 14 & -0.420540(4) & 0.60 \\
$\frac{7}{17}$ & $[1,1,[1]_2]_2$ & 17 & 108 & 14 & -0.442992(4) & 0.77 \\
$\frac{10}{23}$ & $[1,1,1,[1]_2]_2$ & 24 & 154 & 14 & -0.453649(5) & 0.24 \\
\hline
$\frac{7}{19}$ & $[1,[2]_2]_2$ & 25 & 67 & 7 & -0.42258(4) & 0.09 \\
             & $[1,[1,1]_2]_2$ & 19 & 61 & 7 & -0.42264(7) & 0.12 \\
$\frac{12}{29}$ & $[1,1,[2]_2]_2$ & 42 & 102 & 6 & -0.44388(5) & 0.04 \\
              & $[1,1,[1,1]_2]_2$ & 32 & 92 & 6 & -0.44388(1) & 0.31 \\
$\frac{17}{39}$ & $[1,1,1,[2]_2]_2$ & 59 & 144 & 6 & -0.45417(5) & 0.25 \\
              & $[1,1,1,[1,1]_2]_2$ & 45 & 130 & 6 & -0.454135(9) & 0.74 \\
\hline
$\frac{10}{27}$ & $[1,[3]_2]_2$ & 46 & 96 & 6 & -0.42346(6) & 0.09 \\
              & $[1,[2,1]_2]_2$ & 34 & 84 & 6 & -0.42350(4) & 0.12 \\
            & $[1,[1,1,1]_2]_2$ & 28 & 78 & 6 & -0.42353(4) & 0.10 \\
$\frac{17}{41}$ & $[1,1,[3]_2]_2$ & 77 & 128 & 4 & -0.44427(8) & 0.24 \\
              & $[1,1,[2,1]_2]_2$ & 57 & 108 & 4 & -0.44421(6) & 0.12 \\
            & $[1,1,[1,1,1]_2]_2$ & 47 & 132 & 6 & -0.44427(5) & 0.13 \\
$\frac{24}{55}$ & $[1,1,1,[3]_2]_2$ & 108 & 180 & 4 & -0.45438(8)& 0.09 \\
              & $[1,1,1,[2,1]_2]_2$ & 80 & 152 &4 & -0.45427(9) &0.63 \\
            & $[1,1,1,[1,1,1]_2]_2$ & 6 & 186 & 6 & -0.45437(3) & 0.69 \\
\hline
$\frac{13}{35}$ & $[1,[4]_2]_2$ & 73 & 125 & 5 & -0.42389(7) & 0.44 \\
              & $[1,[3,1]_2]_2$ & 55 & 120 & 6 & -0.42396(5) & 0.01 \\
              & $[1,[2,2]_2]_2$ & 49 & 114 & 6 & -0.42400(6) & 0.15 \\
            & $[1,[2,1,1]_2]_2$ & 43 & 121 & 7 & -0.42398(3) & 1.21 \\
          & $[1,[1,1,1,1]_2]_2$ & 37 & 102 & 6 & -0.42403(4) & 0.12 \\
$\frac{22}{53}$ & $[1,1,[4]_2]_2$ & 122 & 188 & 4 &-0.44441(7) & 0.028  \\
              & $[1,1,[3,1]_2]_2$ & 92 & 180 & 5 & -0.4444(1) & 0.19 \\
              & $[1,1,[2,2]_2]_2$ & 82 & 170 & 5 & -0.44446(5) & 0.12 \\
            & $[1,1,[2,1,1]_2]_2$ & 72 & 182 & 6 & -0.44444(5) & 0.06 \\
          & $[1,1,[1,1,1,1]_2]_2$ & 62 & 172 & 6 & -0.44448(4) & 0.24 \\
$\frac{31}{71}$ & $[1,1,1,[4]_2]_2$ &171&233&3& -0.4545(2) & 0.20 \\
              & $[1,1,1,[3,1]_2]_2$ &129&253&5&-0.45446(6) & 0.58 \\
              & $[1,1,1,[2,2]_2]_2$ &115&239&5&-0.45445(8) & 0.89 \\
            & $[1,1,1,[2,1,1]_2]_2$ &101&256&6&-0.45445(7) & 0.23 \\
          & $[1,1,1,[1,1,1,1]_2]_2$ & 87 & 242 & 6 & -0.45448(2) & 0.98 \\
\hline\hline
\end{tabular}
\end{center}
\caption{\label{tab:multi}
Energy per particle for various CF FQHE states involving only parallel flux attachment.}
\end{table*}

\begin{table*}
\begin{center}
\begin{tabular}{c|c|c|c|c|l|l}
\hline\hline
$\nu$ & Construction & $N_\text{min}$ & $N_\text{max}$ & Data points & energy & $\chi^2_\text{red}$ \\
\hline
$\frac{5}{13}$ & $[1,[2]_{-2}]_2$ & 16 & 41 & 6 & -0.43059(9) & 0.04 \\
               & $[1,[1,1]_{-2}]_2$ & 10 & 50 & 8 & -0.43062(3) & 0.37 \\
\hline
$\frac{8}{21}$ & $[1,[3]_{-2}]_2$ & 34 & 74 & 6 & -0.4287(2) & 0.36 \\
             & $[1,[2,1]_{-2}]_2$ & 22 & 70 & 7 & -0.4286(2) & 0.11 \\
             & $[1,[1,1,1]_{-2}]_2$ & 16 & 80 & 9 & -0.42868(8) & 0.48 \\
\hline
$\frac{11}{29}$ & $[1,[4]_{-2}]_2$ & 47 & 91 & 5 & -0.4281(3) & 0.27 \\
              & $[1,[3,1]_{-2}]_2$ & 40 & 95 & 6 & -0.4279(2) & 0.04 \\
              & $[1,[2,2]_{-2}]_2$ & 34 & 78 & 5 & -0.4278(1) & 0.44  \\
            & $[1,[2,1,1]_{-2}]_2$ & 28 & 72 & 5 & -0.4280(2) & 0.44 \\
            & $[1,[1,1,1,1]_{-2}]_2$ & 22 & 99 & 8 & -0.42786(6) & 0.65 \\
\hline\hline
\end{tabular}
\end{center}
\caption{\label{energies}
Energy per particle for the states with parallel flux attachment in the outer state and reverse flux attachment in the inner state.
}
\end{table*}

\begin{figure*}
\begin{center}
\includegraphics[width=0.9\textwidth]{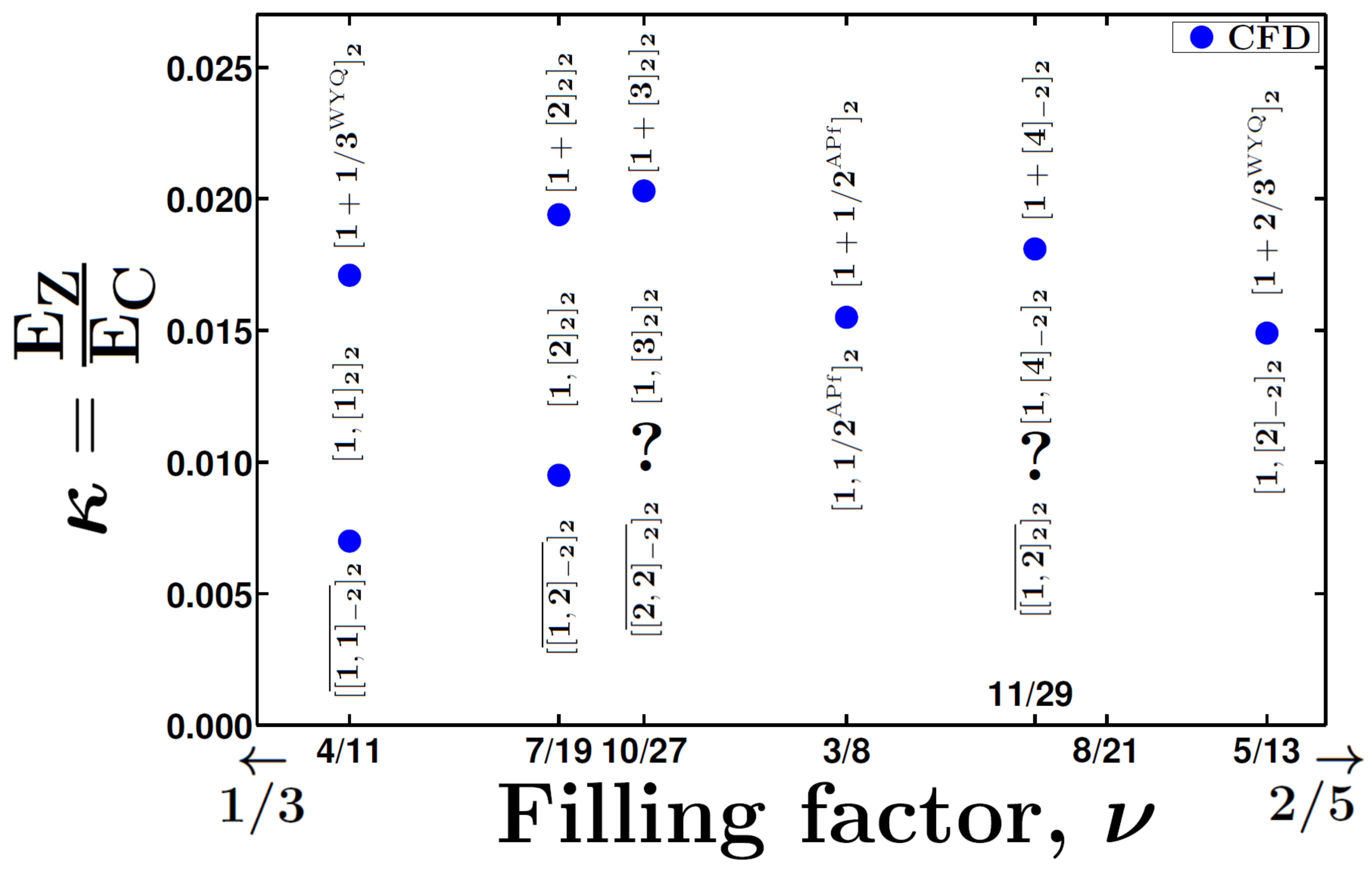}
\includegraphics[width=0.9\textwidth]{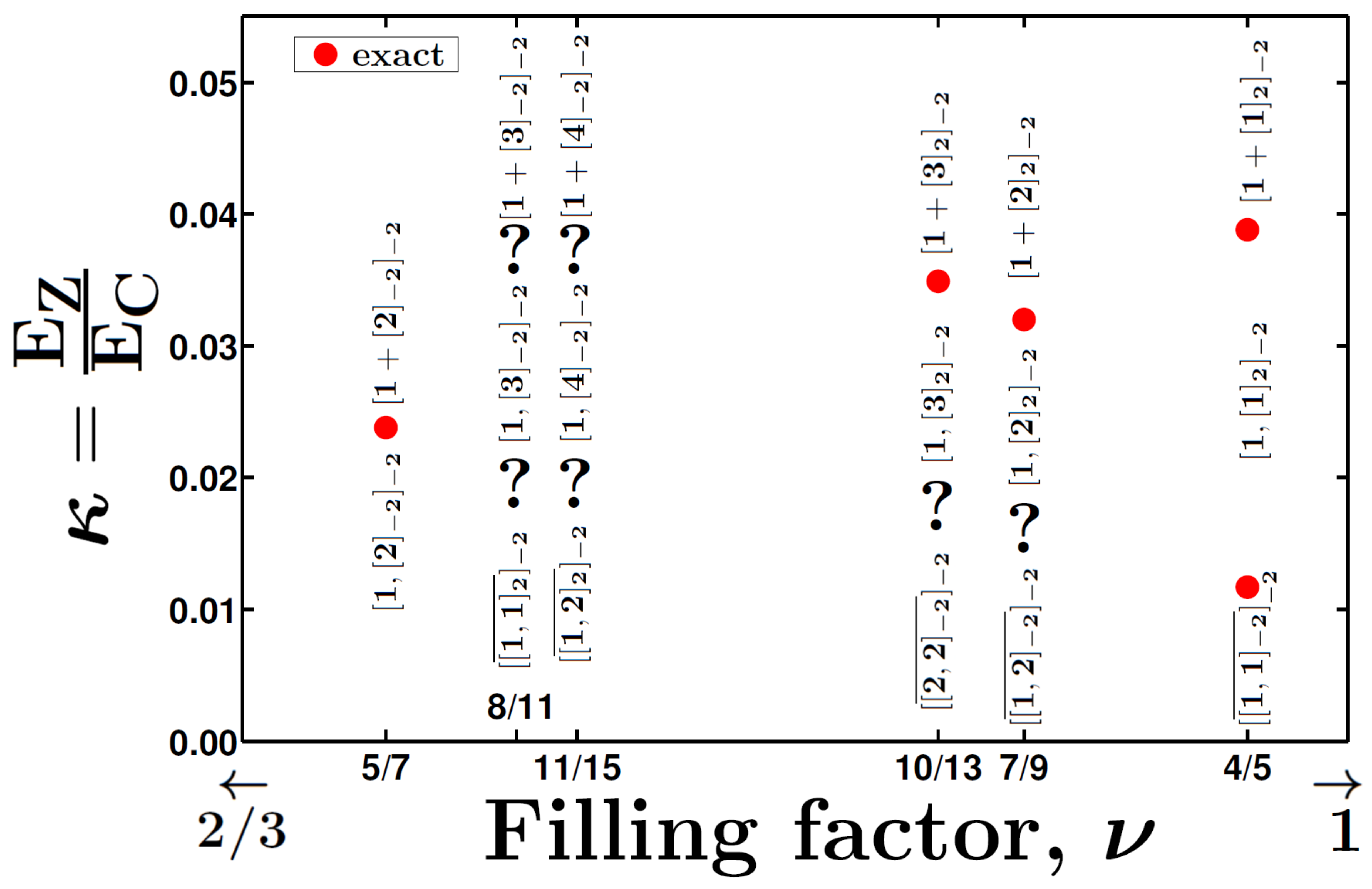}
\caption{(color online) Phase diagram of the CF FQHE states in the filling factor regions $1/3<\nu<2/5$ (upper panel) and $2/3>\nu>1$ (lower panel). The various possible states are shown for several fractions, along with the theoretical critical Zeeman energies where transitions between them are expected. (The ``?" symbol is used to represent transitions that are expected to occur but for which the critical Zeeman energies have not been estimated.)  The spin contents and polarizations of the states can be found in the main text, but lowest state at even-{\em numerator} fractions is spin singlet and the highest state is fully spin polarized. The states in the top (bottom) panel are obtained from states of CF in the filling factor range $1<\nu^{*}<2$ by parallel (reverse) flux attachment. The dots in the upper panel are obtained with the help of CF diagonalization, whereas those in the lower panel are estimated from exact diagonalization studies. In both cases the thermodynamic energies are obtained by extrapolation of finite system results to obtain the critical Zeeman energies.}
\label{fig:2}
\end{center}
\end{figure*}

The phase diagram of various states at many fractions is shown in Fig. \ref{fig:2}. We again stress that the results are obtained for a system with zero thickness, no LL mixing, and no disorder. Also, the critical Zeeman energies from exact diagonalization are not expected to be very accurate because in many cases, the extrapolation to the thermodynamic limit has been performed from only two points as the Hilbert space grows very rapidly for unpolarized systems. The ``?" symbol indicates a transition for which we are not able to estimate the critical Zeeman energy based on the current calculational methods.

\section{Comparisons with experiments}
\label{six}
Only a limited amount of experimental information is currently available for spin transitions involving FQHE states of composite fermions. A comparison of our calculated critical Zeeman energies with those measured in experiments is shown in Table \ref{tab:expt_EZ}. We stress that the theoretical numbers do not include corrections due to finite width, LL mixing, and disorder, which are all expected to affect the observed critical Zeeman energies \cite{Liu14}. This affects the degree of agreement we expect between theory and experiment. The best comparison is with heterostructure samples, as these systems have the smallest effective width of the transverse wave function. Indeed a satisfactory agreement is seen between our predicted critical Zeeman energies at 4/5, 5/7, 7/9 with those measured in the heterostructure sample studied by Yeh {\em et al.}\cite{Yeh99}. (We  mention that the spin-transitions were not interpreted as FQHE of CFs in Ref.~\onlinecite{Yeh99}; the correct understanding in terms of spin transitions involving  FQHE of CFs was given in Ref. \cite{Liu14a}.) Finite transverse width in general softens the interaction, which suggests that the critical Zeeman energies decrease with increasing width (as has been confirmed by explicit calculation; see for example Ref.~\onlinecite{Liu14}). This is consistent with the fact that the observed critical Zeeman energies are smaller than the theoretically predicted ones. Overall, these comparisons confirm CF-FQHE nature of the observed states. 

\begin{figure}
\begin{center}
\includegraphics[width=0.45\textwidth]{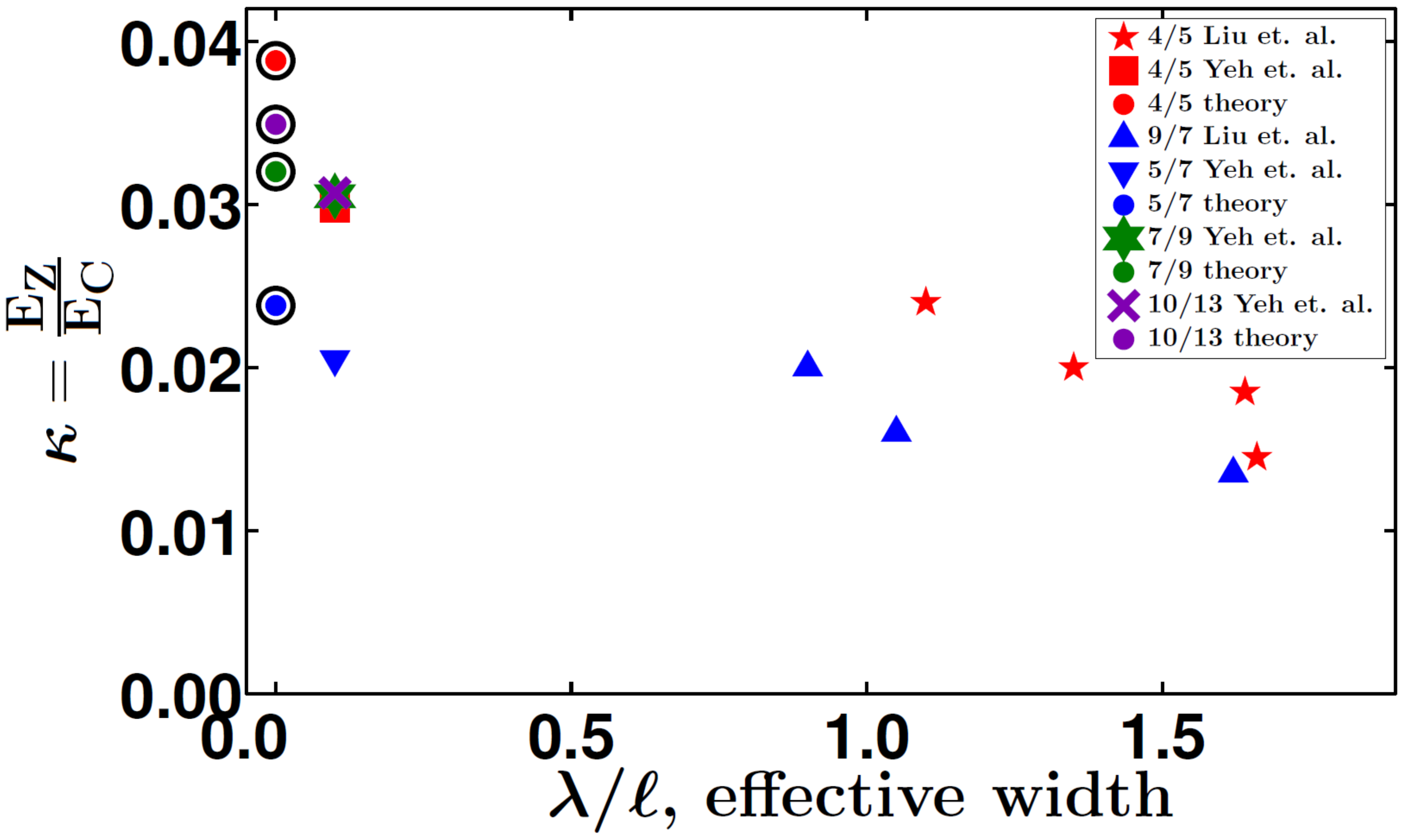}
\caption{(color online) Critical Zeeman energies for transitions at 4/5, 9/7, 5/7, 7/9 and 10/13 as a function of the effective transverse width $\lambda$ (see text for definition) in units of the magnetic length $\ell$. The results are taken from Liu {\em et al.}\cite{Liu14a} and Yeh {\em et al.} \cite{Yeh99}, as indicated on the figure. The experiment of Yeh {\em et al.} employed heterostructures, which correspond to very small transverse widths. Theoretical values at zero width are encircled in black. They are in good agreement with zero width limits of the experiments.}
\label{fig:3}
\end{center}
\end{figure}

Fig.~\ref{fig:3} shows the width dependence of the experimentally measured critical Zeeman energies. The effective width $\lambda$ is defined as the expectation value of $\lambda=\sqrt{\langle w^2\rangle}$, where $w$ is the coordinate in the transverse direction and the expectation value is obtained with respect to a transverse wave function determined from local density approximation. For heterojunctions this is typically of order 0.1 in units of the magnetic length. For the quantum well samples, we have only included results from phase transitions seen as a function of the variation of the density; the phase transitions in which the Zeeman energy is varied by application of an additional magnetic field parallel to the layer also require a consideration of mixing of electric subbands, which can be a significant effect for wide quantum wells. We thank Mansour Shayegan and Yang Liu for these data \cite{Liu14b}. As expected, the critical Zeeman energies decrease with increasing width. The reason is that finite width softens the interaction and thus reduces the interaction energy difference between differently spin polarized states, which therefore requires a smaller Zeeman energy to cause the transition. The zero thickness limits of the critical Zeeman energies are in surprisingly good agreement with our theoretical estimates.

The critical Zeeman energies quoted in Table \ref{tab:expt_EZ} for the spin transitions for $\nu=4/11,~5/13$ and $3/8$ are inferred from the excitations of the respective states. In Ref. \cite{Balram14} it was shown that for a quantum well of width $w=33$ nm and density $n=5.6\times10^{10}$ cm$^{-2}$, certain excitations that appear at $\theta=30^{o}$ tilt are absent at $\theta=50^{o}$ tilt, which was taken as an evidence of a spin transition somewhere between these two tilts; i.e., the ground state is fully polarized state at a tilt of $\theta=50^{o}$ whereas at a tilt of $\theta=30^{o}$ it is partially polarized. Hence we specify a range for  $E_{Z}^{c}$ in Table \ref{tab:expt_EZ} for these filling factors. We should make a note of the fact that the experiment of Ref. \cite{Balram14} does not include transport, and thus does not show direct evidence for FQHE at these fractions.

One puzzle should be mentioned here. In the experiment of Liu {\em et al.}\cite{Liu14a}, they observe two transitions at 5/7 (and its hole partner at 9/7). This is inconsistent with our expectation of a single transition for a two-component system. We speculate on the possible causes. While only two states, $[1+[2]_{-2}]_{-2}$ and $[1,[2]_{-2}]_{-2}$, are possible when we allow FQHE in only one component, another candidate FQHE states of the form $[1+1/3, 1/3]_{-2}$, where one or both of $1/3$ states can be replaced by $1/3^{\rm WYQ}$, becomes possible when we allow FQHE in both spin components. These states satisfy the Fock conditions, but involve much more delicate physics than the two states considered above. Further, we have found that these states are not stabilized by the Coulomb interaction in a sample with zero width (see Appendix \ref{57FQHE}). With three components, more states become possible, such as $[1,[1,1]_{-2}]_{-2}$, but the experimentalists have argued that the subband degree of freedom is suppressed for their experimental parameters (i.e. the separation between the symmetric and antisymmetric subbands is very large compared to the Zeeman energy). We thus do not at present have an explanation for the experimental observation of two transitions at 5/7.

In graphene, many spin transitions have been seen for IQHE states of composite fermions \cite{Feldman12,Feldman13} at fractions of the form $n/(2pn\pm 1)$ but none so far involving FQHE of composite fermions. The two-component systems in AlAs quantum wells, where the two components are valleys, are also understood in terms of IQHE of composite fermions\cite{Bishop07,Padmanabhan09}. 

\begin{table*}
\begin{center}
\begin{tabular}{|c|c|c|c|c|c|c|c|}
\hline
\multicolumn{1}{|c|}{$\nu$} & \multicolumn{1}{|c|}{Transition} & \multicolumn{2}{|c|}{Theoretical $E_{Z}^{c}$}	 & \multicolumn{4}{|c|}{Experimental $E_{Z}^{c}$} \\ \hline
	&						       &	exact	& CFD	 &Reference & $w$ (nm)&$n$ ($\times$ $10^{11}$ cm$^{-2}$)& $E_{Z}^{c}$\\ \hline
4/11	& $[1+1/3^{\rm WYQ}]_{2} \leftrightarrow [1,[1]_{2}]_{2}$ 							& 	0.0208		&	0.0171		
&\onlinecite{Balram14}	&33	&0.55	&0.0172-0.0225\\ \hline
4/5	& $[1+[1]_{2}]_{-2}\equiv \overline{[1]_{4}} \leftrightarrow [1,[1]_{2}]_{-2}$ 					& 	0.0388					&-	&\onlinecite{Liu14a}	&65	&1.13	&0.0145\\ \hline
	&														& 	0.0388			&-	&\onlinecite{Liu14a}	&65	&1.00	&0.0157\\ \hline
	&														& 	0.0388			&-	&\onlinecite{Liu14a}	&60	&0.44	&0.0177\\ \hline
	&														& 	0.0388			&-	&\onlinecite{Yeh99}	&heterostructure	&1.13	&0.0298\\ \hline
6/5	& $[1+[1]_{4}] \leftrightarrow [1,[1]_{4}]$				& 	0.0388					&-	&\onlinecite{Liu14a}	&65	&1.13	&0.0149\\ \hline	
5/13	& $[1+2/3^{\rm WYQ}]_{2} \leftrightarrow [1,[2]_{-2}]_{2}$ 							& 	0.0183		&	0.0149		
&\onlinecite{Balram14}	&33	&0.55	&0.0167-0.0218\\ \hline
5/7	& $[1+[2]_{-2}]_{2} \equiv \overline{[2]_{-4}} \leftrightarrow [1,[2]_{-2}]_{-2}$ 				& 	0.0238			&-	&\onlinecite{Liu14a}	&60	&0.44	&0.0150\\ \hline
	&														& 	0.0238			&-	&\onlinecite{Yeh99}	&heterostructure	&1.13	&0.0205\\ \hline
9/7	& $[1+[2]_{-4}] \leftrightarrow [1,[2]_{-4}]$ 									& 	0.0238		&-		&\onlinecite{Liu14a}	&60	&0.44	&0.0175\\ \hline	
7/9	& $[1,[2]_{2}]_{-2} \equiv [\overline{[3]_{-2}}]_{-2} \leftrightarrow [\overline{[1,2]_{-2}}]_{-2}$				&	-	&	
&\onlinecite{Yeh99}	&heterostructure	&1.13	&0.0251\\ \hline
7/9	& $[1+[2]_{2}]_{-2} \equiv \overline{[2]_{4}} \leftrightarrow [1,[2]_{2}]_{-2} \equiv [\overline{[3]_{-2}}]_{-2}$& 	0.0320		&	-	
&\onlinecite{Yeh99}	&heterostructure	&1.13	&0.0305\\ \hline
8/11	& $[1,[3]_{-2}]_{-2} \equiv [\overline{[2]_{2}}]_{-2} \leftrightarrow [\overline{[1,1]_{2}}]_{-2}$			& 	-		&	-	
&\onlinecite{Yeh99}	&heterostructure	&1.13	&0.0204\\ \hline
8/11	& $[1+[3]_{-2}]_{-2} \equiv \overline{[3]_{-4}}\leftrightarrow [1,[3]_{-2}]_{-2} \equiv [\overline{[2]_{2}}]_{-2}$	& 	-		&	-	
&\onlinecite{Yeh99}	&heterostructure	&1.13	&0.0260\\ \hline
10/13	& $[1,[3]_{2}]_{-2} \equiv [\overline{[4]_{-2}}]_{-2} \leftrightarrow [\overline{[2,2]_{-2}}]_{-2}$		& 	-		&	-	
&\onlinecite{Yeh99}	&heterostructure	&1.13	&0.0276\\ \hline
10/13	& $[1+[3]_{2}]_{-2} \equiv \overline{[3]_{4}} \leftrightarrow [1,[3]_{2}]_{-2} \equiv [\overline{[4]_{-2}}]_{-2}$& 	0.0349		&	-	
&\onlinecite{Yeh99}	&heterostructure	&1.13	&0.0307\\ \hline
3/8	& $[1+1/2^{\rm APf}]_{2} \leftrightarrow [1,1/2^{\rm APf}]_{2}$							& -	&	0.0183	
&\onlinecite{Balram14}	&33	&0.55	&0.0169-0.0223\\ \hline
\end{tabular}
\end{center}
\caption {Comparison of theoretical (zero width $w=0$, no LL mixing, zero disorder) critical Zeeman energies with experimental results. }  
\label{tab:expt_EZ} 
\end{table*}

\section{Conclusion}
\label{seven}

We have carried out an extensive theoretical study of {\em fractional} QHE of composite fermions in multi-component systems. We have explicitly listed a large number of prominent fractions, identifying the possible CF states at each fraction, along with an estimate of their thermodynamic energies. This has allowed us to make predictions regarding the critical values of the Zeeman energy (used in a general sense) where transitions between different states take place. We have compared our predictions to the experimental studies currently available, and found very good qualitative and semi-quantitative agreement. We have also mentioned experimental features that are not understood.

\begin{appendices}
\section{Polarization}

The polarization $\gamma$ of a state is defined as:
\begin{equation}
\gamma=\frac{\nu_{\uparrow}-\nu_{\downarrow}}{\nu_{\uparrow}+\nu_{\downarrow}}
\label{def_pol}
\end{equation}
In this section we state how the polarization of a state changes under the CF transformation and particle hole conjugation. We restrict ourselves to two-component states (this for example describes the spin-polarization of states in GaAs or valley polarization in the case of graphene) and denote the states by $(\nu_{\uparrow},\nu_{\downarrow})$.

Under the CF transformation $[\nu^{*}_{\uparrow},\nu^{*}_{\downarrow}]_{\pm 2p} \rightarrow (\nu_{\uparrow},\nu_{\downarrow})$ the polarization of the state does not change. This is proved by noting the fact that under the CF transformation:
\begin{eqnarray}
[\nu^{*}_{\uparrow},\nu^{*}_{\downarrow}]_{\pm 2p} & \rightarrow & \Bigg(\frac{\nu^{*}_{\uparrow}}{2p(\nu^{*}_{\uparrow}+\nu^{*}_{\downarrow})\pm 1},\frac{\nu^{*}_{\downarrow}}{2p(\nu^{*}_{\uparrow}+\nu^{*}_{\downarrow})\pm 1}\Bigg) \nonumber \\ 
&=&(\nu_{\uparrow},\nu_{\downarrow}) 
\end{eqnarray}
Using the definition of polarization from Eq. \ref{def_pol}, we see that the polarization of the state $(\nu_{\uparrow},\nu_{\downarrow})$ is identical to that of $(\nu^{*}_{\uparrow},\nu^{*}_{\downarrow})$.

Under particle-hole conjugation $\overline{[\nu_{\uparrow},\nu_{\downarrow}]} \rightarrow [1-\nu_{\uparrow},1-\nu_{\downarrow}]$ the polarization of the state changes from $\gamma$ to $-\gamma\nu/(2-\nu)$ where $\nu=\nu_{\uparrow}+\nu_{\downarrow}$.
This can be seen from the definition of Eq. \ref{def_pol}. Let us denote by $\overline{\gamma}$ the polarization of the state obtained from the particle-hole conjugation.
\begin{eqnarray}
 \gamma=\frac{\nu-2\nu_{\downarrow}}{\nu}&\implies& \nu_{\downarrow}=\frac{\nu}{2}(1-\gamma) \\
 \overline{\gamma}=-\frac{\nu-2\nu_{\downarrow}}{2-\nu}&\implies& \overline{\gamma}=-\frac{\gamma\nu}{2-\nu}
\end{eqnarray}

\label{appendix:A}

\section{Results}
In Tables \ref{tab:Table1}, \ref{tab:Table2} and \ref{tab:Table3} we show results for the ground state energies per particle for individual systems obtained from exact and CF diagonalization. In some cases we also include results for the unperturbed CF wave function. The energies listed in these tables include background subtraction and density correction. 
\begin{table*}
\begin{center}
\begin{tabular}{|c|c|c|c|c|c|c|c|}
\hline
\multicolumn{1}{|c|}{$\nu$} & \multicolumn{1}{|c|}{state} & \multicolumn{1}{|c|}{$N$} & \multicolumn{1}{|c|}{$2Q$} & \multicolumn{1}{|c|}{$S$} & \multicolumn{3}{|c|}{Energy ($e^2/\epsilon\ell$)}  \\ \hline
	&   							& 		& 		& 	 	&          exact		&      CFD   		&      CF w. f. 		\\ \hline
4/11	& $[1+1/3^{\rm WYQ}]_{2}$ 					& 	8	& 	17	& 	 4	& 	*  		&  	*	&	\\ \hline
	&   							& 	12	& 	28	& 	 6	& 	-0.41135  	&  -0.41101(1)	     	&\\ \hline
	&   							& 	16	& 	39	& 	 8	& 	-0.41266  	&  -0.41246(3)	     	&\\ \hline
	&   							& 	20	& 	50	& 	 10	& 	-  		&  -0.41299(0)	     	&\\ \hline
	&   							& 	24	& 	61	& 	 12	& 	-  		&  -0.41350(2)	     	&\\ \hline
	&   							& 	28	& 	72	& 	 14	& 	-  		&  -0.41395(1)	     	&\\ \hline
4/5	& $[1+[1]_{2}]_{-2} \equiv \overline{[1]_{4}}$		& 	8	& 	10	& 	 4	& 	-0.56493  	&  -		     	& -0.564933(4) \\ \hline
	&   							& 	12	& 	15	& 	 6	& 	-0.56072  	&  -		     	& -0.560534(8) \\ \hline
4/11	& $[1,[1]_2]_{2}$					& 	6	& 	13	& 	 1	& 	-0.42170  	&  -0.42070(0) & -0.42069(1) \\ \hline
	&   							& 	10	& 	24	& 	 2	& 	-0.42174  	&  -0.42056(0) &-0.42057(1) \\ \hline
	&   							& 	14	& 	35	& 	 3	& 	-  		&  -0.42059(0) &-0.42056(2) \\ \hline
	&   							& 	18	& 	46	& 	 4	& 	-  		&  -0.42053(4) &-0.42055(8) \\ \hline
4/5	& $[1,[1]_{2}]_{-2}$					& 	6	& 	7	& 	 1	& 	-0.58100  	&  -		     	&\\ \hline
	&   							& 	10	& 	12	& 	 2	& 	-0.57264  	&  -		     	&\\ \hline
4/11	& $[\overline{[1,1]_{-2}}]_2$				& 	8	& 	19	& 	 0	& 	-0.42356  	&  -0.42305(1)		&\\ \hline
	&							& 	12	& 	30	& 	 0	& 	-	  	&  -0.42272(2)		&\\ \hline
	&							& 	16	& 	41	& 	 0	& 	-   	  	&  -0.42257(2)		&\\ \hline
4/5	& $[\overline{[1,1]_{-2}}]_{-2}$			& 	8	& 	9	& 	 0	&	-0.57579  	&  -			&\\ \hline
	&							& 	12	& 	14	& 	 0	& 	-0.57221  	&  -			&\\ \hline
5/13	& $[1+2/3]_{2}$ (WYQ)					& 	6	& 	13	& 	 3	& 	-0.42386  	&  -0.42375(0)	     	&\\ \hline
	&							& 	11	& 	26	& 	 5.5	& 	-0.42362  	&  -0.42339(2)	     	&\\ \hline
	&							& 	16	& 	39	& 	 8	& 	-0.42440  	&  -0.42418(3)	     	&\\ \hline
	&							& 	21	& 	52	& 	 10.5	& 	-	  	&  -0.42422(2)	     	&\\ \hline
	&							& 	26	& 	65	& 	 13	& 	-	  	&  -0.42416(3)	     	&\\ \hline
	&							& 	31	& 	78	& 	 15.5	& 	-	  	&  -0.42428(2)	     	&\\ \hline
5/7	& $[1+[2]_{-2}]_{-2} \equiv \overline{[2]_{-4}}$	& 	9	& 	12	& 	 4.5	& 	-0.53788  	&  -		     	& -0.53788(2) \\ \hline
5/13	& $[1,[2]_{-2}]_{2}$ 					& 	6	& 	13	& 	 1	& 	-0.43369  	&  -0.43266(0)     	& $\dag$ \\ \hline
	&							& 	11	& 	26	& 	 1.5	& 	-0.43277	&  -0.43156(0)	     	& -0.43157(4)\\ \hline
	&							& 	16	& 	39	& 	 2	& 	-	  	&  -0.43118(0)	     	& -0.43132(5)\\ \hline
	&							& 	21	& 	52	& 	 2.5	& 	-	  	&  -0.43096(1)	     	& -0.43113(3)\\ \hline
5/7	& $[1,[2]_{-2}]_{-2}$ 					& 	6	& 	7	& 	 1	& 	-0.54899  	&  -		     	&\\ \hline
	& 		 					& 	11	& 	14	& 	 1.5	& 	-0.54442  	&  -		     	&\\ \hline	
\end{tabular}
\end{center}
\caption {The ground state Coulomb energies of $N$ particles with spin $S$ at flux $2Q$ at various filling factors. The energies include background subtraction and density correction. The number shown in the parentheses indicates the error from Monte-Carlo calculation. The symbol $*$ marks systems for which the ground state does not occur at $L=0$; the symbol - indicates systems for which results are not available; and $\ddagger$ indicates systems for which calculations were carried out in only the $L=0$ sector. The symbol $\dag$ indicates that the CF variational wave function cannot be constructed with the given $N$.} 
\label{tab:Table1} 
\end{table*}

\begin{table*}
\begin{center}
\begin{tabular}{|c|c|c|c|c|c|c|c|}
\hline
\multicolumn{1}{|c|}{$\nu$} & \multicolumn{1}{|c|}{state} & \multicolumn{1}{|c|}{$N$} & \multicolumn{1}{|c|}{$2Q$} & \multicolumn{1}{|c|}{$S$} & \multicolumn{3}{|c|}{Energy ($e^2/\epsilon\ell$)}  \\ \hline
	&   							& 		& 		& 	 	&          exact		&      CFD   		&      CF w. f. 		\\ \hline
7/19	& $[1+[2]_{2}]_{2}$ 					& 	9	& 	20	& 	 4.5	& 	*  		&  *		     	&\\ \hline
	& 		 					& 	16	& 	39	& 	 8	& 	-0.41537	&  -0.41516(3)	     	&\\ \hline
	& 		 					& 	23	& 	58	& 	 11.5	& 	-		&  -0.41570(4)$^{\ddagger}$&\\ \hline
7/9	& $[1+[2]_{2}]_{-2} \equiv \overline{[2]_{4}}$ 		& 	9	& 	12	& 	 4.5	& 	-0.56128	&  -		     	& -0.56128(2) \\ \hline
	& 		 					& 	16	& 	21	& 	 8	& 	-0.55402	&  -		     	& -0.55396(1) \\ \hline
	& 		 					& 	23	& 	30	& 	 11.5	& 	-0.55126	&  -		     	& -0.551215(8) \\ \hline
	& 		 					& 	30	& 	39	& 	 15	& 	-0.54995	&  -		     	& -0.549861(8) \\ \hline
7/19	& $[1,[2]_{2}]_{2} \equiv [\overline{[3]_{-2}}]_{2}$ 	& 	11	& 	26	& 	 1.5	& 	-0.42356	&  -0.42238(0)	     	& -0.42238(5)\\ \hline
	&							& 	18	& 	45	& 	 3	& 	-  		&  -0.42241(0)$^{\ddagger}$& -0.42243(4)\\ \hline
	& 							&	25	&	64	&	 4.5	&	-		&  -0.42262(4)& -0.42244(4)\\ \hline
7/9	& $[1,[2]_{2}]_{-2} \equiv [\overline{[3]_{-2}}]_{-2}$  & 	11	& 	14	& 	 1.5	& 	-0.56810  	&  -		     	&\\ \hline
7/19	& $[\overline{[1,2]_{-2}}]_{2}$ 			&	6	& 	13	& 	 1	& 	-0.42446  	&  -0.42346(0)	     	&\\ \hline
	&							&	13	& 	32	& 	 1.5	& 	-	  	&  -0.42365(0)	     	&\\ \hline
	&							&	20	& 	51	& 	 2	& 	-	  	&  -0.42411(8)	     	&\\ \hline
7/9	& $[\overline{[1,2]_{-2}}]_{-2}$			&	6	& 	7	& 	 1	& 	-0.57287  	&  -		     	&\\ \hline
8/21	& $[1+[3]_{-2}]_{2}$ 					& 	8	& 	18	& 	 4	& 	*  		&  *		     	&\\ \hline
	&							&       16	&	39	&	 8	&	-0.42237	&  -0.42216(3)		&\\ \hline
	&							& 	24	& 	60	& 	 12	& 	-  		&  -0.42188(2)$^{\ddagger}$&\\ \hline
8/11	& $[1+[3]_{-2}]_{-2} \equiv \overline{[3]_{-4}}$ 	& 	8	& 	10	& 	 4	& 	-0.53864	&  -		     	& $\dag$\\ \hline
	&							&       16	&	21	&	 8	&	-0.53573	&  -			& $\dag$\\ \hline
8/21	& $[1,[3]_{-2}]_{2} \equiv [\overline{[2]_{2}}]_{2}$	& 	10	& 	24	& 	 2	& 	-0.43166 	&  -0.43046(0)	     	& $\dag$\\ \hline
        &							& 	18	& 	45	& 	 3	& 	-  		&  -0.42952(0)$^{\ddagger}$& $\dag$ \\ \hline
        &							& 	26	& 	66	& 	 4	& 	-  		&  -0.42924(2)$^{\ddagger}$& -0.4292(1)\\ \hline
8/11	& $[1,[3]_{-2}]_{-2} \equiv [\overline{[2]_{2}}]_{-2}$	& 	10	& 	12	& 	 2	& 	-0.54600 	&  -		     	&\\ \hline
	& 							& 	18	& 	23	& 	 3	& 	-	 	&  -		     	&\\ \hline
8/21	& $[\overline{[1,1]_{2}}]_{2}$				& 	12	& 	29	& 	 0	& 	-	 	&  -0.43075(0)$^{\ddagger}$&\\ \hline
        &							& 	20	& 	50	& 	 0	& 	-	 	&  -0.43003(2)$^{\ddagger}$&\\ \hline
8/11	& $[\overline{[1,1]_{2}}]_{-2}$				& 	12	& 	15	& 	 0	& 	-0.54897 	&  -		     	&\\ \hline
	&							& 	20	& 	26	& 	 0	& 	-	 	&  -		     	&\\ \hline
\end{tabular}
\end{center}
\caption {The ground state Coulomb energies of $N$ particles with spin $S$ at flux $2Q$ at various filling factors. The energies include background subtraction and density correction. The number shown in the parentheses indicates the error from Monte-Carlo calculation. The symbol $*$ marks systems for which the ground state does not occur at $L=0$; the symbol - indicates systems for which results are not available; and $\ddagger$ indicates systems for which calculations were carried out in only the $L=0$ and the relevant $S$ sector. The symbol 
$\dag$ indicates that the CF variational wave function cannot be constructed with the given $N$.} 
\label{tab:Table2} 
\end{table*}

\begin{table*}
\begin{center}
\begin{tabular}{|c|c|c|c|c|c|c|c|}
\hline
\multicolumn{1}{|c|}{$\nu$} & \multicolumn{1}{|c|}{state} & \multicolumn{1}{|c|}{$N$} & \multicolumn{1}{|c|}{$2Q$} & \multicolumn{1}{|c|}{$S$} & \multicolumn{3}{|c|}{Energy ($e^2/\epsilon\ell$)}  \\ \hline
	&   				& 		& 		& 	 	&          exact		&      CFD   		&      CF w. f. 		\\ \hline
10/27	& $[1+[3]_{2}]_{2}$ 					& 	14	& 	33	& 	 7	& 	-0.41500&  -0.41487(0)		     	&\\ \hline
	&							& 	24	& 	60	& 	 12	& 	-  	&  -0.41599(3)$^{\ddagger}$    	&\\ \hline
10/13	& $[1+[3]_{2}]_{-2} \equiv \overline{[3]_{4}}$ 		& 	14	& 	19	& 	 7	& 	-0.55480&  -		   	        & $\dag$ \\ \hline
	&							& 	24	& 	32	& 	 12	& 	-0.54972&  -			   	& -0.549679(7) \\ \hline
	&							& 	34	& 	45	& 	 17	& 	-0.54774&  -			   	& -0.54766(1) \\ \hline
10/27	& $[1,[3]_{2}]_{2} \equiv [\overline{[4]_{-2}}]_{2}$ 	& 	6	& 	12	& 	 0	& 	-0.42391&  -0.42336(0)		     	& $\dag$\\ \hline
	& 							& 	16	& 	39	& 	 2	& 	-  	&  -0.42311(0)		     	& $\dag$\\ \hline
	& 							& 	26	& 	66	& 	 4	& 	-  	&  -0.42323(3)		     	& -0.42317(4)\\ \hline
10/13	& $[1,[3]_{2}]_{-2} \equiv [\overline{[4]_{-2}}]_{-2}$ 	& 	6	& 	8	& 	 0	& 	-0.58170&  -			     	&\\ \hline
	& 							& 	16	& 	21	& 	 2	& 	-0.56307&  -			     	&\\ \hline	
10/27	& $[\overline{[2,2]_{-2}}]_{2}$				& 	12	& 	29	& 	 0	& 	-	&  -0.42472		     	&\\ \hline
	& 							& 	22	& 	56	& 	 0	& 	-	&  -			     	&\\ \hline	
10/13	& $[\overline{[2,2]_{-2}}]_{-2}$			& 	12	& 	15	& 	 0	& 	-0.56459&  -			     	&\\ \hline
11/29	& $[1+[4]_{-2}]_{2}$ 					& 	12	& 	29	& 	 6	& 	-0.42243&  -0.42219(0)		     	&\\ \hline
 	&							& 	23	& 	58	& 	 11.5	& 	-	&  -0.42176(9)$^{\ddagger}$	&\\ \hline
11/15	& $[1+[4]_{-2}]_{-2} \equiv \overline{[4]_{-4}}$ 	& 	12	& 	15	& 	 6	& 	-0.53684&  -			     	& $\dag$ \\ \hline
	&							& 	23	& 	30	& 	 11.5	& 	-0.53527&  -				& $\dag$ \\ \hline
11/29	& $[1,[4]_{-2}]_{2} \equiv [\overline{[3]_{2}}]_{2}$ 	& 	14	& 	35	& 	 3	& 	-	&  -0.42956(0)		     	& $\dag$ \\ \hline
	&							& 	25	& 	64	& 	 4.5	& 	-	&  -0.42882(2)$^{\ddagger}$	& $\dag$ \\ \hline
11/15	& $[1,[4]_{-2}]_{-2} \equiv [\overline{[3]_{2}}]_{-2}$ 	& 	14	& 	17	& 	 3	& 	-0.54520&  -			     	&\\ \hline
	&							& 	25	& 	32	& 	 4.5	& 	-	&  -				&\\ \hline
11/29	& $[\overline{[1,2]_{2}}]_{2}$				& 	13	& 	32	& 	 1.5	& 	-	&  -0.42987(0)		     	&\\ \hline
	& 							& 	24	& 	61	& 	 2	& 	-	&  -			     	&\\ \hline	
11/15	& $[\overline{[1,2]_{2}}]_{-2}$				& 	13	& 	16	& 	 1.5	& 	-0.54850&  -			     	&\\ \hline
	& 							& 	24	& 	31	& 	 2	& 	-	&  -			     	&\\ \hline	
3/8	& $[1+1/2^{\rm APf}]_{2}$				& 	6	& 	13	& 	 3	& 	-0.41852&  -0.41844(0)		     	&\\ \hline
	& 							& 	12	& 	29	& 	 6	& 	-0.42002&  -0.41978(5)		     	&\\ \hline
	& 							& 	18	& 	45	& 	 9	& 	-	&  -0.41953(1)		     	&\\ \hline
	& 							& 	24	& 	61	& 	 12	& 	-	&  -0.41991(2)		     	&\\ \hline
3/8	& $[1,1/2^{\rm APf}]_{2}$				& 	8	& 	19	& 	 2	& 	-0.42914&  -0.42833(0)		     	&\\ \hline
	&							& 	14	& 	35	& 	 3	& 	-	&  -0.42711(0)		     	&\\ \hline
	&							& 	20	& 	51	& 	 4	& 	-	&  -0.42669(2)		     	&\\ \hline
	&							& 	26	& 	67	& 	 5	& 	-	&  -0.42640(0)		     	&\\ \hline
\end{tabular}
\end{center}
\caption {The ground state Coulomb energies of $N$ particles with spin $S$ at flux $2Q$ at various filling factors. The energies include background subtraction and density correction. The number shown in the parenthisis indicates the error from Monte-Carlo calculation. The symbol $*$ marks systems for which the ground state does not occur at $L=0$; the symbol - indicates systems for which results are not available; and $\ddagger$ indicates systems for which calculations were carried out in only the $L=0$ sector. The symbol $\dag$ indicates that the CF variational wave function cannot be constructed with the given $N$.}
\label{tab:Table3} 
\end{table*}

\label{appendix:B}

\section{Calculating energy eigenstates from non-orthogonal basis}
The set of CF states that are constructed by composite fermionization of simple IQHE states do not form an orthogonal set. In this situation, one can define the matrix representation $H$ of the Hamiltonian operator $\mathcal{H}$ in the usual way. However the eigenvalues and eigenvectors of this matrix are not the eigenvalues of the Hamiltonian operator. In this section we show how the correct quantities can be obtained by diagonalizing the matrix $O^{-1}H$ where $O$ is the overlap matrix.

Let $\mathcal{H}$ be an operator which we intend to diagonalize within the space $\mathcal{V}$ spanned by the non-orthonormal set of vectors $\left\{ \left| 1 \right\rangle, \left| 2 \right\rangle,\left| 3 \right\rangle,\dots, \left| n \right\rangle \right\}$. The quantities $H$ and $O$ defined as 
\begin{eqnarray}
H_{ij}&=&\left\langle i \right | \mathcal{H} \left|j\right\rangle \nonumber \\
O_{ij}&=&\left\langle i \right |\mathbb{I} \left|j\right\rangle 
\end{eqnarray}
are the quantities numerically computed through Monte Carlo integrations. $\mathbb{I}$ is the identity operator. Let $\left|\phi\right\rangle$ and $\epsilon$ be an eigenvector and corresponding eigenvalue of $\mathcal{H}$ in the Hilbert space $\mathcal{V}$. By using the completeness relation, this eigenstate can be expanded in the non-orthogonal basis.
\begin{eqnarray}
	\left \vert \phi \right \rangle &=& \left[\sum_{i,j=1}^{n} \left \vert i \right\rangle [O^{-1}]_{ij} \left \langle j \right \vert \right] \left \vert \phi \right \rangle \nonumber\\
	&=& \sum_{i=1}^n c_i \left \vert i \right \rangle \text{ where } c_i=\sum_{j=1}^n [O^{-1}]_{ij}\left\langle j|\phi\right\rangle
\end{eqnarray}
We prove below that the column vector containing the expansion coefficients $c_i$ is a \emph{right} eigenvector of the \emph{non}-Hermitian matrix $O^{-1}H$ with an eigenvalue $\epsilon$.

Since $\left |\phi \right \rangle$ is an eigenstate, $\mathcal{H}\left |\phi \right \rangle=\epsilon\left |\phi \right \rangle$. Inserting the completeness relations we get
\begin{eqnarray}
\left[\sum_{k,l=1}^{n} \left \vert k \right\rangle [O^{-1}]_{kl} \left \langle l \right \vert \right] \mathcal{H}\left[\sum_{i,j=1}^{n} \left \vert i \right\rangle [O^{-1}]_{ij} \left \langle j \right \vert \right]\left |\phi \right \rangle \nonumber\\ =\epsilon \left[\sum_{a,b=1}^{n} \left \vert a \right\rangle [O^{-1}]_{ab} \left \langle b \right \vert \right]\left |\phi \right \rangle\nonumber
\end{eqnarray}
After rearranging the summations and using the definition of $H$ and $c_i$, the above equality can be re-written as 
\begin{equation}
\sum_{k=1}^n \left \vert k \right \rangle [O^{-1}Hc]_k = \epsilon \sum_{k=1}^n c_k \left \vert k \right \rangle\nonumber
\end{equation}
Since the basis vectors are linearly independent, this can be true only if the coefficients of the vectors are equal on both sides, which implies
\begin{equation}
O^{-1}H c = \epsilon c
\end{equation}
Every eigenvalue of the operator $\mathcal{H}$ in $\mathcal{V}$ is therefore an eigenvalue of matrix $O^{-1}H$. The corresponding eigenvector of the matrix  provides the coefficients for expanding the state in terms of the non-orthogonal basis. These eigenvalues exhaust all the possible eigenvalues of $O^{-1}H$.

\label{appendix:C}

\section{Double FQHE of composite fermions}
\label{57FQHE}

In this article we have considered only the states in which FQHE occurs in no more than one component of composite fermions. These are expected to be the most prominent FQHE states of composite fermions. In this section we discuss the simplest state which involves FQHE in {\em two} components of composite fermions and also satisfies the Fock condition, namely the two-component 5/3 state formed as $(1+1/3,1/3)$.  (Note that a state of the type $(1/3, 1/3)$ does not satisfy the Fock condition.)

The simplest possibility is $[1+[1]_2, [1]_2]_2$, in which we fill the spin-up lowest $\Lambda$L (L$\Lambda$L) completely and construct $[1]_2$ states in the spin-up second $\Lambda$L and spin-down lowest $\Lambda$L. This gives us the following relations between the flux and the number of particles in each $\Lambda$L.
\begin{eqnarray}
 2Q^{*}+1&=&N_{0\uparrow} \\
 2Q^{*}+1&=&3N_{0\downarrow}-2 \\
 2Q^{*}+3&=&3N_{1\uparrow}-2
\end{eqnarray}
From the last two equations we obtain $(N_{1\uparrow}-N_{0\downarrow})=\frac{2}{3} \notin \mathbb{Z}$. Therefore this state cannot be constructed. A similar argument shows that the corresponding 5/3 state formed by 1/3 WYQ states in both spin-up second $\Lambda$L and spin-down L$\Lambda$L with spin-up L$\Lambda$L full cannot be constructed. \\

Next, we consider $[1+1/3^{\rm WYQ}, [1]_2]_2$, in which 
fill the spin-up L$\Lambda$L completely and construct $1/3^{\rm WYQ}$ and $[1]_2$ states in the spin-up second $\Lambda$L and spin-down lowest $\Lambda$L respectively. This gives us the following set of relations:
\begin{eqnarray*}
2Q^{*}+1&=&N_{0\uparrow}  \\ \nonumber
2Q^{*}+1&=&3N_{0\downarrow}-2\\ \nonumber
~2Q^{*}+3&=&3N_{1\uparrow}-6 \\ \nonumber
N&=&N_{0\uparrow}+N_{0\downarrow}+N_{1\uparrow},					 \\ \nonumber
N_{0\uparrow}=3k-2,~N_{0\downarrow}&=&k,~N_{1\uparrow}=k+2,~~~~(k \in \mathbb{N}) \\ \nonumber
N=5k,~2Q^{*}&=&3k-3,~S=\frac{3k}{2}\\ \nonumber
\end{eqnarray*}
From this state at $\nu^{*}=5/3$, we can construct states at $\nu=5/13$ and $\nu=5/7$ (by reverse flux attachment). Doing exact diagonalization at the corresponding flux, we find that the lowest energy state does not have $L=0$ for all allowed values of $N$ and hence is unlikely to be incompressible.

The third possibility is $[1+[1]_2,1/3^{\rm WYQ}]_2$, in which one constructs the $1/3^{\rm WYQ}$ state in the spin-down lowest $\Lambda$L and $[1]_2$ in second $\Lambda$L. This implies the following relations:
\begin{eqnarray}
2Q^{*}+1&=&N_{0\uparrow} \\
2Q^{*}+1&=&3N_{0\downarrow}-6 \\
2Q^{*}+3&=&3N_{1\uparrow}-2 
\end{eqnarray}
From the last two equations we find that $(N_{1\uparrow}-N_{0\downarrow})=-\frac{2}{3} \notin \mathbb{Z}$. Hence this state cannot be constructed. 

\section{Spin-singlet FQHE at 4/11}
\label{4_11_SS}
In this section we provide details of the excitation spectrum of the 4/11 spin-singlet state. The spectra of partially polarized and fully polarized 4/11 states have been discussed previously in Refs.~\cite{Chang03b,Wojs04,Wojs07,Mukherjee14,Mukherjee14b}. The 4/11 FQHE has been observed but no spin phase transition has yet been observed. This also serves as an illustration of how we have obtained various energies by the method of CFD.

\begin{figure}
\begin{center}
\includegraphics[width=0.45\textwidth]{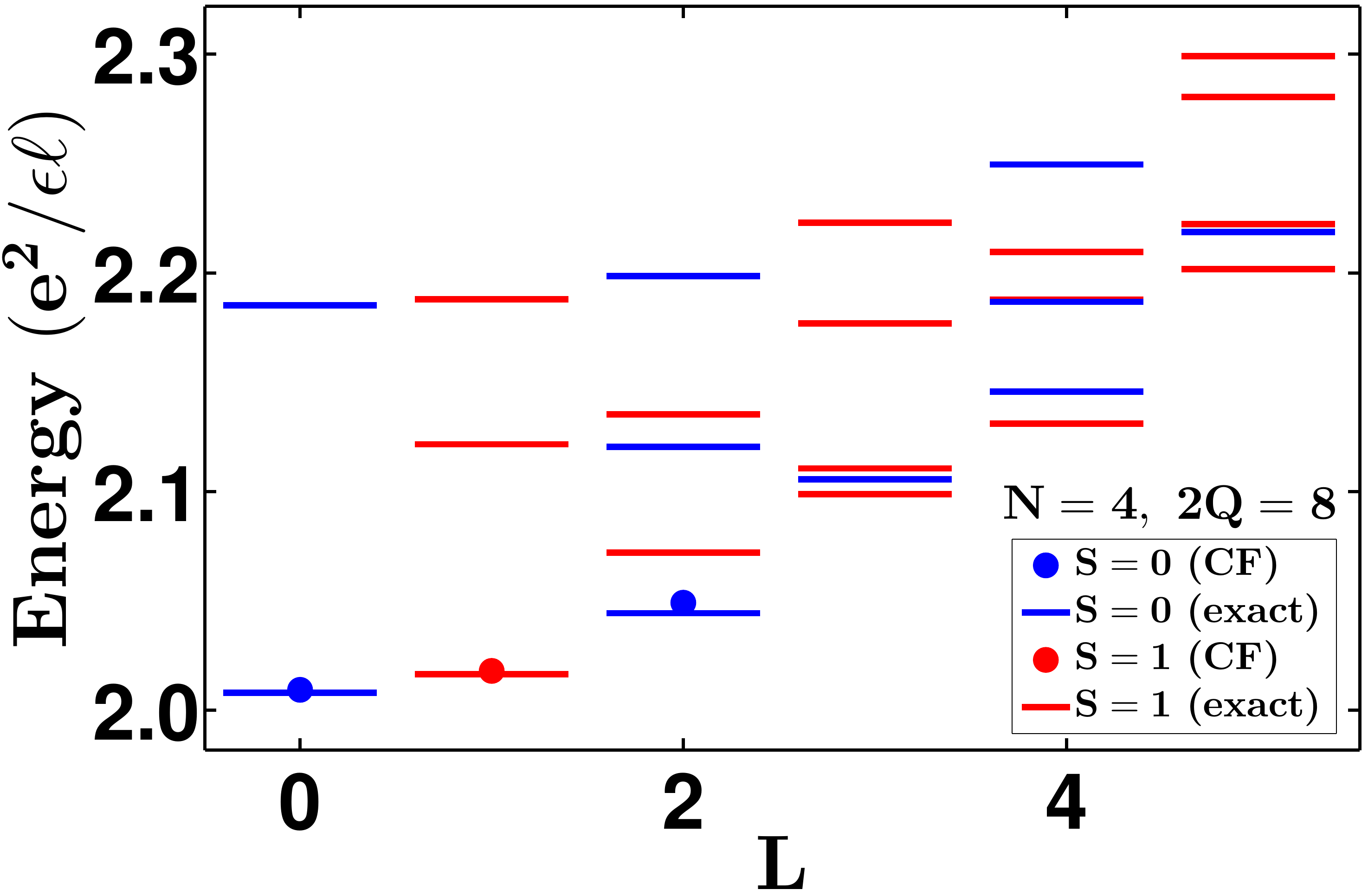}
\includegraphics[width=0.45\textwidth]{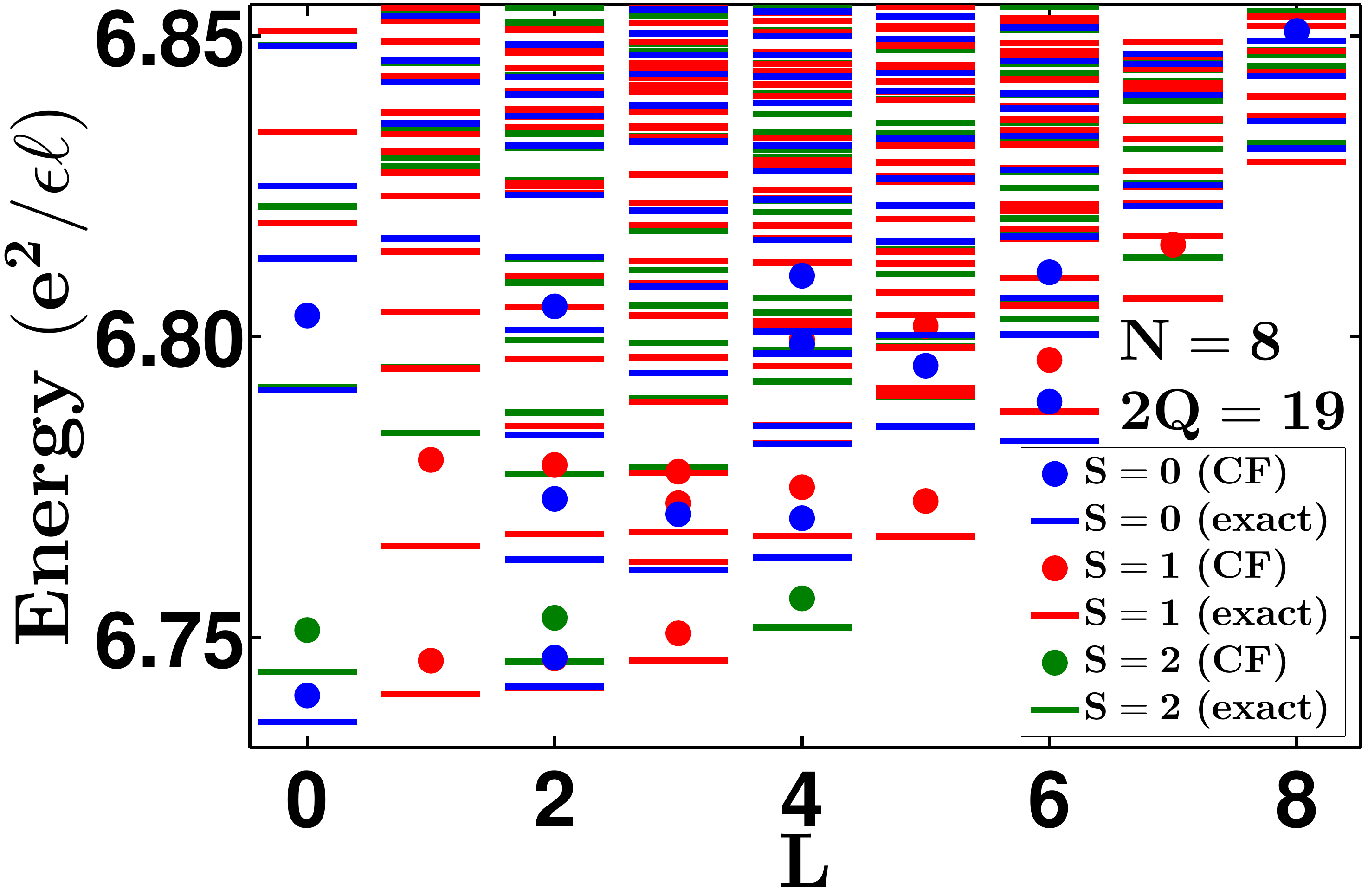}
\includegraphics[width=0.45\textwidth]{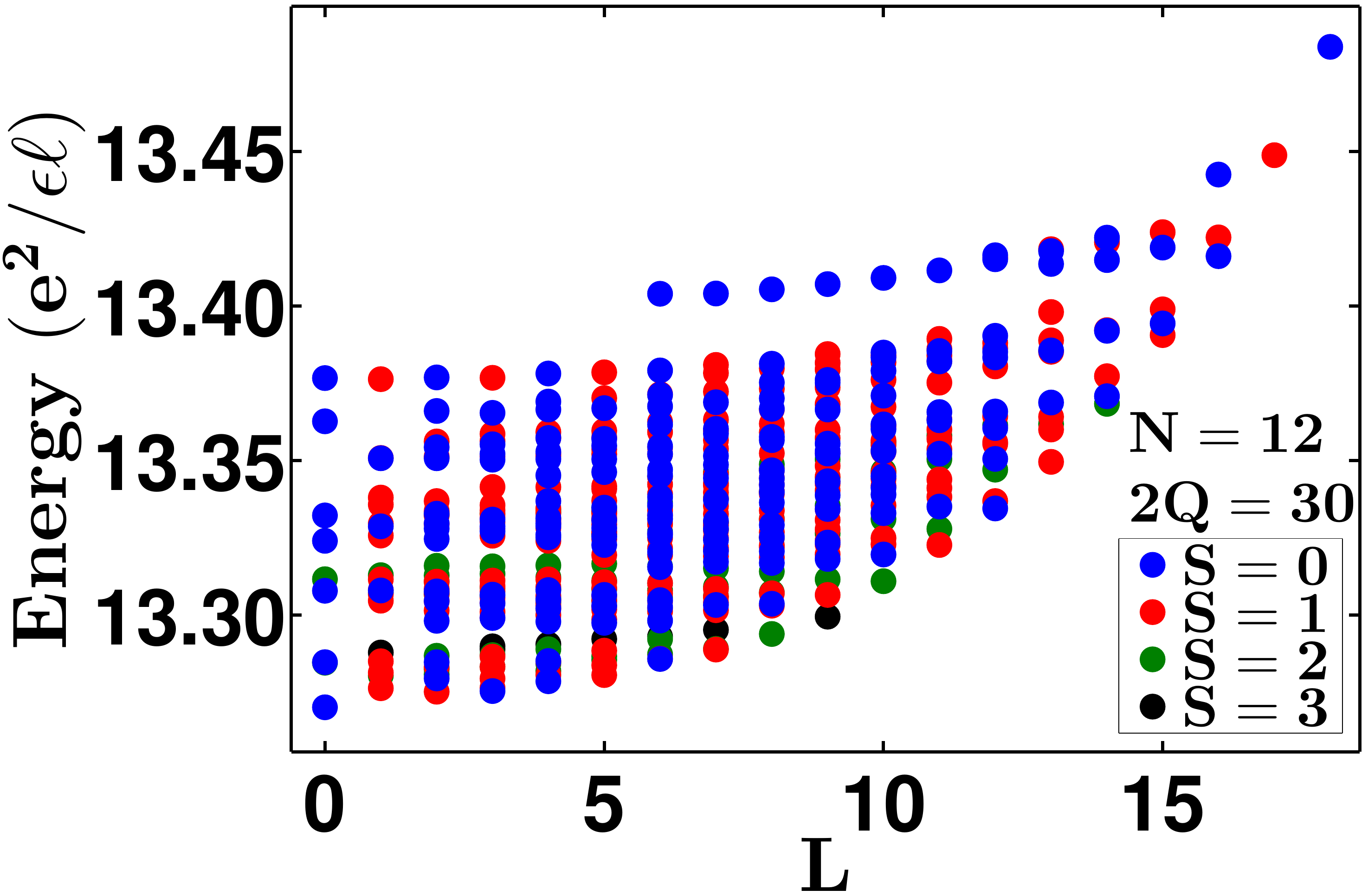}
\includegraphics[width=0.45\textwidth]{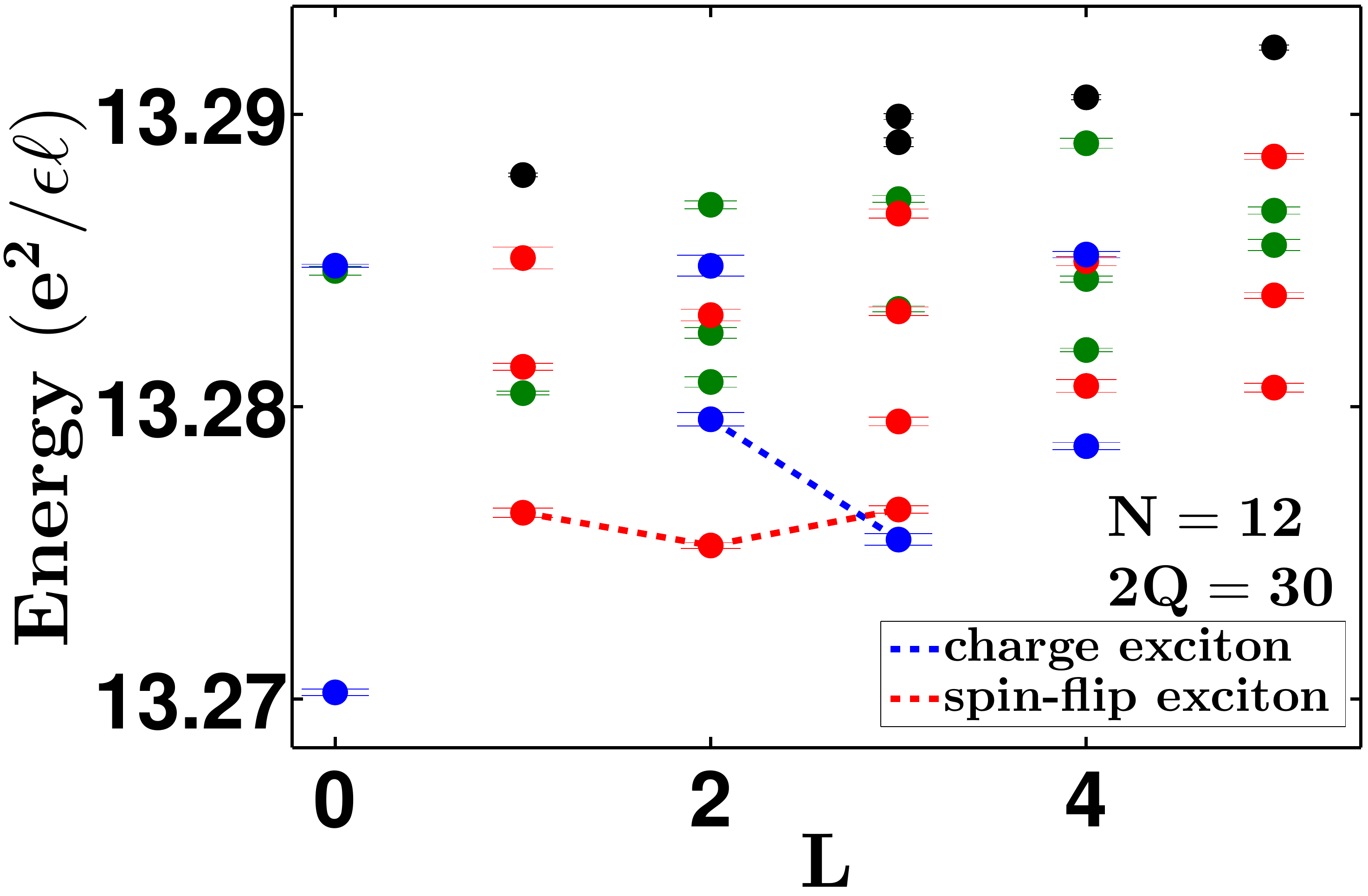}
\caption{(color online) Energy spectrum of the 4/11 spin singlet FQHE state, where composite fermions form a spin singlet 4/3 FQHE state. The spectra shown by dots (different colors representing different spin quantum numbers $S$) obtained by CF diagonalization. The dashes in the top two panels show the exact spectrum obtained by exact diagonalization in the full Hilbert space. The spherical geometry is used for the calculation. $L$ is the total orbital angular momentum, $2Q$ is the flux through the sphere in units of the flux quantum, and $N$ is the total number of electrons. The bottom panel shows the two lowest energy collective modes, namely the charge exciton and the spin-flip exciton.}
\label{fig:4}
\end{center}
\end{figure}

As stated above, the $\nu=4/11$ spin-singlet state is obtained from the $\nu^{*}=4/3$ spin-singlet state which is the particle-hole conjugate state to the 2/3 spin-singlet. The 2/3 spin-singlet state is obtained by filling the lowest $\Lambda$L of spin-up and spin-down completely and composite-fermionizing this state by reverse flux attachment. A straightforward calculation shows that the spin singlet 4/11 state occurs at flux 
\begin{equation}
2Q=\frac{11}{4}N-3
\label{4112Q}
\end{equation}
Therefore we must choose the particle number $N$ to be a multiple of 4. \\
Fig. \ref{fig:4} shows the Coulomb spectra obtained from CF diagonalization in the spherical geometry for three systems, namely $N=4, 8$ and $12$. The spectrum is obtained by constructing all $S_z=0$ and $L_z=0$ states at 4/3 at the effective flux $2Q^*=2Q-2(N-1)$, composite fermionizing them by the standard method to obtain the CF basis at the desired $2Q$ given by Eq.~\ref{4112Q}, and then diagonalizing the Coulomb interaction in that subspace by the method of CF diagonalization. (Here $S_z$ and $L_z$ are the z components of the total spin and orbital angular momentum.) For 4 and 8 particles, we also display the exact spectrum (obtained by a diagonalization of the Coulom interaction in the full LLL Hilbert space), and a comparison shows the accuracy with which the CF theory captures the low energy physics. In particular, the ground state is seen to have $S=0$ and $L=0$, consistent with a a spin-singlet FQHE state of composite fermions here. The thermodynamic limit of the ground state energy per particle is obtained by an extrapolation of the three energies. We find that even with two points (with 4 and 8 particles) we get a reasonably good approximation for the ground state energy, which is why we have used for some states that involve reverse flux attachment the exact energies only from two systems to obtain the thermodynamic energy.

It is interesting to ask if we can make quantitative predictions about the nature of the charge and spin collective modes at 4/11. As known from previous studies \cite{Dev92,Scarola00}, the collective modes are excitons of composite fermions that involve either a spin flip or no spin flip, called spin-flip excitons or charge excitons, respectively. They can be constructed up to a maximum $L$ in the spherical geometry. The CF theory predicts this value as follows. The 12 particle 4/11 state maps into 12 particle 4/3 state at flux $2Q^*=8$, which, by particle hole symmetry, is equivalent to 6 particle spin-singlet 2/3 state at $2Q^*=8$. This state, in turn, maps into 6 particle spin singlet state at filling factor 2.  The lowest energy CF exciton of the spin singlet 4/11 state are thus derived from the lowest energy exciton of the 6 particle $\nu^{**}=2$ state at flux $2|Q^{**}|=2$. The latter exciton corresponds to a particle with angular momentum 2 and a hole with angular momentum 1, giving a spin-polarized exciton at $L=2,~3$ ($L=1$ exciton is annihilated \cite{Dev92,He94}) and a spin-flip exciton at $L=1,~2,~3$. This shows that larger systems would be required for bringing out the full nature of the excitonic collective mode dispersion. Nonetheless, from the spectra, it is clear that both the charge and the spin-flip exciton modes are gapped, as expected for a spin singlet FQHE state, and that both of them have at least one roton minimum. Rotons in the charge mode were first predicted by Girvin, MacDonald and Platzman \cite{Girvin85} and spin-flip rotons have also been predicted \cite{Mandal01} and observed \cite{Wurstbauer11} for fully spin polarized 2/5 and 3/7. We note that the roton gaps for the spin singlet 4/11 are on the order of $0.01$ $e^2/\epsilon \ell$ for both charge and spin-flip exciton modes. 

\end{appendices}

{\bf Acknowledgment}
We thank M. Shayegan and Y. Liu for numerous discussions, and for sharing their data with us given in Fig. \ref{fig:3}. We acknowledge financial support from the DOE Grant No. DE-SC0005042 (ACB, JKJ), Hungarian Scientific Research Funds No. K105149 (CT), the Polish NCN grant 2011/01/B/ST3/04504 and the EU Marie Curie Grant PCIG09-GA-2011-294186 (AW). We thank Research Computing and Cyberinfrastructure at Pennsylvania State University, the HPC facility at the Budapest University of Technology and Economics and Wroc{\l}aw Centre for Networking and Supercomputing and Academic Computer Centre CYFRONET, both parts of PL-Grid Infrastructure. Csaba T\H oke was supported by the Hungarian Academy of Sciences.

\bibliography{biblio_FQHE}
\bibliographystyle{apsrev}
\end{document}